\documentclass[sn-basic]{sn-jnl}
\usepackage{amsmath,amssymb,amsfonts}

\usepackage{graphicx}
\usepackage{textcomp}

\usepackage[table,xcdraw]{xcolor}
\usepackage{array}
\usepackage{makecell}
\usepackage{graphicx}
\usepackage{hhline}
\usepackage{mathtools}
\usepackage{multirow}
\usepackage{slashbox}
\usepackage{dblfloatfix}
\usepackage{enumitem}
\usepackage{algorithm, float}
\usepackage{wrapfig}
\usepackage{tabularray}
\DeclarePairedDelimiter\floor{\lfloor}{\rfloor}
\DeclareMathOperator{\Buff}{Buff\_op}
\DeclareMathOperator{\Op}{Loc\_op}

\def\BibTeX{{\rm B\kern-.05em{\sc i\kern-.025em b}\kern-.08em
    T\kern-.1667em\lower.7ex\hbox{E}\kern-.125emX}}
    
\jyear{2021}%

\theoremstyle{thmstyleone}%
\theoremstyle{thmstyletwo}%

\theoremstyle{thmstylethree}%

\begin{document}

\title[RAMP]{RAMP: A Flat Nanosecond Optical Network and MPI Operations for Distributed Deep Learning Systems}


\author*[1]{\fnm{Alessandro} \sur{Ottino}}\email{alessandro.ottino.16@ucl.ac.uk}

\author[1]{\fnm{Joshua} \sur{Benjamin}}\email{joshua.benjamin.09@ucl.ac.uk}

\author[1]{\fnm{Georgios} \sur{Zervas}}\email{g.zervas@ucl.ac.uk}

\affil*[1]{\orgdiv{Optical Networks Group}, \orgname{University College London (UCL)}, \orgaddress{\street{Roberts Engineering Building, Torrington Place}, \city{London}, \postcode{WC1E 7JE}, \country{United Kingdom}}}




\abstract{Distributed deep learning (DDL) systems strongly depend on network performance. Current electronic packet switched (EPS) network architectures and technologies suffer from variable diameter topologies, low-bisection bandwidth and over-subscription affecting completion time of communication and collective operations.

We introduce a near-exascale, full-bisection bandwidth, all-to-all, single-hop, all-optical network architecture with nanosecond reconfiguration called RAMP, which supports large-scale distributed and parallel computing systems (12.8~Tbps per node for up to 65,536 nodes). 

For the first time, a custom RAMP-x MPI strategy and a network transcoder is proposed to run MPI collective operations across the optical circuit switched (OCS) network in a schedule-less and contention-less manner. 
RAMP achieves 7.6-171$\times$ speed-up in completion time across all MPI operations compared to realistic EPS and OCS counterparts. It can also deliver a 1.3-16$\times$ and 7.8-58$\times$ reduction in Megatron and DLRM training time respectively} while offering 38-47$\times$ and 6.4-26.5$\times$ improvement in energy consumption and cost respectively.




\maketitle

\section{Introduction}
In recent years, there has been tremendous growth in distributed deep learning (DDL)~\cite{amodei_hernandez_2018}. As the computational complexity of DDL is increasing at a faster rate than hardware improvements, the performance requirement needs to be matched by distributing such jobs to more nodes. As all of these compute resources need to be interconnected, a significant portion of the performance-scaling responsibilities resides in the network. It has been argued~\cite{khani_mehrdad_2021}~\cite{Strong_scaling} that Tbps communication between compute nodes is required to maximise the operational throughput and accommodate future applications. As most of the communications are dominated by MPI (Message Passing Interface) collective operations~\cite{mpi}~\cite{mpi_util} (sec.\ref{MPI}),
the combination of higher node capacity and better collective communication strategies could lead to lower idling time and higher operational throughput, resulting in lower neural network model training times.

Current HPC systems (e.g.~\cite{superpod}) used for DDL training use electronic packet switching (EPS) with Tbps communication (e.g. 2.4Tbps per GPU \cite{dgxa100}) at the intra-system (8-16 GPUs)~\cite{superpod} level and are limited to communicate at hundreds of Gbps regime (e.g. 200Gbps per GPU \cite{dgxa100}) for the inter-rack/inter-system level, leading to high intra-to-inter system oversubscription (e.g. 12:1 \cite{dgxa100}). This is due to network cost and power consumption constraints \cite{khani_mehrdad_2021, ballani2020sirius}. The network over-subscription forces the use of Ring-based MPI collective operations \cite{NCCL} that take full advantage of the available bandwidth but, at large scale, they lead to significant network overheads and slow down the training time of large models. For these reasons, it is critical to co-design novel Tbps all-to-all network architectures together with MPI strategies. This is required to achieve better application performances and meet future High-Performance Computing (HPC) and Data Center Network (DCN) job requirements.

Motivated by these observations, we propose RAMP, a large-scale Deep Neural Network (DNN) training system using a co-optimised nanosecond re-configurable Optical Circuit Switching (OCS)-based network, MPI strategies and network scheduling. The following are the novel aspects of this paper: 

\begin{enumerate}
\item{The co-design of four aspects: optical/opto-electronic network technologies (\textit{physical layer}), network architecture (\textit{network-level}), MPI-x communication strategy (\textit{kernel-level}) and network transcoder (\textit{system-level}).} 
\item \textit{Network level}: RAMP OCS network architecture that supports all-to-all, single hop, full-bisection bandwidth (not possible with prior architectures) communication with \textit{ramped up} (increased) end-node capacity (12.8~Tbps), scale (65,536~nodes) and system capacity (0.84~Ebps).
\item \textit{Kernel level}: We introduce custom MPI RAMP-x strategies  (x = gather, scatter, reduce, broadcast, etc.) for the proposed OCS-architecture and compare it against Ring-x, Torus-x, Hierarchical-x strategies. The strategies can be applied to any fully-connected network.
\item \textit{System level}: We develop a novel system-level Network Transcoder that maps all MPI collective operations in a schedule-less and contention-less manner to optical network configurations.
\item We analyse the scalability, cost, power consumption and computational speed-up of the proposed system/algorithms and compare it with EPS counterparts.
\item {We assess  Megatron  \cite{megatron-lm_2019} and DLRM training times with the RAMP system for different target losses and model sizes and compare them with EPS and OCS baselines.}
\end{enumerate}

\noindent We perform extensive simulations to compare RAMP with a plethora of EPS and OCS network topologies and technologies at the network level, MPI collective operations and application level with two representative large-scale distributed neural network models, Megatron and DLRM. We show that the co-design of network technology and architecture of RAMP leads to a 38-42$\times$ reduction in energy consumption and up to 12.4$\times$ reduction in cost per bit with respect to current EPS systems. RAMP's architecture offers both significantly higher node-to-node I/O capacity and full bisection bandwidth with inherent broadcast and nanosecond topology reconfiguration capability. Compared to current electronic switched systems, the RAMP system achieves an improvement from a factor of 7.6$\times$ (for reduce-scatter) to 171$\times$ (for all-to-all) in completion times across different MPI operations. The values reflect the improvement against the best-performing strategy for the best EPS and OCS topologies at maximum scale using a 1GB message size. We assess the individual contributions of network architectures, network bandwidth and MPI operations on MPI collective completion times (in sec.\ref{result}). The proposed system is able to achieve a 1.01-16.7$\times$ speed-up and 7.8-58$\times$ faster iteration time in Megatron and DLRM DDL training time, respectively, as well as 23.8-85 percentage points decrease in communication time contribution when compared to the OCS and EPS baselines for maximum scalability systems. 

\section{Background and motivation}
\noindent Network and interconnect technologies are limiting factors for the performance of High-Performance Computing (HPC) systems. Recent benchmarks showed that only 5\% of peak compute performance of HPC systems can be achieved on practical workloads and network overhead represents one of the main bottlenecks \cite{hpcg-sc-21, hpcg-next-platform}. Considering that improvements in hardware and application requirements are growing at a faster pace than interconnect technologies \cite{scaling_models_moores, ballani2020sirius}, the network overhead becomes the significant bottleneck in the system.

\label{sec:background}
\begin{figure}[t]
    \centering
    \includegraphics[width=0.8\linewidth]{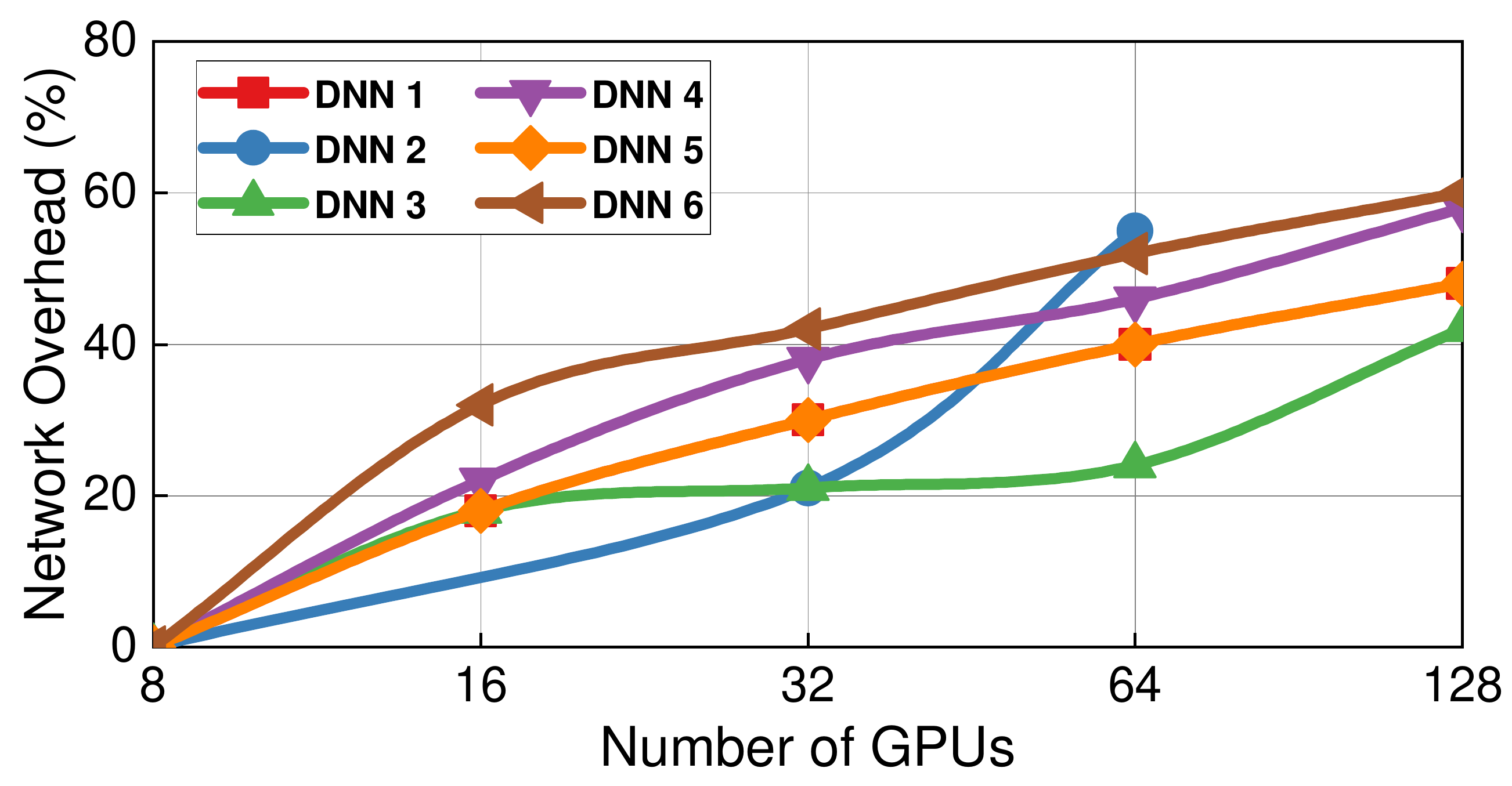}
    \caption{Network overhead of different large-scale Deep Neural Networks with a different number of devices. Figure taken from \cite{TopoOpt_2022}}
    \label{fig:DNN_overhead}
\end{figure}

Previous research has shown that training large-scale deep neural network (DNN) models lead to significant network overhead (40-60\% even for using as few as 128 nodes) \cite{TopoOpt_2022}, Fig.\ref{fig:DNN_overhead}.
The demand for bigger and more predictive models is increasing at an exponential rate; the number of neural network parameters is doubling every 3-6 months. This forces the use of more workers and, in turn, leads to higher network overheads on training times. Two key factors that affect the communication/network overhead are the DDL partitioning strategy (sec.\ref{partitioning}) and the DDL scaling method (sec.\ref{scaling}) used.

At the core of these approaches are the MPI collective operations, which are a set of operations needed to perform data processing across multiple devices sec.\ref{MPI}. The network overheads are caused by the use of over-subscribed and low-bandwidth EPS networks that also force the use of sub-optimal strategies across a range of MPI collective operations (e.g all-gather, all-to-all). Therefore, the network bandwidth, network connectivity and MPI strategies of current systems must significantly improve in order to unlock the scalability and performance of DDL systems. 

Limitations of current EPS systems and proposed OCS architectures will be explained in sec.\ref{EPS_limitation} and sec.\ref{OCS_limitation} respectively.




\subsection{DDL partitioning methods}
\label{partitioning}
\noindent There are three main methods for partitioning a DL job: Data Parallelism (DP), Model Parallelism (MP) and Hybrid Parallelism (HP), which is a combination of DP and MP. 
\noindent\textbf{DP} consists of replicating the same model on multiple workers, where each worker processes a different set of data (local batch). In this way, the overall system can process a larger global batch while keeping the iteration time of each worker constant. At the end of each training step, the workers must share their weight updates (gradients).  Using larger batch sizes reduces the number of iterations required to converge, reducing the overall training time.

\noindent\textbf{MP} consists of partitioning the DNN model between workers who act on the same batch of data. This requires multiple activation/gradient communications between workers within a training iteration. The partitioned DNN model generates multiple different local ML computations for each worker (partitioned local computational graph). When dealing with large models, MP is required as it reduces the memory footprint for a single worker.

\subsection{Bandwidth Requirements for DDL applications}
\label{scaling}
There are two main ways to scale DDL jobs: weak and strong scaling.

\noindent\textbf{Weak scaling} aims at increasing the throughput per training iteration in terms of samples/sec by increasing the number of workers. This is mainly performed using Data Parallelism (DP), which ensures a constant computation and communication time with scaling. These properties make it the most commonly chosen distribution technique, as it does not require an increase in node interconnect bandwidth at scale as the network overhead is approximately constant with the number of workers. This means that, when keeping batch size per worker large, Gbps per second communication might suffice to handle the partitioning, making the weak scaling approach a feasible option for current oversubscribed EPS HPC/DDL systems. However, this approach cannot scale indefinitely as the number of iterations to accuracy does not always decrease with the increase in global batch size \cite{shallue2018data_parallelism_limit}, leading to inefficient training. In addition, the memory footprint of the model does not decrease with scale, making the technique unfeasible to be used alone for large models, which are the main drive for application growth. 

\noindent\textbf{Strong scaling} is decreasing the overall time per iteration. It is mostly achieved through Model Parallelism (MP). Achieving strong scaling is fundamental for the development of large models as it is capable of reducing the memory footprint of DNNs \cite{khani_mehrdad_2021, TopoOpt_2022}. Achieving strong scaling is challenging because increasing the number of workers working on a specific batch leads to a decrease in computation time, which in turn, leads to more frequent communication steps involving the transmission of constant/increasing messages. This requires a super-linear increase of network bandwidth (multiple Tbps per node) and low latency communication. For this reason, current EPS systems can efficiently support strong scaling only within a single node that hosts 8-16 devices (without incurring significant network overhead). 

\begin{figure}
    \centering
    \includegraphics[width=\linewidth]{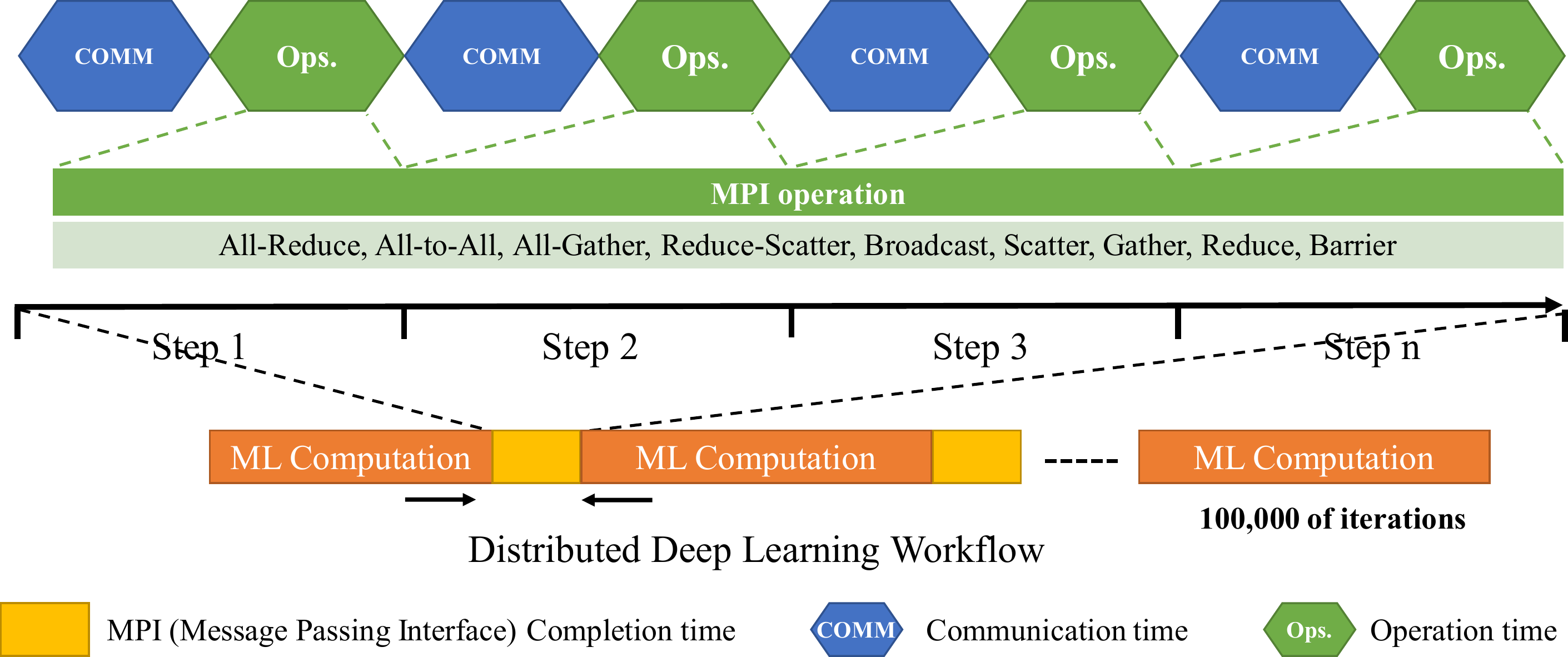}
    \caption{Example of DDL process, involving computation and MPI operations.}
    \label{fig:MPI-ML}
\end{figure}

\subsection{MPI operations for DDL}
\label{MPI}
Both scaling approaches for distributed training rely on MPI collective operations. MPI operations correspond to a series of communication steps between multiple workers followed by local operations with the goal of performing a distributed task. Depending on the strategy selected (e.g. Ring-based strategies), it is possible to (partly) overlap computation and communication to decrease completion time. A DDL workload, as shown in Fig.\ref{fig:MPI-ML}, can be viewed as a sequence of local ML computations (local partitioned computational graph) and MPI operations between the workers to share information. Overlap between MPI communication and computation, when application logic permits, leads to higher operational efficiency, however, for strong scaling scenarios this is unlikely. For example, in most MP cases, there is data dependency between subsequent operations (layers). In weak scaling, the weight update information sharing is represented by an all-reduce operation between workers.  In strong scaling, different models require different types of activation sharing depending on their architecture and partitioning strategy, leading to different MPI operations for different models. The most widely used are all-reduce (e.g. Megatron \cite{megatron-lm_2019}), all-to-all (e.g DLRM \cite{naumov2019DLRM}, Switch Transformer \cite{switch_transformer_2021}) and all-gather (MoE \cite{MoE_1991}). Different collective operations have different requirements on the network. Operations such as all-reduce and all-gather, allow effective communication on oversubscribed and limited connectivity systems (sec.\ref{sec:architectural_comp}). On the other hand, data-intensive operations such as all-to-all benefit from full bandwidth connectivity between all devices (sec.\ref{sec:architectural_comp}). For this reason, high and full-capacity connectivity between all device pairs is needed to minimize collective completion times.

\subsection{Limitation of EPS systems}
\label{EPS_limitation}
 One possible solution to achieve strong scaling and meet the network requirements of fast-growing applications consists in increasing the capacity of EPS systems. However, this approach leads to serious challenges.
 Increasing the capacity of EPS in terms of I/O bandwidth and transistor density is becoming hard to sustain due to physical limitations\cite{ballani2020sirius}. It has been argued that the cost and power of switches are unlikely to stay constant above two generations, resulting in an increase in cost and power and eventually hitting a wall in capacity with standard EPS \cite{khani_mehrdad_2021}. 
 Higher capacity switches may still be created by hierarchical/Clos construction of smaller ASICs, at the expense of cost, power and complexity \cite{ballani2020sirius}.
 Another option is creating parallel networks by replicating copies of EPS systems to increase the overall bandwidth. However, this type of solution leads to unsustainable power and cost as shown in section \ref{power_sec}.
 To allow higher capacity systems, limited connectivity networks, such as Toruses and Meshes (e.g. Google TPU Pod \cite{TPU_pod}), have been proposed. However, these approaches lead to high-diameter topologies and inefficient node bandwidth and resource utilisation.
 
\begin{figure}[h]
    \centering
    \includegraphics[width=\linewidth]{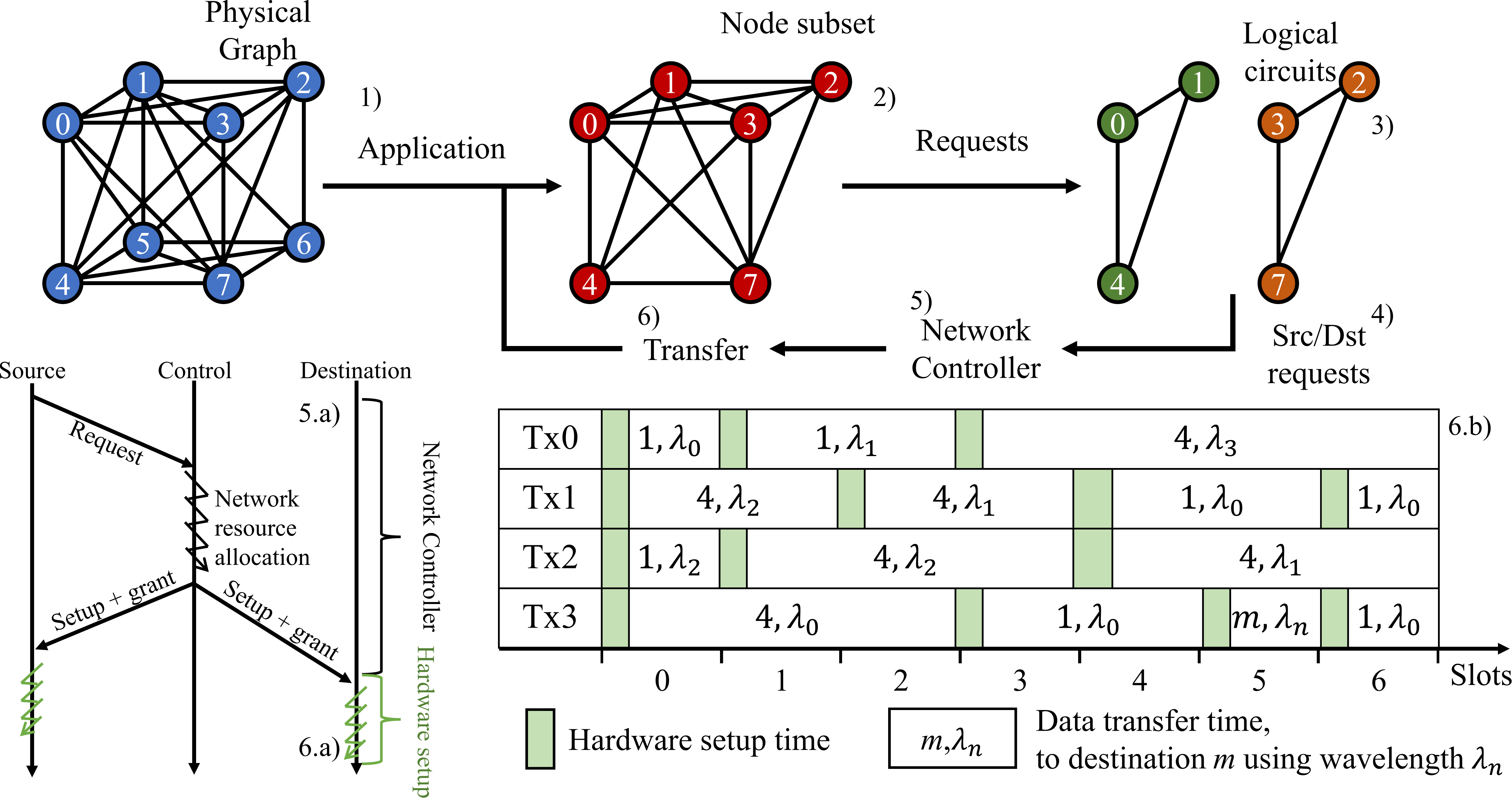}
    \caption{Diagram representing the process to accommodate an OCS request, with transmission example by the different transceivers of node 0 for the described logical circuit.}
    \label{fig:OCS diagram}
\end{figure}

\subsection{OCS for Distributed Deep Learning (DDL) systems}
\label{OCS_limitation}
Optical interconnects are capable of achieving high bandwidth low latency communication at scale and are a promising technology to meet the future requirements for DDL and HPC jobs. 

Packet switching following one-way reservation (in the form of EPS) is the underlying transportation method used in DCNs and HPCs. This is due to the fact that it allows the creation of highly scalable networks, distributes its forwarding rules to each node, handles diverse flows due to variable packet size structures and has the ability to queue and manage flows along the path \cite{Zervas_Pulse}. However, packet-switched networks require complex control methods such as admission and congestion, buffer/queue management and complex addressing \cite{Zervas_Pulse}. While these can be handled by electronic chips such as ASICs in EPS systems (with the limitations described in sec.\ref{EPS_limitation}, the equivalent is not possible in optical networks. This is due to the fact that optical technologies are not able to replicate such functionalities due to limited and rigid data processing capabilities (limited optical computing performance \cite{Zervas_Pulse}) and storage (lack of high-capacity and high-bandwidth enough photonic memories \cite{optical_memory}) \cite{Zervas_Pulse}. For these reasons, Optical Packet Switching (OPS) cannot be considered a feasible solution for HPC and DCN applications. Circuit switching, which was widely used in electronic networks before EPS \cite{benes_cs8}, can deliver guaranteed and deterministic traffic without the need for any of the packet-switching complex methods \cite{Zervas_Pulse, Benjamin:20}, making it compatible with optical technologies.

Optical circuit switching (OCS) has been proposed as a possible replacement for EPS networks for DCN and HPC systems. This type of system decouples the data to the control plane and it could lead to a significant reduction in cost and energy consumption while being capable to scale to high bandwidth and low deterministic latency.

In OCS, the control and scheduling are separate from the data, leading to different operations compared to EPS networks. A flowchart describing the transmission for OCS systems is shown in Fig.\ref{fig:OCS diagram}. As for EPS, the 1) application is placed onto a subset of nodes of the physical graph and during its lifetime will generate 2) communication requests between one or multiple workers. However, these requests are handled differently, as they are first converted into 3) logical circuits. The logical circuits contain all the sources and destinations to which the requesting node is involved and represent the overall flow of information in discrete time periods. The logical circuits generate a set of 4) source-destination requests to perform the communication. These set of point-to-point requests are sent to a 5) network controller which allocates the network resources in a synchronous manner. All devices have to transmit synchronously, in a discrete period of time called timeslots. The controller allocates the transceivers, path, wavelength and timeslots to allow communication (5.a). Once the resources are allocated, the setup instructions are sent to the source and destination and their physical system 6.a) is to be reconfigured. The time taken for slots and resource allocation by the network controller (5.a) can be pipelined and does not affect the throughput of the communication, as different data for previously scheduled requests are transmitted in parallel. The time taken for the hardware to configure to allow the transmission is called hardware reconfiguration/setup time or switching time (6.a). During hardware setup no data can be transmitted, effectively limiting the throughput. An example of how the communication looks like for multiple nodes is shown as 6.b) in Fig.\ref{fig:OCS diagram}, where the timeslots, path (destinations) and wavelength have been allocated for different transceivers for node 0 for the example logical circuit. The green shaded region represents the hardware reconfiguration time and the white region is the payload transmission time. It needs to be noted that, due to the discrete and synchronous nature of OCS, the overall timeslot might not be fully utilised for transmission. The total number of concurrent transfers per path in an optical network depends on the total amount of wavelengths.

In spite of the multiple advantages that OCS brings, there are significant challenges that prevent the implementation and adoption of these systems for large-scale DCN and HPC. 
\begin{enumerate}
    \item One of the main challenges is the control plane, as it is now separate from the data. Fast large-scale scheduling with low latency and high throughput for the traffic is challenging at a large scale, leading to multiple proposals to use schedule-less systems \cite{ballani2020sirius}. However, some promising results have been shown in the development of large-scale scheduling systems \cite{Benjamin:20}.

    \item Fast hardware setup time is required to allow efficient data transmission. In fact, the timeslot duration and minimum message size are directly dependent on it. The timeslot duration needs to be significantly larger ($>10\times$) than the circuit reconfiguration time to limit the overhead. This affects the utilisation, latency and goodput of the network.

    \item Synchronisation of transmission for all devices is required to avoid contention. Recently proposed systems have been shown to be capable of synchronising thousands of devices for OCS \cite{rabbit}.

    \item Change in paradigm to existing networks is required. Implementing OCS systems would require a significant change in the communication stack of existing systems, by removing all communication protocols and switch-centric network representation.
    
\end{enumerate}

However, due to the previously discussed limitations of EPS, OCS systems need to be explored and these challenges resolved to meet future HPC and DCN application requirements.

\subsection{Previous OCS limitations}

\noindent Recently, there have been other OCS network architectures proposed for HPCs and DCNs.
However, none of the previously proposed systems is capable of meeting all the requirements necessary for HPC and DDL applications.
Between these systems, two architectures based on 3D-MEMS/patch panels OCS have been proposed: TopoOpt \cite{TopoOpt_2022} and SiP-ML OCS \cite{khani_mehrdad_2021}. These topologies might allow full-bandwidth connectivity between nodes at large scale (1024 and 384 directly connected nodes). However, they are affected by long circuit reconfiguration times, defined by the switching technologies ($>$10ms and $>$s for 3D-MEMS and patch panels respectively). These properties make in-application circuit reconfiguration unfeasible as it would lead to significant overhead and they require static circuit pre-allocation \cite{TopoOpt_2022}. 
For this reason, for each logical circuit in the application life span, a physical path should be allocated. This limits the effective bandwidth available at any time between device pairs and limits the number of devices that a node can be connected to (communication degree). This might lead to low bandwidth communication and high-diameter logical and physical topologies, which lead to inefficiencies. To mitigate these problems for DDL applications, \cite{khani_mehrdad_2021, TopoOpt_2022} developed custom partitioning methods which consider these physical limitations. In \cite{TopoOpt_2022}, it has been argued the need for fast circuit reconfiguration networks for DDL applications.

Fast circuit reconfiguration OCS systems have been previously proposed for DCN and HPC applications. 
For HPC and DDL applications SiP-ML Ring \cite{khani_mehrdad_2021} and TeraRack \cite{terarack} have been proposed. These systems use Micro Ring Resonators (MRRs) which lead to $\sim$25$\mu$s reconfiguration delay, which could allow in-application dynamic circuit reconfiguration while allowing high-capacity node-to-node connectivity. However, both of these systems are limited by scale and connectivity. In fact, both of these topologies are realised through a wavelength selection add and drop ring connectivity, which limits the number of nodes available in the system to the number of available wavelengths available for the transceiver (namely 256 \cite{terarack}). In addition due to the optical ring's physical topology (which allows communication with all nodes in the ring without passing through the neighbouring compute hops), the number of optical hops and network components between node pairs increases linearly with the diameter between node pairs, significantly varying the signal quality between source-destination pairs. The signal quality effectively determines the connectivity of the system, limiting the connectivity of a node to its 16 closest neighbours. Due to this property, all-to-all connectivity is not achievable and complex multi-stage control and routing need to be implemented. In addition, these topologies are largely affected by a single point of failure, where if a single hop is malfunctioning all communication passing through it is lost.

For DCN, multiple architectures which allowed large circuit reconfiguration at high scalability have been proposed. Between these, PULSE~\cite{Benjamin:20} and Sirius~\cite{ballani2020sirius} are OCS systems which allow ns-speed switching and large scalability ($>$10,000 nodes for PULSE, not specified for Sirius). While these systems meet the requirements in terms of scalability and reconfiguration, they are limited by node-pair capacity. In fact in these networks, each node pair is connected by a single transceiver, which limits the node-to-node capacity. In addition, these networks suffer from a single point of failure, where a transceiver or network component malfunctioning leads to one or multiple nodes being unable to communicate to racks and/or clusters. Moreover, the single transceiver connectivity for a rack of devices limits the devices each node can communicate with at the same time. In these types of networks, another main challenge is scheduling. PULSE demonstrated a promising scheduler that could handle thousands of devices while being reliable to skewed and varied traffic \cite{Benjamin:20}, but it is not capable to manage deterministic, long-lasting patterns such as the ones for collective operations. Sirius, on the other hand, uses a scheduless round-robin approach for communication \cite{ballani2020sirius}, with limited transmission epoch duration, which makes it unsuitable for skewed and large flow traffic.

In addition, the MPI collective operation strategies considered by these OCS architectures (\cite{TopoOpt_2022, khani_mehrdad_2021}) have been developed for EPS systems without taking into consideration the network architecture characteristics. MPI collective operations for OCS systems should take into consideration the physical properties of the network (e.g. path, wavelength) such that the number of algorithmic steps and completion time is minimised while avoiding contention. This also synchronous scheduling and communication to take advantage of the deterministic latency properties. These characteristics are not present in EPS systems, as each packet can be sent to any destination at any time, without pre-determined circuits and latency (which both depend on the switch buffering and forwarding) \cite{Zervas_Pulse}.



\section{{RAMP} Architecture}
\label{arch}

We introduce the RAMP architecture --- the first large-scale, high-capacity, full bandwidth architecture for DCN and HPC/DDL systems. It provides
\begin{enumerate}
    \item High-capacity communication between node pairs ($>$12.8Tbps), making it suitable for HPC and DDL application requirements.
    
    \item High scalability ($>$4096 nodes). Capable of handling increasingly complex workloads.
    
    \item Nanosecond level circuit reconfiguration through wavelength switching and broadcast-and-select space switching. The system takes advantage of Time-Division Multiplexing (TDM), Wavelength-Division Multiplexing (WDM) and Space-Division Multiplexing (WDM). This allows each node to communicate to any other node with virtually no communication degree constraints; allows using collective operations with logical graphs with significantly lower diameters without sacrificing bandwidth~\cite{TopoOpt_2022}; allows the proposed architecture to handle fast-changing circuits which are required for DCN traffic.
    
    \item Port-level all-to-all connectivity and re-arrangeable or strictly non-blocking communication. Any transceiver can transmit/receive information to/from any node. Communication blocking probability depends on the selection of the sub-network only.
    
    \item Fully passive interconnect system. Removing complexity from the core of the network and moving it to the edge. 
    
    \item Unrestricted multi-node communication and reliability, without any single point of failure. Every node can talk to every other node using multiple possible paths, and any failure for transceivers/network components still allows all-to-all communication just at a slightly decreased capacity.
\end{enumerate}




\noindent These properties make RAMP the first architecture suitable for both HPC and DCN systems.

\begin{figure}[h]

    \centering
    \includegraphics[width=0.95\textwidth]{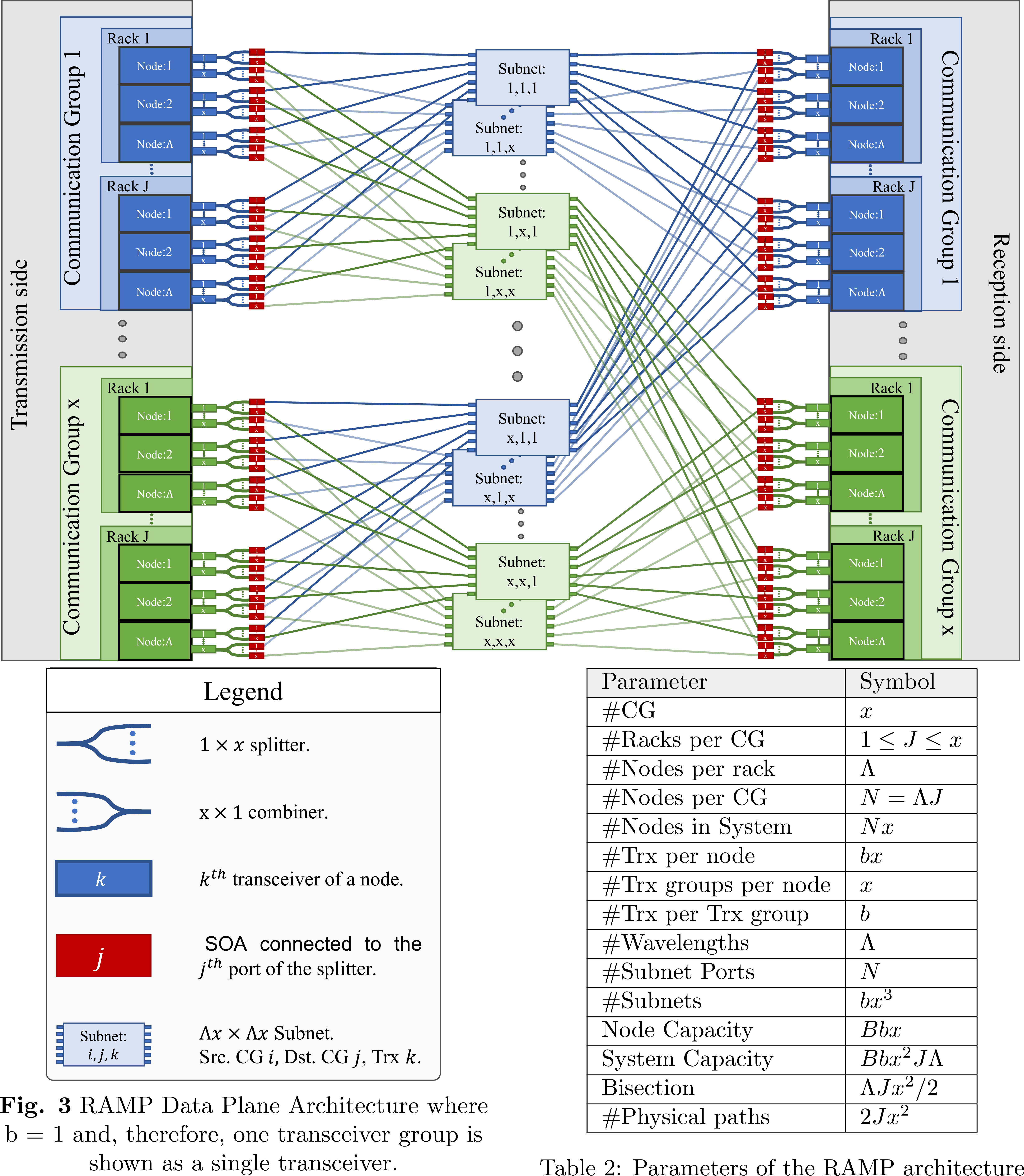}

    \label{fig:Ramp-visual}

\end{figure}
\setcounter{figure}{3}

\subsection{RAMP Data Plane}

\noindent The RAMP data plane consists of parallel subnets arranged in communication groups and transceivers. There are $x$ communication groups, where each group contains $J$ racks. The maximum number of racks per communication group is $J = x$. Each rack contains $\Lambda$ devices or nodes, where $\Lambda$ is the total number of available wavelength channels. Hence, the maximum number of nodes in one communication group is $N=\Lambda x$. Each node is equipped with $x$ transceiver groups, each containing $b$ transceivers sharing the same tunable light source and the same control instruction but different spatial planes. An increase in the $b$ parameter leads to an increase in node capacity for the same control complexity and reduced energy per bit at the cost of a larger number of components. Further explanation in sec.\ref{sec:scalability}. Each transmitter is connected to a $1:x$ splitter, creating $x$ possible paths per transmitter. Each path is selected by activating the Semiconductor Optical Amplifier (SOA) \cite{Alkharsan:22} attached to each port of the $1:x$ splitter and connected to a different sub-net and therefore, a different communication group. In this way, each transmitter is able to communicate with every communication group. Each receiver is connected to a $x:1$ combiner so that each receiver can receive information from every communication group. Under the proposed network configuration, the $i^{th}$ transmitter of any node can send information to the $i^{th}$ receiver of every node, enabling all-to-all transceiver-wise communication. The topology requires a total of $bx^3$ sub-nets, i.e. a sub-net for a communication group pair per transceiver. Each node is equipped with a single NIC handling all transceivers.

As shown in Table II in Fig. \ref{fig:Ramp-visual}, the RAMP architecture scales up to $\Lambda x^2$ nodes, providing a total capacity of $b B \Lambda x^2 $, where $B$ is the effective line rate of each transceiver. The bisection bandwidth is $\Lambda J x^3/2$, the total number of fibres is $2 b Jx^3$ (same as for the  PULSE architecture $2 x^4$ \cite{Benjamin:20}), and the total number of physical links required is $2Jx^2$ (compared to $2 x^3$ of PULSE \cite{Benjamin:20}), as paths can be grouped/ribboned by racks and source-destination communication groups. Source-destination selection and circuit reconfiguration are performed through path/transceiver, wavelength and time-slot mapping. 

There are three possible choices for the subnet: (i) a simple star coupler with $N$ ports (Broadcast and select, B\&S), (ii) $J$ parallel $\Lambda \times \Lambda$ arrayed waveguide gratings (AWGRs) followed by $\Lambda$ parallel $J \times J$ star couplers mixing information between same ports of each AWGRs (Route and Broadcast, R\&B) or (iii) the same AWGRs followed by SOA based $J \times J$ crossbars switch (Route and Switch, R\&S).

\begin{figure}[h]
    \centering
    \includegraphics[width=\linewidth]{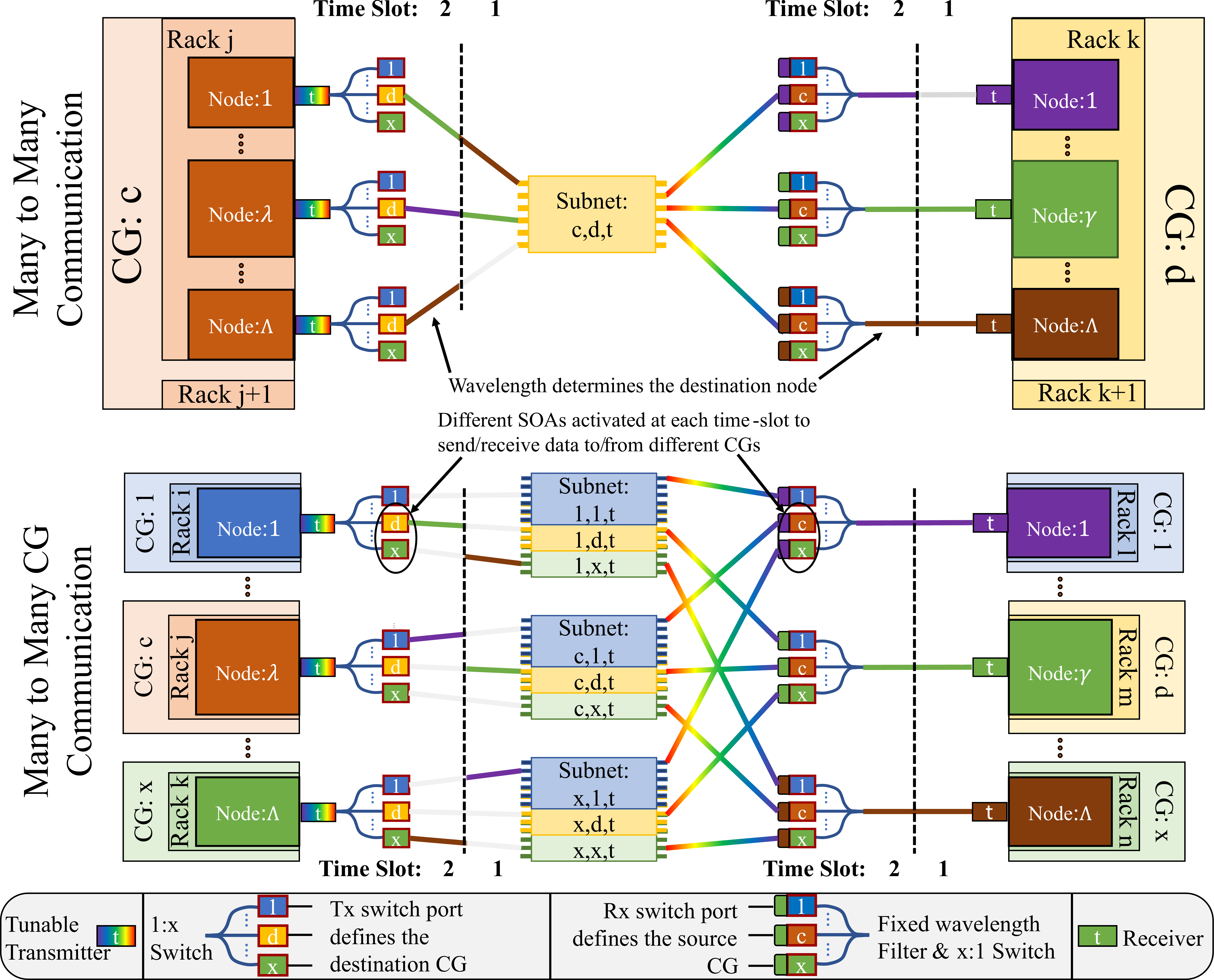}
    \caption{Example of a many-to-many communication pattern across different time slots between nodes of a) same source-destination communication group pairs and b) different source-destination communication group pairs. Showcase the WDM, TDM and SDM (for different communication groups) principles of RAMP). The colour of the line represents the wavelength used to communicate at each time slot (1 \& 2). The color of the line used at the transmission side for each timeslot matches the colour of the destination node and receiver line. This way the colour/wavelength maps source and destination for each node-pair (and transceiver) at any timeslot.}
    \label{fig:many2many}
\end{figure}

\begin{figure}[t]
    \centering
    \includegraphics[width=\linewidth]{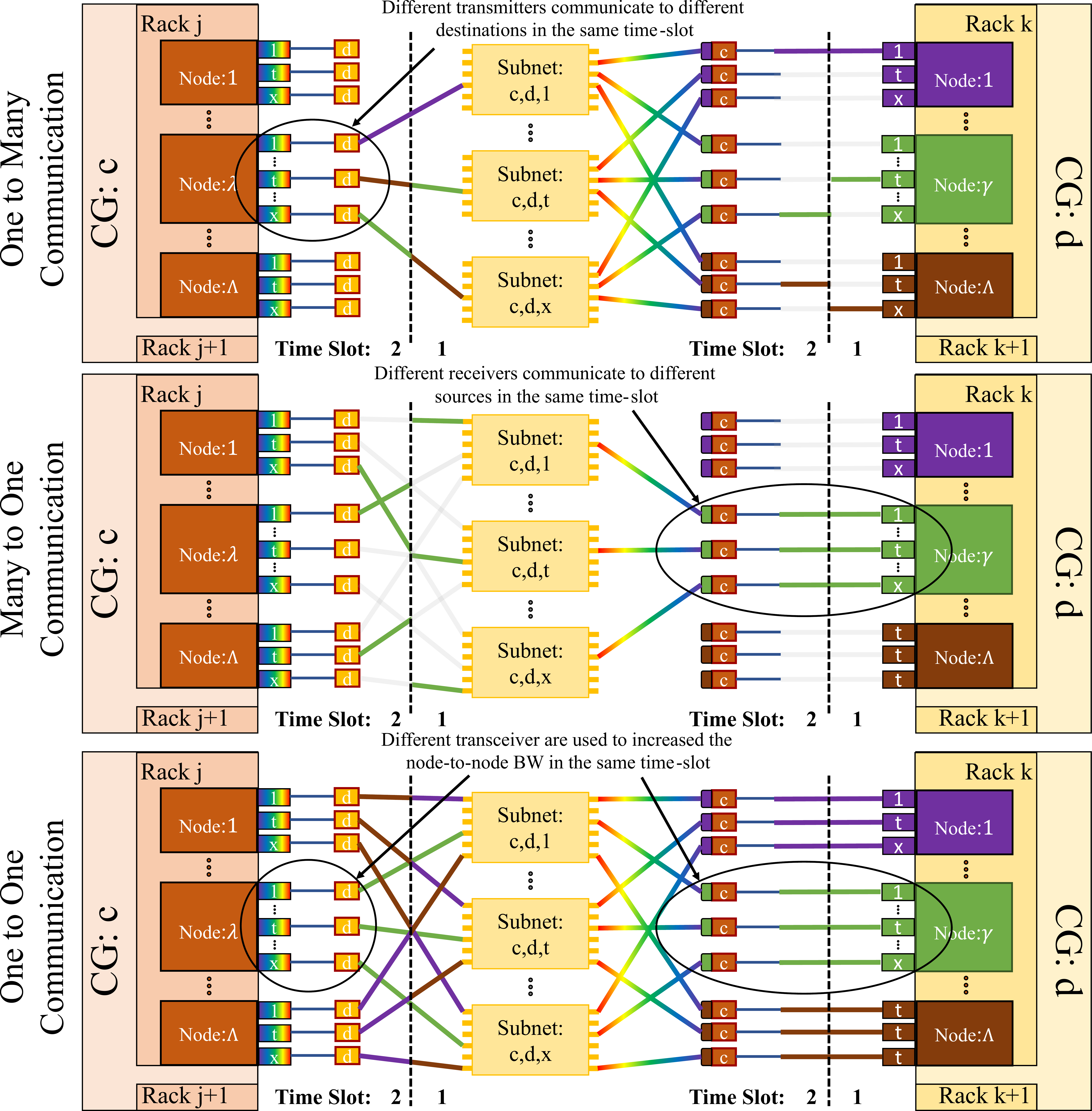}
    \caption{Example a) one-to-many, b) many-to-one and c) one-to-one communication patterns at each time slot between nodes with same source-destination communication group pairs. It shows the WDM, TDM and SDM (across multiple transceivers) principles of RAMP, therefore allowing high bandwidth (up to full capacity) communication between one or multiple node-pairs or sets.}
    \label{fig:one2many}
\end{figure}

Figures Fig.\ref{fig:many2many} and Fig.\ref{fig:one2many} show how the RAMP architecture handles different communication patterns. It needs to be noticed that in both Figures the considered configuration of the RAMP architecture is a fixed receiver Broadcast \& Select (B\&S). 

In Fig.\ref{fig:many2many} the many-to-many communication pattern in multiple time-slots across multiple sources and destinations within a) single source-destination communication groups pair and b) multiple communication groups are shown. For Fig.\ref{fig:many2many}.a) communication between multiple source nodes ($1, \lambda$, $\Lambda$) of rack $j$ and communication group $c$ and destination nodes ($1, \gamma, \Lambda$) of rack $k$ and communication group $d$ by using the $t^{th}$ transceiver. At the transmission side each node has a tunable transmitter followed by a $1:x$ space switch (implemented by an SOA gated splitter), whereas at the reception side each receiver is preceded by a filtered (single wavelength) $x:1$ switch (SOA gated coupler), making it fixed receiver. Each node in a rack receives at different wavelengths represented in both Fig.\ref{fig:many2many} and Fig.\ref{fig:one2many} by receiving node, receiver and filter colour. Between the communication group pairs ($c-d$) for the $t^{th}$ transceiver exists the single subnet: $c,d,t$ which allows communication between all transmitter $t$ of all source nodes in communication group $c$ and all destination nodes of communication group $d$. To perform the communication and transmit through the correct subnet the correct port of the switches need to be selected at both the transmission and reception side. At transmission, the switch port corresponds to the destination communication group (port $d$ is used to communicate to the $d^{th}$ communication group) and at reception the source destination group. For both Fig.\ref{fig:many2many} and Fig.\ref{fig:one2many}, the colour of the transmission switch port and subnet matches the one of the destinations communication group, and similarly, the colour of the receiving switch port matches one of the source communication groups which the ports receive from.

At each Time Slot, each node set its destination by selecting its receiving wavelength, as shown at the transmitting side of Fig.\ref{fig:many2many}.a) where the transmitting node ($c, j, \lambda$) sends info to node ($d, k, \gamma$) and ($d, k, 1$) by choosing wavelength $\gamma$ and $1$ for time slots $1$ and $2$ respectively. In each subnet, due to the broadcast principle, each active wavelength is available at each output port (represented by the rainbow colour in Fig.\ref{fig:many2many} and Fig.\ref{fig:one2many}), the correct for each destination is recovered by the filter before each port of the $1:x$ switch. For both time slots as the communication group pair of the source and destination, nodes is constant, the ports $d$ and $c$ of the transmission and reception side switches respectively are selected. In a similar fashion, node ($d, k, \gamma$) receives from nodes ($c, j, \lambda$) and ($c, j, 1$) in different time slots have been tuned their transmitter at the $\gamma^{th}$ wavelength.

Fig.\ref{fig:many2many}.b) shows a similar many-to-many pattern between different nodes ($1, \lambda, \Lambda$ for tx and $1, \gamma, \Lambda$ for rx) in different racks ($i,j,k$ for tx and $l,m,n$ for rx) of different communication groups ($1, c, x$ for tx and $1,d,x$ for rx). Each pair of communication groups is connected by a subnet, accessed through a specific source and destination switch port selection. As in Fig.\ref{fig:many2many}.a) the node selection in a rack is performed through wavelength selection for every time slot whereas different communication groups are accessed by gating different ports of the transmission and reception side switch. In the figure, node ($c, j, \lambda$) communicates to nodes ($d,m,\gamma$) and ($1,l,1$) in different time slots by selecting wavelengths $1, \gamma$ and gating the ports $d, 1$ and $c, c$ for transmission and reception side switches respectively in each time slot. Different switch port pairs selection at each time slot lead to different communication group communication allowing effective port-level all-to-all communication with fast reconfiguration.

Fig.\ref{fig:one2many} shows different communication patterns per same time-slot: a) one-to-many, b) many-to-one and c) one-to-one.
For all the communication patterns Fig.\ref{fig:one2many} depicts the communication between multiple source nodes ($1, \lambda$, $\Lambda$) of rack $j$ and communication group $c$ and destination nodes ($1, \gamma, \Lambda$) of rack $k$ and communication group $d$ by using multiple transceivers. 

Fig.\ref{fig:one2many}.a) shows the one-to-many communication pattern from the source node ($c,j,\lambda$) to all the nodes of communication group $d$ rack $k$. Each transceiver of the source node transmits in the same time slot to different destinations by selecting different wavelengths. If the destinations would have been in different communication groups different transmission and destination switch ports would have been selected for each time slot, similarly to Fig.\ref{fig:one2many}.b).

Fig.\ref{fig:one2many}.b) shows the many-to-one communication pattern, where the destination node ($d, k, \gamma$) receives at the same time from multiple destinations by using different transceivers.

Fig.\ref{fig:one2many}.c) shows multiple one-to-one communication patterns between different source pair destinations. In this figure, all transmitters of each source node are used to communicate to all receivers of the same destination node, such that full-capacity communication between node pairs is used at any time slot. It needs to be noticed that also only a subset o transceivers can be used between node pairs depending on the application requirements.

The described principles can be used at the same time to adapt the network requests and they are extensively used together for collective operations. It needs to be noted that in both Fig.\ref{fig:many2many} and Fig.\ref{fig:one2many} rack selection has not been performed. This is due to the fact that the signal between nodes with the same node number of different racks is coupled together, broadcasting the same information to all racks. This effectively creates contention in each subnet, however, the multiple paths between each source-destination pair allow communication to be re-arrangeably non-blocking, and when correctly scheduled up to full bandwidth.

It needs to be noted that both in Fig.~\ref{fig:Ramp-visual}, Fig.~\ref{fig:many2many} and Fig.\ref{fig:one2many} we show the architecture with $b=1$, so the case when a transceiver group is equivalent to one transceiver. 

It is possible to create an equivalent Electrical Circuit Switched (ECS) RAMP architecture by replacing each subnetwork with a $\Lambda J \times \Lambda J$ electrical switches and increasing the total amount of transceivers to $bx^2J\Lambda(1+x)$. However, this approach would require the use of high-capacity switches, it would incur electro-optical conversion which would increase energy consumption and increase the cost (number of transceivers and more expensive switches) and inefficiencies as only a portion of the transceivers could be active at all times. These characteristics make an electrical version of RAMP over-provisioned and cost-ineffective. Transforming the ECS to EPS switches and transceivers would further increase costs and power consumption (explained in sec.\ref{EPS_limitation}).

\subsection{Dynamic Traffic}
\label{sec:dynamic_traffic}
While the focus of this work is on the deterministic traffic of MPI collective operations and the co-design of architecture and collective strategies (described in sec.\ref{sec:ramp_engine}), it is important to note that different types of traffic can also be handled. In fact, dynamic traffic is prevalent in DCN and HPC applications and secondary for DDL workloads. The RAMP architecture has been designed such that it is compatible with previously proposed OCS scheduling approaches such that the dynamic nanosecond level circuit scheduler presented in PULSE \cite{Scheduler-PULSE}. This scheduler proved to deliver skew-tolerant performances whilst achieving above 90\% throughput and tens of millisecond tail latency, making it suitable for DCN and HPC applications. However, to allow this scheduler to work on a RAMP architecture, we need to limit the connectivity such each transceiver is mapped to a specific rack and therefore limiting the effective node-to-node capacity. A novel scheduler which uses the multi-path and full-capacity capabilities of RAMP whilst taking into consideration the sub-network contention is currently under development. 


\section{Optical Technologies}
\label{technologies}
In this section, we introduce the optical technologies powering the proposed architecture describing the operating principles of the system. These technologies have been experimentally demonstrated in prior art. The components, scalability and operating properties (cost and power consumption) are discussed. Further, a more detailed explanation, with in-depth optical properties analysis and architectural features and characteristics will be discussed in a follow-up optical manuscript.

\subsection{Components}
\label{sec:components}
\noindent Switching in the RAMP networks is achieved by configuring the wavelength/time-slot/path at the end-node transceivers. 
For wavelength switching, at the transmitter side, we assume the employment of wavelength tunable sources (WTSs). WTSs composed of time-interleaved tunable lasers (spanning a wide range of 122~wavelength channels) with gated SOAs capable of achieving $<1ns$ wavelength switching \cite{Gerard:21} have been previously demonstrated.
On the destination side, the receiver can be either tunable or fixed depending on the subnetwork choice. If the B\&S network is selected, the receiver can operate at a fixed wavelength by the use of passive filters. However, wavelength tunability is required when considering sub-networks with wavelength routing functionalities. The tunability can either be implemented by a wavelength filter gated by SOAs or by the use of an additional tunable laser for coherent detection.

For space switching, broadcast and select filter-based SOA-gated couplers and combiners are used. Using SOA-based gating as a space-switching mechanism allows sub-nanosecond path selection \cite{chris_soa}. In addition, SOAs are also used for amplification.

Time-division multiplexing is achieved by using pre-defined timeslots. The synchronisation and Clock Data Recovery (CDR) uses the same principle as the one proposed in PULSE~\cite{PULSE-scale} and Sirius~\cite{ballani2020sirius} and described in~\cite{kari}. The duration of the timeslot has been selected such that the maximum reconfiguration overhead is 5\%, leading to a minimum data-transfer slot of 20ns.

Transceiver node capacity of ($B=$) 400~Gbps can be achieved using low-energy silicon-organic hybrid (SOH) modulators~\cite{Wolf:18}, which is the assumed line-rate for the analysis in this paper. Using this data-rate the minimum message size that can be transmitted in a timeslot per transceiver is 950B. Such small messages are common in DCN traffic and HPC MPI collective operations at large scale. 
Nanosecond circuit reconfiguration time is fundamental for HPC application as it allows effective transmission of small message sizes and the use of dynamic collective strategies for MPI operations (sec.\ref{coll_ops}). When the circuit reconfiguration time is smaller than the node I/O time (transceiver and computation delay), it will not create any overhead in the transmission time. Since transceiver (I/O) delays can be as low as a few tens of ns~\cite{ANDREADES201951}, switching reconfiguration times should follow suit.

Star-couplers are used as broadcast technology at both the edge and core of the network. At the edge, they are used in the form of SOA gated splitters and combiners to create 1:N and N:1 switches. At the core, we propose the use of N:N star-couplers, which have been shown to scale to 1024~ports~\cite{adam} as an individual component and larger when using a cascaded approach. This approach makes the network passive and cost-effective.

The wavelength routing component considered for the network core is AWGR, which has been proven to scale to 100s of ports with low loss \cite{ballani2020sirius}.

The combination of these technologies allows the RAMP transceiver and network of achieving ns circuit reconfiguration while achieving high node capacity.



\begin{figure}[htp]
    \centering
    \includegraphics[width=\linewidth]{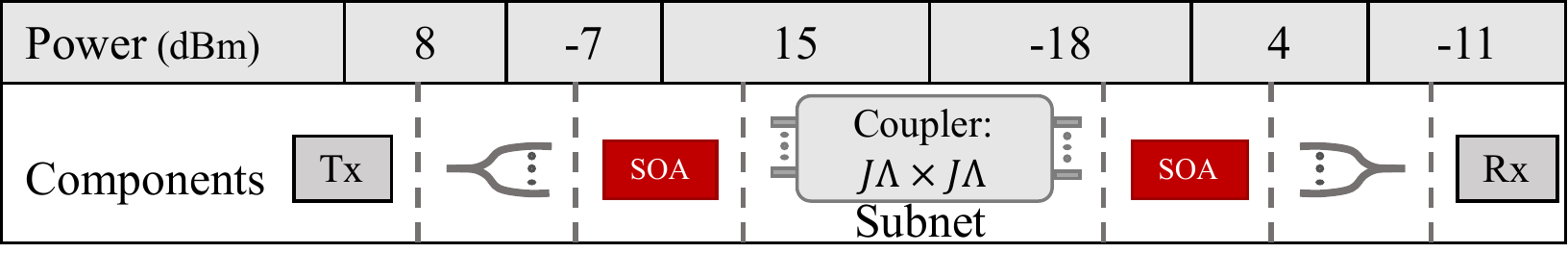}

    \centering
    \caption{Power budget after each component in the path for most constrained RAMP architecture at maximum scalability.}
    \label{fig:Powerbudget2}

\end{figure}

\subsection{Scalability}
\label{sec:scalability}
 \begin{figure}[!t]
    \centering
    \includegraphics[width=0.9\linewidth]{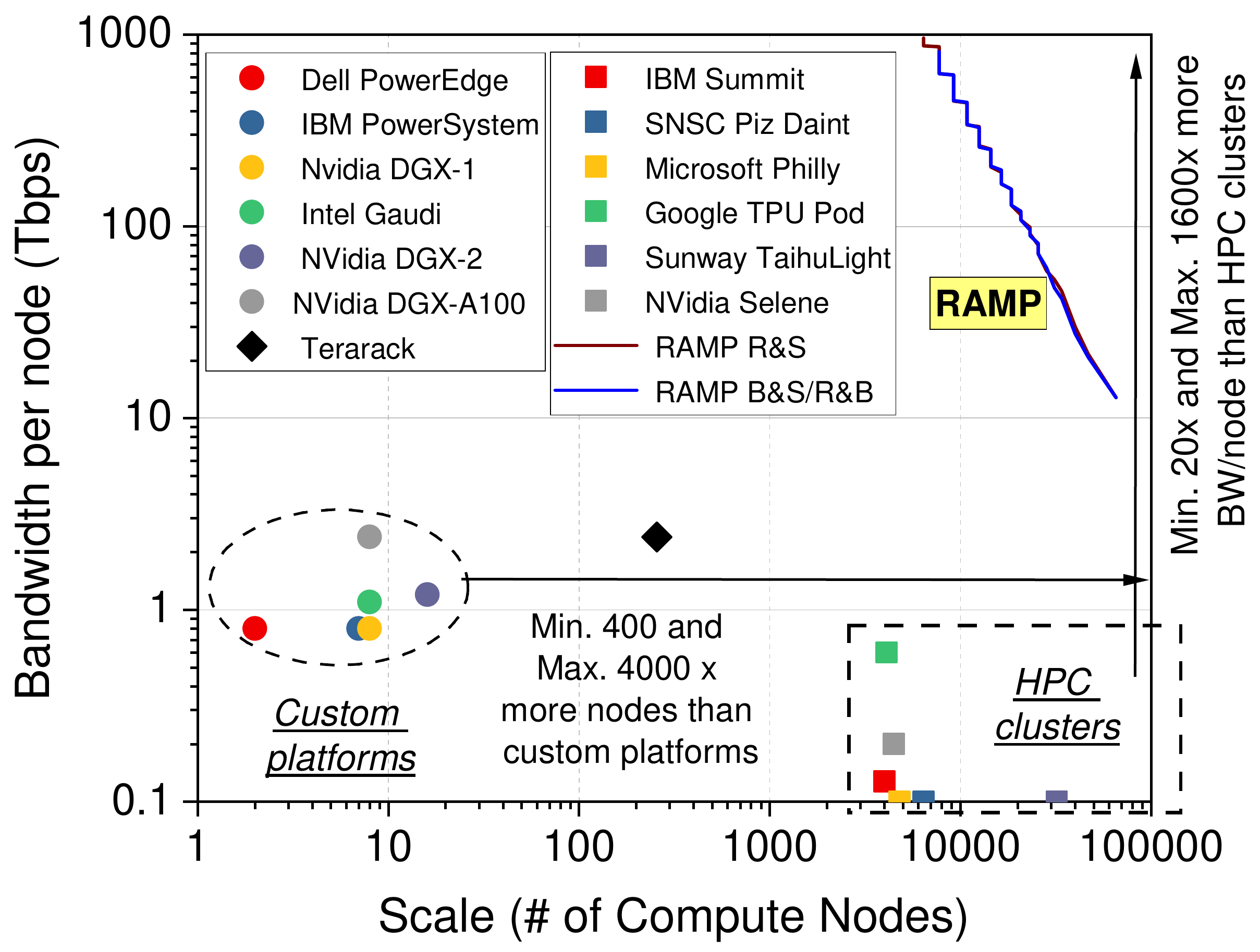}

    \caption{\centering Comparison in bandwidth per node and system scale between RAMP and current or proposed systems~\cite{terarack, DELL, IBM, nvidia_dgx1, habana_labs_ltd__gaudi_2019, nvidia_dgx2, dgxa100, summit, Piz_Daint, Philly, TPU_pod, TaihuLight, selene}. Figure adapted from~\cite{terarack}.}
    \label{fig:sota}

\end{figure}

The scalability of the system in terms of end nodes is limited by the optical properties of the network. In fact, the limiting factor is the optical Signal-to-Noise Ratio (SNR) degradation and attenuation that the optical signal incurs at the output of each network component. To allow direct detection the optical power budget at the receiver side photodetector has to be above -15dBm and the minimum optical power across the path have to be above -20dBm to maintain good SNR. Using the architecture proposed and the components described in sec.\ref{sec:components} under the lossiest configuration (B\&S), this leads to a maximum scale of 65,536 nodes each having 12.8~Tbps node capacity. These results are obtained considering the following configuration which is going to be further analysed in sec.\ref{result}: $\Lambda=64, x=J=32, B=400Gbps, b=1$. The power budget under the described configuration after each network component is shown in Fig.\ref{fig:Powerbudget2}.

Under the maximum scale conditions, the proposed system is capable of achieving larger scalability than current SoTA HPC clusters ($>5.5\times$) while allowing significantly higher effective node-to-node bandwidth ($>20\times$) than custom small-scale platforms. These characteristics give the RAMP architecture the upper hand with respect to SoTA HPC and AI system architectures \cite{terarack}. A comparison between current and proposed HPC systems and RAMP in terms of bandwidth per node and number of compute nodes is displayed in Fig.\ref{fig:sota} using curves representing the different RAMP configurations. For this analysis, the number of communication groups $x$ and the number of transceivers per transceiver group $b$ have been modified while the number of racks has been kept equal to the number of communication groups ($J=x$) and the number of nodes per rack has been kept constant ($\Lambda=64$). By varying $x$ from 32 to 10 and $b$ from 1 to 256, the scalability in terms of nodes reduces to 4096 whereas the node capacity supported increases to 960Tbps. This shows the capability of the RAMP system to scale to future bandwidth requirements. Considering current SoTA technology, a RAMP system could allow full-capacity (288~Tbps), all-to-all communication between 12,544 Tesla DOJO tile~\cite{tesla} accelerators. Currently, such accelerators can only be connected by using limited degree connectivity topologies such as meshes and toruses and limit the effective node-to-node bandwidth. 

Similar scalability analysis can be performed while keeping constant other metrics such as cost and power. Multiple configurations will be explored in future work.

\subsection{Cost and Power Consumption}
\label{power_sec}
\setcounter{table}{1}
\setcounter{table}{2}
\definecolor{Gallery}{rgb}{0.937,0.937,0.937}
\begin{table}[b!]
\centering
\caption{Cost estimated for RAMP OCS network compared to current state-of-the-art EPS HPC (SuperPod\cite{superpod}) and DCN (Fat-Tree) networks scaled to 65,536 nodes with matched bandwidth of 12.8Tbps. $\sigma$ represents the intra-to-inter system oversubscription ratio. To match the EPS system node to the RAMP one at 12.8Tbps ($\sigma=1:1$), 64-128 transceivers (200-100 Gbps) per node and 64-128 copies of separate 3-tier Fat-Tree networks are presented for HPC and DCN, respectively. The OCS system uses $x=32$ transceivers each operating at 400Gbps.}
\label{tab:cost}
\resizebox{\linewidth}{!}{%
\begin{tblr}{
  row{2} = {Gallery,c},
  row{3} = {Gallery,c},
  column{2} = {Gallery},
  cell{1}{1} = {c=2,r=3}{},
  cell{1}{3} = {c=2,r=2}{Gallery,c},
  cell{1}{5} = {c=7}{Gallery,c},
  cell{2}{5} = {c=3}{},
  cell{2}{8} = {c=3}{},
  cell{2}{11} = {r=2}{},
  cell{4}{1} = {r=2}{Gallery},
  cell{4}{3} = {c=2}{c},
  cell{4}{5} = {c},
  cell{4}{6} = {c},
  cell{4}{7} = {c},
  cell{4}{8} = {c},
  cell{4}{9} = {c},
  cell{4}{10} = {c},
  cell{4}{11} = {c},
  cell{5}{3} = {c=2}{c},
  cell{5}{5} = {c=6}{c},
  cell{5}{11} = {c},
  cell{6}{1} = {c=2}{Gallery},
  cell{6}{3} = {c},
  cell{6}{4} = {c},
  cell{6}{5} = {c},
  cell{6}{6} = {c},
  cell{6}{7} = {c},
  cell{6}{8} = {c},
  cell{6}{9} = {c},
  cell{6}{10} = {c},
  cell{6}{11} = {c},
  cell{7}{1} = {c=2}{Gallery},
  cell{7}{3} = {c=2}{c},
  cell{7}{5} = {c=6}{c},
  cell{7}{11} = {c},
  cell{8}{1} = {c=4}{Gallery},
  cell{8}{5} = {c=3}{c},
  cell{8}{8} = {c=3}{c},
  cell{8}{11} = {c},
  cell{9}{1} = {c=4}{Gallery},
  cell{9}{5} = {c},
  cell{9}{6} = {c},
  cell{9}{7} = {c},
  cell{9}{8} = {c},
  cell{9}{9} = {c},
  cell{9}{10} = {c},
  cell{9}{11} = {c},
  cell{10}{1} = {c=4}{Gallery},
  cell{10}{5} = {c=3}{c},
  cell{10}{8} = {c=3}{c},
  cell{10}{11} = {c},
  vlines,
  hline{1,4,6-11} = {-}{},
  hline{2} = {5-11}{},
  hline{3} = {3-10}{},
  hline{5} = {2-11}{},
}
                                       &               & {\textbf{Component}\\\textbf{ Cost (\$) }} &              & \textbf{\#items / network} &               &               &                       &               &               &                \\
                                       &               &                                            &              & \textbf{HPC SuperPod \cite{superpod}}         &               &               & \textbf{DCN Fat-Tree} &               &               & \textbf{RAMP } \\
                                       &               & \textbf{EPS}                               & \textbf{DCN} & \textbf{1:1}               & \textbf{10:1} & \textbf{64:1} & \textbf{1:1}          & \textbf{10:1} & \textbf{64:1} &                \\
{\textbf{Transceivers}\\\textbf{ for}} & \textbf{EPS \cite{marc}} & 200                                        &              & 25.2M                      & 2M            & 0.3M          & 50.3M                 & 4.7M          & 0.8M          & -              \\
                                       & \textbf{OCS}  & 600-1200                                   &              & ~-                         &               &               &                       &               &               & 2.1M           \\
\textbf{Switch EPS \cite{mellanox_technologies_qm8790_2021, arista_price_min,arista_price_max}}                  &               & 23,7k                                      & 44k          & 530k                       & 49.7k         & 8.3k          & 655k                 & 61.4k         & 10.2k         & -              \\
\textbf{Coupler OCS \cite{PON_cost}}                  &               & 3000                                       &              & -~                         &               &               &                       &               &               & 32.8k          \\
\textbf{Trxs. : Switches cost ratio}   &               &                                            &              & 25:75                      &               &               & 19:81                 &               &               & 93:7,96:4      \\
\textbf{Total Network cost (B\$)}      &               &                                            &              & 16.8                       & 1.57          & 0.26          & 35.5                  & 3.33          & 0.55          & 1.35-2.61      \\
\textbf{Normalised cost (\$/Gbps)}     &               &                                            &              & 20.02                      &               &               & 42.38                 &               &               & 1.62-3.12      
\end{tblr}
}
\end{table} 
\definecolor{Gallery}{rgb}{0.937,0.937,0.937}
\begin{table}[t]
\centering
\caption{Power Consumption of RAMP, SuperPod\cite{superpod} and DCN Fat-tree \cite{fat-tree}; scaled to 65,536-node networks with 12.8 $\text{Tbps}$/node and all-to-all connectivity. \relax{Component: Comp., Transceiver: Transc.} $\sigma$ represents the intra-to-inter system oversubscription ratio.}
\label{power_breakdown}
\begin{tblr}{
  cells = {c},
  column{2} = {Gallery},
  cell{1}{3} = {c=3}{Gallery},
  cell{1}{6} = {c=3}{Gallery},
  cell{1}{9} = {Gallery},
  cell{2}{1} = {Gallery},
  cell{3}{1} = {Gallery},
  cell{3}{3} = {c=3}{},
  cell{3}{6} = {c=3}{},
  cell{4}{1} = {Gallery},
  cell{4}{3} = {c=3}{},
  cell{4}{6} = {c=3}{},
  cell{5}{1} = {Gallery},
  cell{5}{3} = {c=3}{},
  cell{5}{6} = {c=3}{},
  cell{6}{1} = {Gallery},
  cell{6}{3} = {c=3}{},
  cell{6}{6} = {c=3}{},
  cell{7}{1} = {Gallery},
  cell{7}{3} = {c=3}{},
  cell{7}{6} = {c=3}{},
  cell{8}{1} = {Gallery},
  cell{8}{3} = {c=3}{},
  cell{8}{6} = {c=3}{},
  cell{9}{1} = {Gallery},
  hlines,
  vlines,
}
\textbf{}                               & \textit{\textbf{Units}}           & \textbf{HPC SuperPod}     &      &      & \textbf{DCN Fat-Tree}    &      &      & \textbf{RAMP} \\
\textbf{Oversub. $\mathbf{\sigma}$}     & -                                 & 1:1                    & 10:1 & 64:1 & 1:1                   & 10:1 & 64:1 & -             \\
{\textbf{Switch }\\\textbf{ Comp.}}     & \textit{-}                        & {NVidia\\QM8790\cite{mellanox_technologies_qm8790_2021}} &      &      & {Arista\\7170\cite{arista}} &      &      & {SOA\\\cite{soa}}           \\
\textbf{Power/Comp.}                    & \textit{W/Comp.}                  & 404                    &      &      & 320                   &      &      & 0.88          \\
\textbf{Comp./path}                     & \textit{\#/path}                  & 11                     &      &      & 11                    &      &      & 2             \\
{\textbf{Transceiver}\\\textbf{ Power}} & \textit{W}                        & {4.35\\\cite{200g_tx}}                    &      &      & {0.5-3.5\\\cite{short_tx, long_tx}}                &      &      & {3.4-3.8}          \\
{\textbf{Energy/}\\\textbf{bit/path}}   & {\textit{pJ/}\\\textit{bit/path}} & 383                    &      &      & 400                   &      &      & 8.5-9.5           \\
\textbf{Power/Gbps}                     & \textit{mW/Gbps}                   & 365                   &      &      & 400                  &      &      & 85-95         \\
\textbf{Total Power}                    & \textit{MW}                       & 306                    & 28.7 & 4.8  & 336                   & 31.5 & 5.2  & 7.1-8           
\end{tblr}
\end{table}
 In Tables \ref{tab:cost}, \ref{power_breakdown}, the estimated cost and power requirements for the proposed architecture are estimated and compared to EPS systems for HPC and DCN at matched scalability (maximum RAMP scalability of 65,536 nodes) for different intra-to-inter server over-subscriptions (1:1, 10:1, 64:1).

 The HPC system considered is a 3-Tier Fat-Tree SuperPod \cite{superpod} architecture scaled to support 8,192 DGX-A100 \cite{dgxa100} servers, each supporting 8 GPUs (65,536 GPUs) and exposing one 200Gbps HDR Infiniband transceiver per GPU \cite{200g_tx}. This represents the 64:1 over-subscription ratio case. The switch used is the 40-port 200Gbps-per-port NVidia QM8790 InfiniBand switches \cite{mellanox_technologies_qm8790_2021, Nvidia_switch_cost}, as described in the reference architecture \cite{superpod}. It needs to be noted that intra-server (DGX-A100) connectivity is discarded (this leads to an underestimation of the effective network cost and power consumption).

  A 3-Tier Fat-Tree system interconnecting 65,536 nodes (servers) is considered as DCN system. The system is based on Arista 7170-64c \cite{arista} 64-port 100Gbps-per-port switches. As transceivers, 100Gbps QSFP optical transceivers \cite{long_tx} for inter-rack communication and 100Gbps copper twinax cables for intra-rack \cite{short_tx}. This choice has been performed due to the different power drawn by the different transceivers (as shown in Table.\ref{power_breakdown})  

 For both EPS networks, the cost of \$1/Gbps is assumed \cite{marc}. To match the node (GPU) I/O bandwidth to the one of the proposed RAMP system (12.8 Tbps, considered as intra-system bandwidth), additional ports per node are exposed and parallel copies of the network are created. At matched bandwidth (1:1 intra-to-inter server oversubscription), in the HPC system, each GPU exposes 64 200Gbps ports ($64\times 8$ for DGX-A100) with 64 independent parallel SuperPod networks. For 1:1 over-subscription in Fat-Tree DCN case, each node of the DCN exposes 128 100Gbps ports and all nodes are connected through 128 parallel independent networks. For the 10:1 oversubscription (1.2Tbps inter-node bandwidth), the number of ports and parallel networks for HPC and EPS is 6 and 12, respectively. Equivalently, for the 64:1 these numbers are 1 for HPC (same as in the SuperPod architecture \cite{superpod}) and 2 for DCN.

 We consider for the RAMP architecture the maximum scalability system with 65,536 nodes ($x=J=32$, $b=1$ and $\Lambda=64$) with 400Gbps modulators (12.8Tbps total capacity), as described in sec. \ref{sec:components}, \ref{sec:scalability}. For the OCS networks, we propose the use of integrated transceivers (with the laser, modulator, SOAs) and assume cost of 1.5-3$\times$ greater than that of EPS transceivers. The cost of the optical coupler for OCS is assumed to be \$3000, estimated from \cite{PON_cost}. For RAMP, the power consumption of the transceiver and switching elements is estimated by considering the individual power consumption of the relevant components described in sec. \ref{sec:components} and reported in~\cite{Benjamin:20, non-tunable, Dsdbr, mod1, mod2, pd_Yoshimatsu:12, soa}. In the power analysis, we consider the power consumption for transceivers with both fixed and tunable wavelength reception.

Table.\ref{tab:cost} shows that the normalised cost (\$/Gbps) of EPS networks is significantly higher than the RAMP counterparts. For the EPS networks, the cost is switch-dominant, having a transceiver:switch cost ratio of 25:75 and 19:81 for HPC and DCN respectively. On the other hand, for the OCS-based RAMP architecture, we nearly eliminate the switching cost and significantly reduce the number of transceivers required per connection. Hence, RAMP, the transceiver:switch cost ratio is 93:7 - 98:2, with the overall normalised network cost reduced by a factor 6.4-26.5$\times$ to 1.62-3.12\$/Gbps. This also applies to the overall network cost at matched bandwidth (1:1) where the budget for EPS networks is $\geq16.8$B\$ whereas the RAMP network cost is $\leq2.16$B\$. A similar cost to the proposed RAMP architecture is reported for EPS networks with 10:1 over-subscription (1.57-3.33\$ for HPC and DCN respectively) while offering 10 times lower capacity.

 It is important to note that, compared to EPS systems, it might seem that the RAMP architecture is over-provisioned, due to the large number of sub-networks. However, differently to electrical switches, the coupler-based sub-networks are passive, inexpensive devices whose cost will further reduce with mass production, making the proposed interconnect feasible for large-scale systems. When compared to PULSE \cite{PULSE-scale}, the number of subnets required by RAMP to accommodate the same number of nodes is smaller (assuming the maximum scale network, PULSE requires 32$\times$ more subnets than RAMP). It also needs to be noted that, due to the passive properties of the sub-networks, the core of the interconnects does not need to be changed when the computational nodes are upgraded. Accommodating workers with higher data-rate in EPS systems requires substituting all network switches to models capable of handling the new line-rate leading to a short core network life cycle (around 4 years in DCNs). Similarly, in traditional active OCS switching systems such as 3D-MEMS, to take advantage of higher data-rate faster circuit reconfiguration time is needed, requiring the core switching infrastructure to be replaced periodically. On the other hand, RAMP, by using a passive core, does not require any substitution in the network with a higher data-rate, as all switching and line-rate dependent technologies reside at the edge. This means that, when upgrading to higher capacity nodes, the only networking component which requires change is the transceiver. This property significantly decreases recurrent costs.

In Table.\ref{power_breakdown}, the power consumption of the EPS and proposed OCS architecture is compared.
Our study shows that RAMP consumes as low as 8.5-9.5~pJ/bit/path, whereas the HPC and DCN counterparts consume 383 and 400~pJ/bit/path respectively. It is important to note that while the number of paths of the RAMP system is significantly higher than the EPS counterparts, all subnets are passive consuming no power and each communication considers only a single chain of active devices. Therefore, the total amount of active paths at any time step is equal to the number of transceivers in the system ($bx^2J\Lambda$). The overall network energy consumption only depends on the active paths. For this reason, the difference in energy per bit is matched by the overall energy consumption. EPS systems at matched scale (65,536 nodes) and matched bandwidth (12.8Tbps) would consume 306-336 MW, which is 10$\times$ larger than the upper DCN network power budget of $\sim30$MW~\cite{ballani2020sirius}. In contrast, RAMP consumes 7.1-8 MW leading to a reduction by a factor of 38-47$\times$. When compared to the similar cost 10:1 oversubscribed EPS systems, the proposed architecture leads to a reduction in energy consumption $\geq3.6\times$ for $10\times$ increase in bandwidth. The proposed system energy consumption is similar (36-66 \% increase) to the EPS networks for 64:1 oversubscription (current SuperPod node-to-node capacity and 2 copies of DCN system) while allowing 64 times higher network communication.

\section{RAMP Collective Operations}
\label{coll_ops}

We propose a set of collective communication algorithms valid for the proposed architecture, arranged in a way such that contention is avoided and collective completion time is minimised. Each RAMP-x (x$=$MPI operation) collective operation follows a set of schedule-less reconfiguration steps through: a) parallel subgroup mapping (devices performing a subset of collective operations in parallel); b) information/message per node mapping at communication step; c) wavelength selection and subnet selection; d) time-slot mapping.

\begin{figure}[h]

    \centering
    \includegraphics[width=0.82\textwidth]{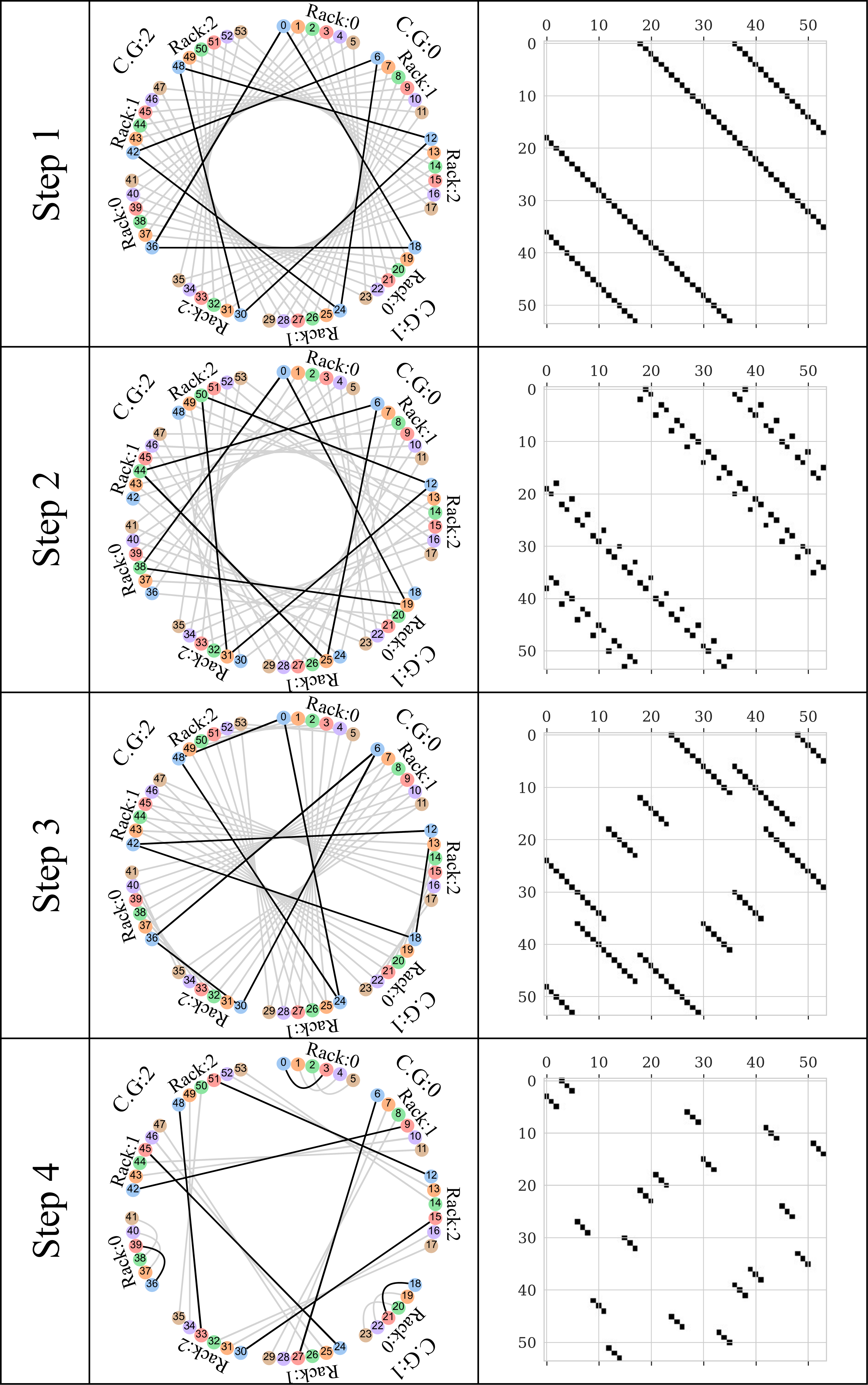}

    \caption{\centering Visualisation of algorithmic steps subgroups for a 54-node ($x=3, J=3, \Lambda = 6$) network example. Top-to-bottom: Communication steps 1-4. 
    }
    \label{fig:All Reduce}

\end{figure}

The proposed strategy could be implemented on any port-level all-to-all large-scale network without over-subscription. However, creating such networks using EPS or previous OCS networks would be expensive in terms of cost, power~\cite{fat-tree, superpod}, connectivity~\cite{Benjamin:20, ballani2020sirius} and reconfiguration time~\cite{khani_mehrdad_2021}. The strategy, topology and scheduling have been co-designed to maximise performance and minimise power and cost to enable high capacity and scalability.

In the following sections, $0\leq g \leq x-1$, $0\leq j \leq J-1$, and $0 \leq \lambda \leq \Lambda-1$ correspond to the local communication group, rack and device number (represented by colour in Fig.~\ref{fig:All Reduce}) respectively.

The strategy of MPI operations is performed in RAMP-x using 3-4 sequential algorithmic steps.
An example that works as a visual aid for the reconfiguration steps for a strategy (e.g reduce-scatter)
is shown in Fig.\ref{fig:All Reduce}, where $\Lambda=6$ and $J=x=3$. 

In this figure, the four rows represent steps 1-4.
At each algorithmic step, parallel logical graphs, called subgroups, are created between a unique subset of devices, represented in Fig.\ref{fig:All Reduce} as a line. 
The left side of Fig.\ref{fig:All Reduce} represents the chord diagram of the RAMP network for each step, with nodes grouped in communication groups, racks and device IDs. 
The right-hand side of the figure represents the connectivity matrix for each node at each step. 
The number representation of each node for the connectivity matrix is shown as the number inside each vertex of the chord diagram. 
It needs to be noted that, while the graph is sparse, the network resources are maximised as each node uses $x-1$ transceivers for the first 3 steps and $x$ for the last.
In Fig.\ref{fig:All Reduce}, an example of 3 subgroups is shown with black lines and the others are greyed out in the background. 
The devices in each subgroup will perform a partial collective operation, depending on the MPI operation. In the first step of the reduce-scatter operation (Step 1), for each node, the overall message is divided in three portions and sent to different destinations in the subgroup. Then the information received is summed (reduced) in each node. The information portion that needs to be sent/received to/by each node is determined by the information map, and the transformation operations (e.g. summation) are dictated by the MPI operation. Each node now contains the sum of a unique 1/3 of the information of the message in each subgroup. 

Note that we track the location of the information portion  (Table \ref{tab:reduce-info}) in every node after each communication step. For the following steps, the subgroups are selected such that they include only nodes with the same information portion combinations. In the second step (Step 2), the message is further partitioned into 3 parts (1/9 of the original message), transmitted to the correct node in each subgroup and processed. In the same way, the third step (Step 3) is performed, such that each device contains the sum of a unique 1/27 of the original information (global reduce-scatter). In the fourth step (Step 4), the information is exchanged between pairs of nodes to complete information updates across all 54 devices (i.e. nodes). This step might have some variations depending on the formulation chosen for subgroup selection.

A similar process, performed backwards (Steps 4 to 1), is valid for all-gather, where unique portions of information are shared and gathered (concatenated) at each algorithmic step in every subgroup. In this way, starting with having 1/54 of the overall message, each node will contain a full 1/27, 1/9, 1/3 and whole information after Step 4, Step 3, Step 2 and Step 1 respectively. In general, the number of steps required by the algorithm can be described as $\text{log}_x(N)$ which, for the maximum scalability case of RAMP, is equal to 4. It is important to note that in cases where $x=2$, the algorithm effectively becomes equivalent to a recursive halving/doubling \cite{recursive_halving}. Recursive halving doubling could be used as a collective operation, however, at the maximum data rate, would lead to high network contention. On the other hand, the proposed strategies have been co-designed such that bandwidth is maximised and contention is avoided by selecting different physical paths and time slots and wavelengths for communication.

The following sections provide further explanation on subgroup (sec.\ref{sec:comm_subgroup}), wavelength (sec.\ref{wavelength_map}), information map (sec.\ref{sec:info_map}), transformation (sec.\ref{sec:buf_op},sec.\ref{sec:loc_op}), transceiver and path (sec.\ref{sec:subnet_trx}) and time-slot (sec.\ref{sec:timeslot_map}) selection.

\subsection{Overall RAMP-MPI procedure}
\begin{figure*}[h]

    \centering
    \includegraphics[width=0.9\linewidth]{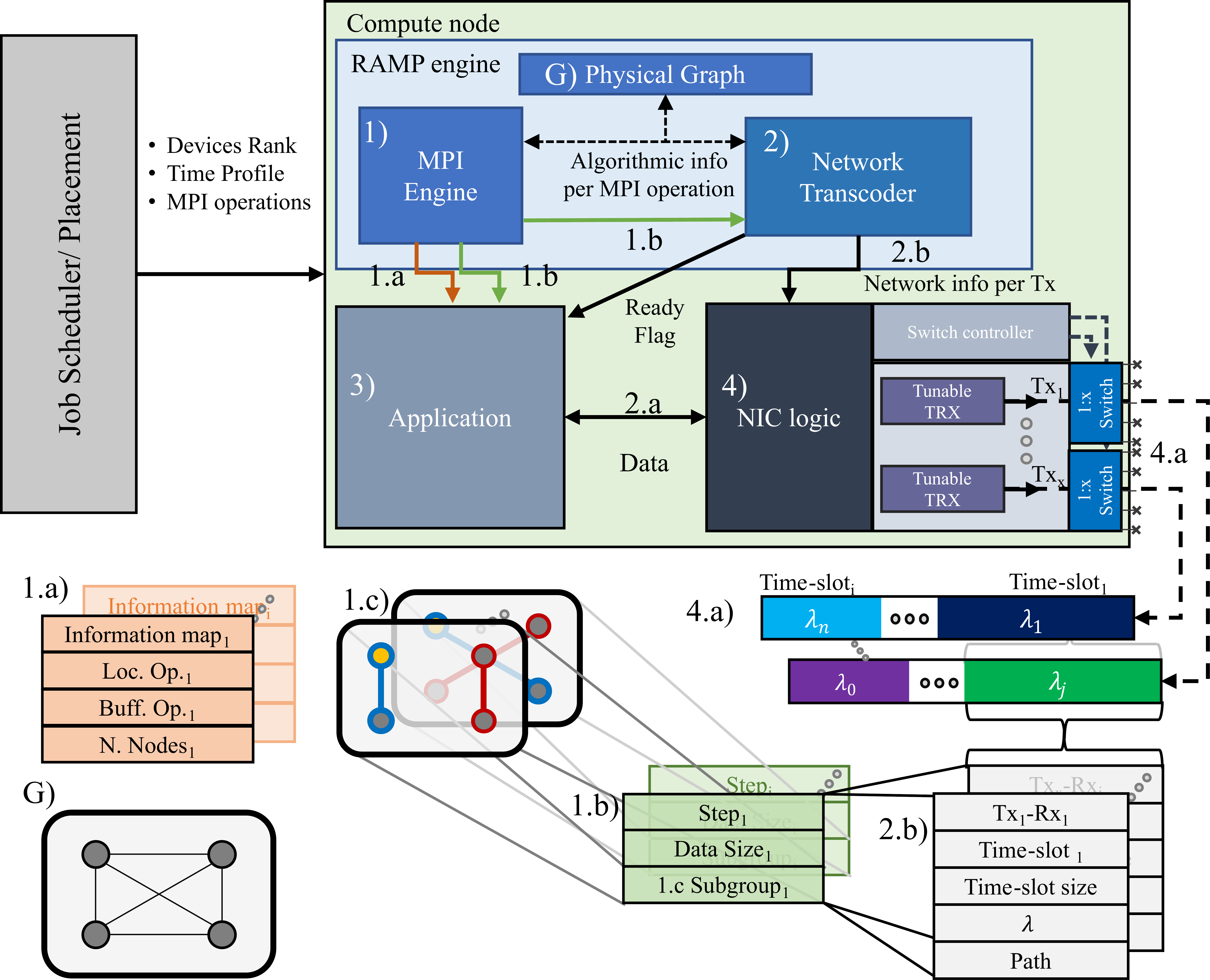}

    \caption{Birdview of the MPI operational process from application to physical implementation.}
    \label{fig:birdview}

\end{figure*}
The core of the proposed research is the combination/co-design of architecture (physical graph sec.\ref{arch}), MPI strategy and scheduling algorithm. An overview of how these components come together to perform an MPI operation is shown in Fig.\ref{fig:birdview}. The process can be viewed in subsequent stages. 

After a distributed task/job is placed by the job scheduler, the information about the ranks of the devices and the MPI operations are shared to all devices/nodes involved. This information is processed by the RAMP engine. The RAMP engine consists of two main components: 1) the MPI Engine (section sec.\ref{sec:mpi_engine}) and 2) the network transcoder (section sec.\ref{sec:network_transcoder}). In the first stage, as shown in Fig. \ref{fig:birdview}, the MPI Engine (1) uses the  physical graph (G) and the MPI operation information to calculate the number of algorithmic steps required and generates information 1.a) and 1.b). 1.a) contains only the information required by 3) the Application to process and retrieve the data correctly for every step. 1.b) represents the algorithmic information required by the network transcoder and consists for every step, the data-size and the subgroup. 1.c) subgroup information represents the logical graph (a derivative of the physical graph G) of devices performing a partial MPI operation at each step. Note that in 1.c), the current node is represented as the yellow dot and the current subgroup as the blue connection in Fig. \ref{fig:birdview}.

The network transcoder gets the information of 1.b) and G) and translates (trans-codes) it into instructions for the Network Interface Card (NIC). For each algorithmic step, the network transcoder generates instruction 2.b) for each individual transceiver to select time-slot size and number, transmitting/receiving wavelength and path. After processing these instructions, the network transcoder sends Ready signal (2.a) to the 3) Application signalling that it is ready for transmission. The Application retrieves and transforms the data using 1.a) such that it could be correctly handled and transmitted by the NIC to perform the MPI operation. The Application shares the processed data to the 4) NIC, which using information 2.b), transforms it into 4.a) signal on the physical system. The NIC Logic tunes the transceiver at the instructed wavelength and selects the correct SOA path (to turn on) for the given time-slot size. 

\section{RAMP Engine}
\label{sec:ramp_engine}
The RAMP Engine is composed of two main blocks: MPI Engine and Network Transcoder. Together they handle the scheduling and communication while the processing is handled by the application.

\subsection{MPI Engine}
\label{sec:mpi_engine}
The MPI Engine uses the physical topology, application, and MPI operation to generate the instructions required by the application and network transcoder to complete the collective operation.

\subsubsection{Communication subgroup map}
\label{sec:comm_subgroup}
\setcounter{table}{4}
\begin{table}[t]
\centering
\caption{Table showing subgroup ID selection. \#SG is the number of subgroups, \#NS is the number of nodes per subgroup. 
}
\begin{tabular}{|l|l|l|l|}
\hline
\hhline{*4{-}}
\rowcolor[HTML]{EFEFEF} 
Step & \#SG          & \#NS & Subgroup ID formula                                        \\ \hline
1    & $\Lambda J $ & $x$  & $\lambda + \Lambda \cdot j$                             \\ \hline

2 & $\Lambda J $ & $x$ & $(\lambda-g)\textrm{ mod } x +  \Lambda j + \floor{\lambda/x}x$ 
\\ \hline
3    & $\Lambda x$  & $J$  & $\left(\lambda + \Lambda (j - g)\right)\textrm{ mod }(\Lambda j)$   \\ \hline
     &                &      & $1) \left(\lambda - \floor{\lambda/x}x \right)\textrm{ mod }x +x^2 j $ \\
     &                &      & $+ \left(\left( g - j \floor{\lambda/x}\right)\textrm{ mod }x \right) x $ \\
     &                &      & or \\
\multirow{-4}{*}{4} & \multirow{-4}{*}{$J x^2$}        & \multirow{-4}{*}{$\Lambda/x$} &  2) $x^2 j + x\left[(g- \floor*{\lambda/x}) \textrm{ mod } x\right] + d \textrm{ mod }x$ \\ \hline
\end{tabular}
\label{tab:subgroup}
\end{table}
\setcounter{table}{5}
\begin{table}[b]
\caption{Table describing the formulas to calculate the RAMP coordinate (communication group, rack and device ID) the other members of the subgroup of the current node (node: $g, j, \lambda$) at any algorithmic step. The Variable column shows the range of the variable to describe all members of the subgroup.}
\label{tab:subgroup_member}
\begin{tabular}{|l|l|l|l|l|}
\hline
\rowcolor[HTML]{EFEFEF} 
Step &
  Communication Group &
  Rack &
  Device ID &
  Variable \\ \hline
1 &
  $(g +\gamma) \text{mod } x$ &
  $j$ &
  $\lambda$ &
  $0\leq \gamma \leq x-1$ \\ \hline
2 &
  $(g +\gamma) \text{mod } x$ &
  $j$ &
  \begin{tabular}[c]{@{}l@{}}$(\lambda \text{mod } x +\gamma)\text{mod }x +$\\ $x \floor{\lambda/x}$\end{tabular} &
  $0\leq \gamma \leq x-1$ \\ \hline
3 &
  \begin{tabular}[c]{@{}l@{}}$[(g - j)\text{mod }x$\\ $+ \gamma] \text{mod }x$\end{tabular} &
  \begin{tabular}[c]{@{}l@{}}$[(j+\gamma)\text{mod }J$\\ $- j]\text{mod 
 }J$\end{tabular} &
  $\lambda$ &
  $0\leq \gamma \leq J-1$ \\ \hline
4.1 &
  \begin{tabular}[c]{@{}l@{}}$(g + j  - $\\  $\floor*{(\floor{\lambda / x} +1)/\floor{\Lambda/x}}$\\  $\times (\floor{\lambda/x}+1)*j ) \text{mod } x$\end{tabular} &
  $j$ &
  
  \begin{tabular}[c]{@{}l@{}}$\lambda \text{mod } x + x \times [$ \\ $ (\floor{\lambda/x}+1) \text{mod} \floor{\Lambda/x}] $ \end{tabular} &
  NA \\ \hline
4.2 &
  \begin{tabular}[c]{@{}l@{}}$[(g - \floor{\lambda/x})\text{mod }x$\\ \\ $+\gamma] \text{mod }x$\end{tabular} &
  $j$ &
  \begin{tabular}[c]{@{}l@{}}$\lambda \text{mod }x + x \times [$\\ $(\gamma + \floor{\lambda/x})\text{mod }\floor{\Lambda/x}$\\ $- \floor{\lambda/x}] \text{mod }\floor{\Lambda/x}$\end{tabular} &
  $0\leq \gamma \leq \floor{\Lambda/x}$ \\ \hline
\end{tabular}
\end{table}
\noindent The subgroup describes the set of devices (logical graph) that each node needs to share information (communicate) with at any algorithmic step.

Summary and formulae describing how each device is mapped to any subgroup at any communication step are shown in Table~\ref{tab:subgroup}.
For this mapping, the nodes in a rack are further divided into groups of $x$ devices called device groups, where each node has a unique device group number from $1$ to $x$.

The communication subgroups at each algorithmic step correspond to communication performed between unique sets of devices in different system dimensions. These consist of:

\textbf{Step 1}: Nodes with the same node number, rack and different communication groups;
 
\textbf{Step 2}: Nodes with sequential node numbers in the same device group, rack and different communication group;

\textbf{Step 3}: Nodes with the same node number, different rack and communication group;

\textbf{Step 4}: Nodes with the same device group number, different device groups, racks and communication groups or nodes in sequential device groups with the same device group number rack and different communication groups.

Depending on the selection of the formulation for the subgroup in Step 4, two different operations will be used. It has to be noted that when the first formulation is selected, the algorithm considered for the last step has to use strategies with one-to-one communication (such as ring, recursive halving/doubling~\cite{recursive_halving} and Bruck's~\cite{bruck}), which might incur additional steps if the number of devices is greater than 2 (value at maximum scale). 

The subgroup selection defines the logical circuit in which each node is part. The number of nodes per subgroup, as shown in Table.\ref{tab:subgroup}, selects which of the four steps is active ($\#NS>1$). From the subgroup information, each node is able to know all sources and destinations active at any algorithmic step as described in Table.\ref{tab:subgroup}.

Using the information provided in Tab.\ref{tab:subgroup}, the members of each subgroup can be found from each algorithmic step by each node. The formulae to find the coordinate of the other members of the same subgroup for the current step of each node is shown in Table.\ref{tab:subgroup_member}.

\subsubsection{Information Map}
\label{sec:info_map}
\begin{table}[t]
    \centering
\caption{Formula describing what portion of the previous message should be received by a node at any algorithmic step.}
\begin{tabular}{|l|l|}
\hline
\hhline{*2{-}}
\rowcolor[HTML]{EFEFEF} Step & Information portion formula                                    \\ \hline
\multirow{2}{*}{1}           & $\left( g  - \lambda - j - \floor{\lambda/x}j \right) $        \\
                             & $\textrm{mod } x$                                              \\ \hline
2                            & $\left(  g  - j - \floor{\lambda/x}j \right) \textrm{ mod } x$ \\ \hline
3                            & $j$                                                            \\ \hline
4                            & $\floor{\lambda/x}$                                            \\ \hline
\end{tabular}
\label{tab:reduce-info}
\end{table}
\noindent The information map consists of a set of formulae describing the portion of the information that should be sent-received and processed by each node at each algorithmic step. The formulae describing the information map at each algorithmic step for data transfer-related strategies are described in Table \ref{tab:reduce-info}. The combination of values generated by the table across each algorithmic step represents the node rank. This also represents either the portion of the original message or the collected information available at the node after the last operation depending on the selected operation. The decimal representation of the information value at all algorithmic steps represents the rank of each node in the collective. 

\subsubsection{Buffer Operation}
\label{sec:buf_op}
The buffer operation ($\Buff$) corresponds to the transformation performed on the message before transmission that is generated by the MPI Engine and defined by the MPI Operation. It takes three arguments: the message that needs to be processed ($DATA$), the number of nodes in the current subgroup ($nodes$) and the information map for the current step ($info$). 

As shown in Table \ref{tab:MPI Operation}, there are three types of operations:
\begin{itemize}
    \item Reshape: the information vector is reshaped such that is divided into $nodes$ addressable contiguous segments of the same size.
    \item Copy: the buffer size is increased by a factor of $nodes$ and reshaped as described above. The original information will be in the segment of the array corresponding to the local rank of the node in the subgroup.
    \item Identity: no transformation is performed.
\end{itemize}

$info$ is used to sort the message in such a way that the correct portion of the information is given to the correct transceiver.

\subsubsection{Local Operation}
\label{sec:loc_op}
The local operation ($\Op(DATA)$) is the transformation performed on the received data after a communication step. There are four operations:
\begin{itemize}
    \item Reduce: associative operation, usually sum, between vectors received from different sources.
    \item Reshape: used only in the all-to-all operation. Transpose the information (considered as a 3D array) in the source, rank dimension and flatten it into a one-dimensional vector. This operation puts the information to be transmitted into a contiguous portion of memory in the correct rank order.
    \item Logical-AND between Booleans representing the presence of a correct message. Only used for barrier operation.
    \item Identity: no transformation is performed.
\end{itemize}

$info$ is used to place in the correct order information coming from the NIC.

\subsubsection{MPI operation algorithm}
\label{sec:mpi_alg}

\begin{table}[t]
 \centering
\caption{Message size and operations per algorithmic step for different collective operations}
\label{tab:MPI Operation}
\begin{tabular}{ll|llll|ll|}
\hhline{*2{~}*6{-}}
 &
   &
  \multicolumn{4}{c|}{\cellcolor[HTML]{EFEFEF}Message size/step ($m=$ original msg size)} &
  \multicolumn{2}{c|}{\cellcolor[HTML]{EFEFEF}Local Ops.} \\ 
  \hhline{*2{~}*6{-}}
 &
   &
  \multicolumn{1}{c|}{\cellcolor[HTML]{C0C0C0}1} &
  \multicolumn{1}{c|}{\cellcolor[HTML]{C0C0C0}2} &
  \multicolumn{1}{c|}{\cellcolor[HTML]{C0C0C0}3} &
  \multicolumn{1}{c|}{\cellcolor[HTML]{C0C0C0}4} &
  \multicolumn{1}{c|}{\cellcolor[HTML]{C0C0C0}Op.} &
  \multicolumn{1}{c|}{\cellcolor[HTML]{C0C0C0}Buff\_Op} \\ \hline
\multicolumn{1}{|l|}{\cellcolor[HTML]{EFEFEF}} &
  \cellcolor[HTML]{C0C0C0}Red.-Scatter &
  \multicolumn{1}{l|}{$m/x$} &
  \multicolumn{1}{l|}{$m/x^2$} &
  \multicolumn{1}{l|}{$m/(Jx^2)$} &
  $m/(J\Lambda x)$ &
  \multicolumn{1}{l|}{Reduce} &
  Reshape \\ \hhline{|>{\arrayrulecolor{gray!15}}->{\arrayrulecolor{black}}|--|->{\arrayrulecolor{black}}|--|->{\arrayrulecolor{black}}|-|}

\multicolumn{1}{|l|}{\cellcolor[HTML]{EFEFEF}} &
  \cellcolor[HTML]{C0C0C0}All-Gather &
  \multicolumn{1}{l|}{$m \cdot \Lambda/x$} &
  \multicolumn{1}{l|}{$m \cdot J \Lambda/x$} &
  \multicolumn{1}{l|}{$m \cdot J \Lambda$} &
  $m \cdot J \Lambda x$ &
  \multicolumn{1}{l|}{Identity} &
  Copy \\ \hhline{|>{\arrayrulecolor{gray!15}}->{\arrayrulecolor{black}}|--|->{\arrayrulecolor{black}}|--|->{\arrayrulecolor{black}}|-|}
\multicolumn{1}{|l|}{\cellcolor[HTML]{EFEFEF}} &
  \cellcolor[HTML]{C0C0C0}Barrier &
  \multicolumn{1}{l|}{$0$} &
  \multicolumn{1}{l|}{$0$} &
  \multicolumn{1}{l|}{$0$} &
  $0$ &
  \multicolumn{1}{l|}{AND} &
  Identity \\ \hhline{|>{\arrayrulecolor{gray!15}}->{\arrayrulecolor{black}}|--|->{\arrayrulecolor{black}}|--|->{\arrayrulecolor{black}}|-|}
\multicolumn{1}{|l|}{\cellcolor[HTML]{EFEFEF}} &
  \cellcolor[HTML]{C0C0C0}All-to-All &
  \multicolumn{1}{l|}{$m/x$} &
  \multicolumn{1}{l|}{$m/x$} &
  \multicolumn{1}{l|}{$m/J$} &
  $m \cdot x/\Lambda$ &
  \multicolumn{1}{l|}{Reshape} &
  Reshape \\ \hhline{|>{\arrayrulecolor{gray!15}}->{\arrayrulecolor{black}}|--|->{\arrayrulecolor{black}}|--|->{\arrayrulecolor{black}}|-|}
\multicolumn{1}{|l|}{\cellcolor[HTML]{EFEFEF}} &
  \cellcolor[HTML]{C0C0C0}Scatter &
  \multicolumn{1}{l|}{$m/x$} &
  \multicolumn{1}{l|}{$m/x^2$} &
  \multicolumn{1}{l|}{$m/(Jx^2)$} &
  $m/(J\Lambda x)$ &
  \multicolumn{1}{l|}{Identity} &
  Reshape \\ \hhline{|>{\arrayrulecolor{gray!15}}->{\arrayrulecolor{black}}|--|->{\arrayrulecolor{black}}|--|->{\arrayrulecolor{black}}|-|}
\multicolumn{1}{|l|}{\cellcolor[HTML]{EFEFEF}} &
  \cellcolor[HTML]{C0C0C0}Gather &
  \multicolumn{1}{l|}{$m \cdot \Lambda/x$} &
  \multicolumn{1}{l|}{$m \cdot J \Lambda/x$} &
  \multicolumn{1}{l|}{$m \cdot J \Lambda$} &
  $m \cdot J \Lambda x$ &
  \multicolumn{1}{l|}{Identity} &
  Copy \\ \hhline{|>{\arrayrulecolor{gray!15}}->{\arrayrulecolor{black}}|--|->{\arrayrulecolor{black}}|--|->{\arrayrulecolor{black}}|-|} 
\multicolumn{1}{|l|}{\multirow{-7}{*}{\cellcolor[HTML]{EFEFEF}{\rotatebox[origin=c]{90}{\footnotesize{MPI Operations}}}}} &
  \cellcolor[HTML]{C0C0C0}Broadcast &
  \multicolumn{4}{l|}{\centering $m/k$ for $s+k-2$s steps} &
  \multicolumn{1}{l|}{Identity} &
  Identity \\ \hline
\end{tabular}
\end{table}

The combination of $\Buff$ and $\Op$ is defined by the MPI operation (Table~\ref{tab:MPI Operation}), which will be performed on the message by the application. The pseudo-code for a single MPI operation running on an individual node is shown in Alg.\ref{alg:cap}. In Alg.\ref{alg:cap} starting with the local message, each node requests information to the MPI Engine given the current and active nodes' rank and the MPI operation (line 2). For each of the steps dictated by the MPI Engine, the DATA is first transformed by the $\Buff$ (line 6) and after receiving confirmation from the Transcoder that the NIC is ready (line 7) pushes/receives data to/from the NIC which will be transformed by local operation ($\Op$, line 9) and will be used as the data for the next step.

The selection of $\Buff, \Op$ for each MPI operation is shown in Table~\ref{tab:MPI Operation}. The message sizes for each step and operation in Table~\ref{tab:MPI Operation} are derived by the combination of $\Buff$ and $\Op$ following Alg.\ref{alg:cap}.

It can be noted that the Reduce and All-Reduce operations have not been included in Table~\ref{tab:MPI Operation}. These are implemented by following an approach similar to Rabenseifner's algorithm~\cite{raben}, where we consider the reduce and all-reduce operations as a reduce scatter followed by a gather and all-gather operation respectively.

For the broadcast operation, we take advantage of the optical property of the systems. Using SOA gating, one device can multi-cast data at full-node capacity to $x^2$ or $x^3$ nodes depending on the selected system configuration. Given this property, a pipelined tree broadcast is created, where a root node can talk up to $x^2$ nodes, $\lambda-1$ of which will transmit to an additional $x^2$ devices each using different wavelengths. This creates a logical tree with a diameter of 3. The number of stages $k$ for the pipeline considered is:
\begin{align}
    k = \sqrt{\frac{m \cdot  (s-2)}{\alpha}\beta},
\end{align}
where $s$ is the diameter of the tree generated to perform the broadcast, $\alpha$ is communication setup latency (propagation and node/software dependent latencies) and $\beta$ is the inverse of the total node capacity.
The total number of steps needed to perform the operation is $k +s -2$ and the message transmitted per stage is $MSG/k$.
\newline
\begin{algorithm}[t]
\caption{Collective operation routine for single node}\label{alg:cap}
\begin{algorithmic}[1]
\STATE $DATA \gets msg$

\STATE $steps, \Op, \Buff, Info\_map, n\_nodes \gets MPI\_Engine(\textrm{my rank, devices rank, MPI Op})$

\FOR{$step=1$ to $steps$}

        \STATE $nodes \gets n\_nodes[step]$
        
        \STATE $info \gets Info\_map[step]$
        
            \STATE $DATA \gets \Buff{(DATA,nodes, info)}$
            \STATE \textrm{WAIT} $Network\_Transcoder \textrm{ready signal}$
            \STATE $DATA \gets NIC(DATA)$
    \STATE $DATA \gets \Op(DATA, info)$ 
\ENDFOR
\end{algorithmic}
\end{algorithm}
\vspace{-2cm}
\newline
\newline
\newline
\newline

\subsection{Network Transcoder}
\label{sec:network_transcoder}
The network transcoder uses the information from the MPI Engine and collective operation and translates them to instructions for the NIC to establish an optical circuit by just configuring the transceiver (wavelength) and the $1:x$ switches (path) of that node (see Fig.\ref{fig:birdview}).

\subsubsection{Wavelength mapping}
\label{wavelength_map}
\noindent Wavelength selection in OCS networks is fundamental to correctly route the information and avoid contention. 
Together with the subgroup selection, colour/wavelength is also assigned for each node to communicate appropriate information at each algorithmic step.

The wavelength mapping varies for the various subnets and it uses a look-up table. Using a subnet with only a star coupler the mapping is dictated by the node receiving wavelength whereas with the AWGR it is forced by the source/destination pair.

\subsubsection{Subnet/Path/transceiver selection}
\label{sec:subnet_trx}
 
\noindent For any source-destination pair, there are $b x$ possible paths and subnets that allow communication. 
Between the parallel subgroups in the first three algorithmic steps, there might be up to $b x$ communications using the same wavelength sharing the same set of subnets. To avoid contention, a wavelength must be used only once in the same subnet. 
 
To minimise control complexity, the transceivers used by any node to perform the collective operations are pre-determined. 
The transceiver groups chosen between any source-destination pair are:

\begin{equation}
\begin{aligned}
    Trx & \left(  d(g_{src},j_{src}, \lambda_{src}), d( g_{dst}, j_{dst}, \lambda_{dst} ) \right) \label{Eq:Trx} \\
    &= \left(g_{src} + g_{dst} +j_{src}\right) \textrm{ mod } x,
\end{aligned}
\end{equation}
where $g_{src}$ - $g_{dst}$,  $j_{src}$ - $j_{dst}$ and $\lambda_{src}$ - $\lambda_{dst}$ are the source and destination communication group, racks and node numbers respectively. The transceiver selection forces the subnet selection as each subnet is defined by the combination of $g_{src}, g_{dst}, Trx$.

In practice, whenever the number of devices per subgroup is smaller than the number of communication groups, multiple transceiver groups might be used to communicate between the same source-destination pair. The additional number of transceiver groups that can be used for each communication in a collective operation is:
\begin{align}
    \textrm{\#TRX}_{\textrm{additional}} =  \floor*{\frac{x - \floor{x/d}(d-1)}{d-1} }, 
\end{align}
where $d$ is the number of devices in the active subgroup. If $\textrm{\#TRX}_{\textrm{additional}}$ is different than 0, then the additional transceiver groups are used for communication. The transceiver groups used for any communication pair is:
\begin{equation}
\begin{aligned}
    TRX(d_{src}, d_{dst}) &=  [Trx(d_{src}, d_{dst})   \\
     + \{0, 1, & ..., \textrm{\#TRX}_{\textrm{additional}} -1\}] \textrm{ mod } x \cdot d,   \label{Eq:transceivers_all}
\end{aligned}
\end{equation}
 where $Trx(d_{src}, d_{dst})$ is the original transceiver group described in Eq. \ref{Eq:Trx}. 
 
 From Eq. \ref{Eq:transceivers_all} the effective I/O unidirectional bandwidth of a node can be defined as:
 \begin{align}
     B_{\textrm{IO Eff}} = B\cdot b \cdot(1+ \textrm{\#TRX}_{\textrm{additional}}) (d-1).
     \label{Eq:eff_bw}
 \end{align}
 
For the fourth step, the transceiver selection may vary depending on the sub-groups formula selected (Table \ref{tab:subgroup}). For the first formula, the number of transceiver groups used per communication is $x$ as there would not be any contention for a single job. Selecting the second formula, the transceiver mapping follows Eq. \ref{Eq:transceivers_all}.
 
\subsubsection{Time-slot mapping}
\label{sec:timeslot_map}
\noindent The time-slot map is given by the data-transmitted per step (Tab.\ref{tab:reduce-info}) and the effective bandwidth per transceiver (Eq. \ref{Eq:eff_bw}), and gives deterministic communication latency.

It is possible to further increase the number of parallel jobs by selecting different subnets (e.g. AWGR-based subnets allow support for different device number sets, same reason as for the communication groups set).
 



\subsection{MPI Operation workflow}
\begin{figure}
    \centering
    \includegraphics[width=\linewidth]{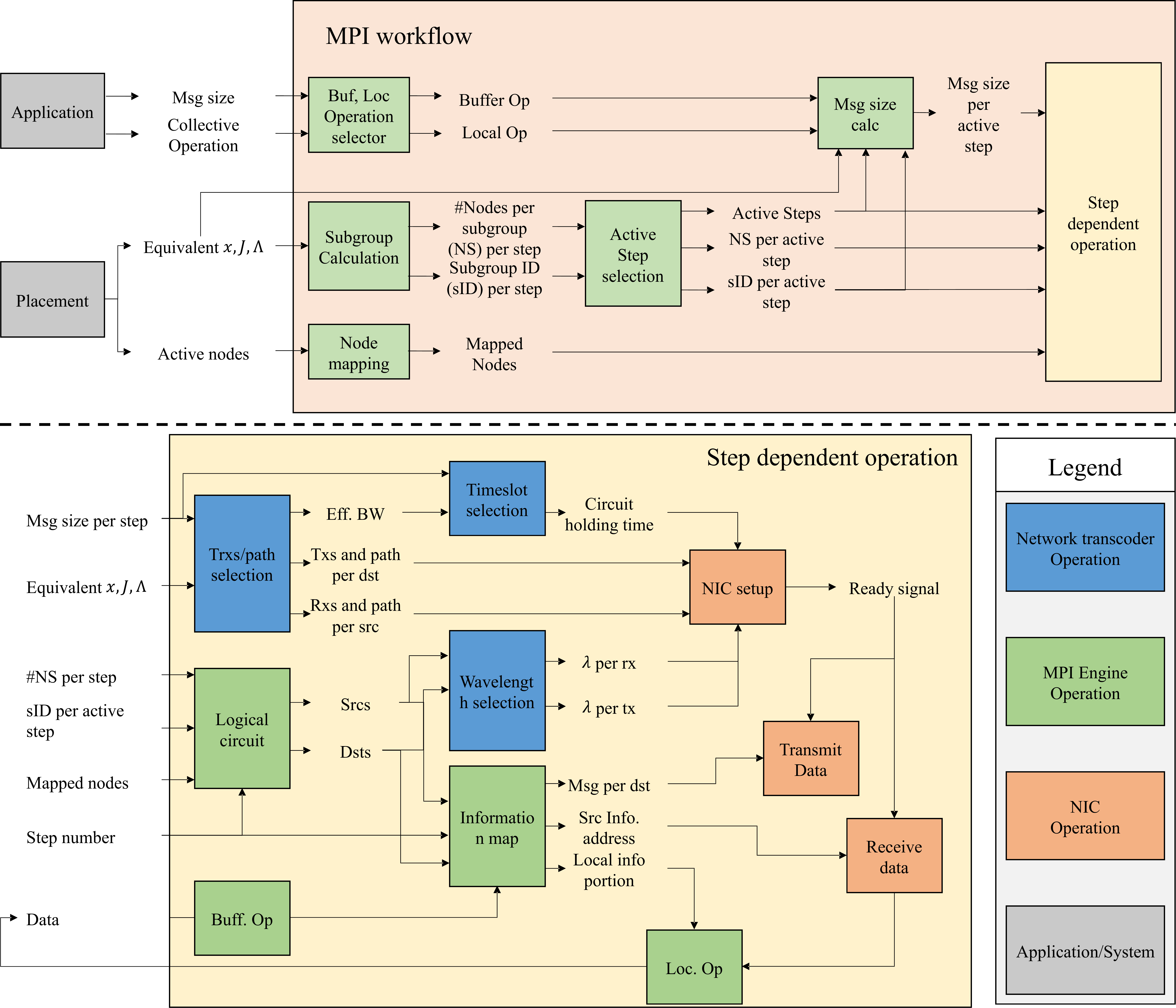}
    \caption{Workflow detailing MPI operation.}
    \label{fig:MPI workflow}
\end{figure}
To perform a complete MPI operation, each node has to perform the following operations as described in Fig.\ref{fig:MPI workflow}. 
Each node first receives from the job allocator/scheduler the collective operation, the message size, the active nodes for the collective and equivalent RAMP architecture parameters in terms of used communication groups ($x$), racks ($J$) and node numbers ($\Lambda$). Using this information, each node calculates its subgroup ID and the number of nodes in each subgroup for each algorithmic step, using the principles described in sec.\ref{sec:comm_subgroup}. While these are calculated, also the active steps (the steps of the collective operation that have to be run) are selected, as they will have a number of nodes $>1$. Then for each active step, the logical circuits (nodes with the same subgroup ID) are found using the formulations described in sec.\ref{sec:comm_subgroup}. Once the logical circuits have been identified the information portion that needs to be sent to each of them is calculated as described in sec.\ref{sec:info_map} and stored in a lookup table. From the information portion and the buffer operation, the message size per source-destination pair is calculated. Using the topological and logical circuit information the transceivers for each source-destination pair are selected, which determine the effective bandwidth of the node pair communication. From the message size and effective bandwidth, the number of time-slots per communication is determined and the wavelength and path per active transceiver are selected. The received data is processed by the local operation and considered as the message for the next active step.

All the information is deterministic and pre-computed at application setup, such that it can be used as a lookup table at runtime following the principles described in sec.\ref{sec:mpi_alg}.

\subsubsection{Alternative strategies}
Strategies different to the one described in sec.\ref{sec:ramp_engine} could be implemented either by considering them as input to the scheduler described at sec.\ref{sec:dynamic_traffic} or pre-determined by mapping the strategy to the RAMP network configurations. As an example, both (up-to) full-capacity ring-based and recursive halving/doubling strategies can be implemented for the RAMP architecture. However, these strategies lead to poorer performance than the ones which have been co-designed for the architecture and require complex network control. 

\section{Simulation Methodology}

\label{sec:methods}
We develop a simulator in Python that is provided as open source to evaluate collective completion times of various collective operations using different strategies and topologies. The main elements of the simulator are described in sec.\ref{simulator}. The network setup and strategies description are described in sec.\ref{sec:eps} and sec.\ref{sec:mpi_strat_comp} respectively. The simulator has been made open source and is accessible online (see \textit{Artifact Description}).

\begin{figure}[h]
    \centering
    \includegraphics[width=\linewidth]{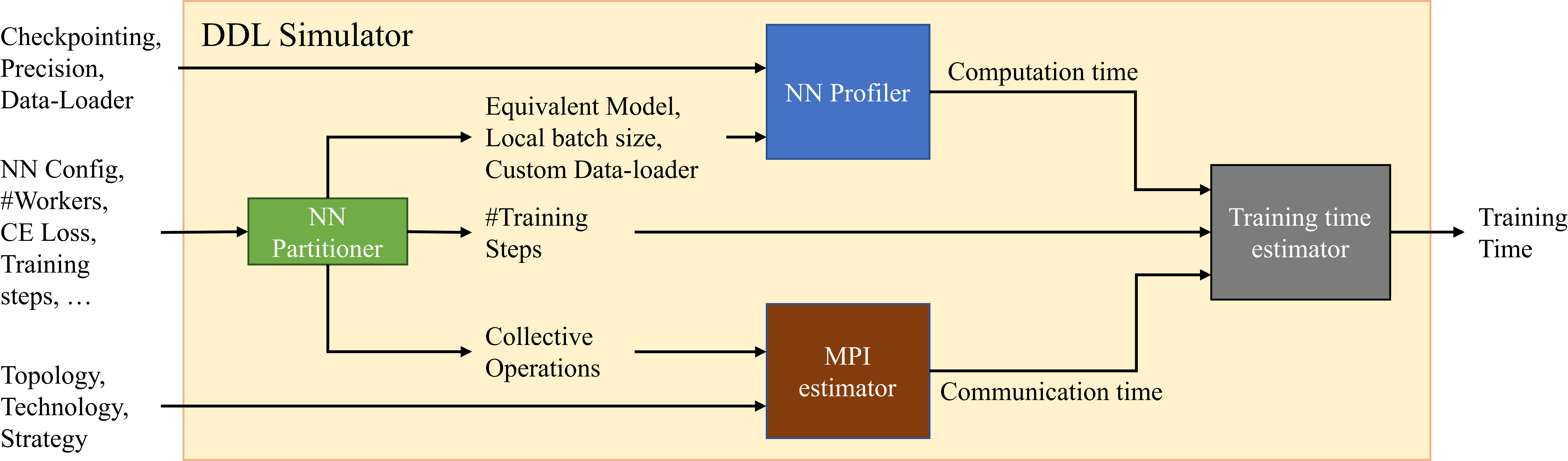}
    \caption{DDL Simulator Diagram}
    \label{fig:DDL simulator diagram}
\end{figure}
\subsection{DDL Simulator}
\label{simulator}
To evaluate the performance of the proposed system and strategies and compare it to traditional EPS and proposed OCS systems a custom DDL simulator has been developed in Python. Fig.\ref{fig:DDL simulator diagram} describes the main building blocks and functionalities of the simulator. The goal of the simulator is to determine training time in either terms of time-to-accuracy or training iteration time for different Deep Neural Networks, partitioned at different scales onto different systems.
The Simulator consists of three main building blocks: the NN Partitioner (sec.\ref{sec:NN_partitioner}), the NN Profiler (sec.\ref{sec:NN_Profiler}) and MPI Estimator (sec.\ref{sec:MPI Estimator}).

To evaluate the overall training time, the input neural network is processed by the partitioner, which, using the configuration parameters, the number of workers and additional constraints, generates the equivalent partitioned model which consists of the computation graph present on a single GPU when running a distributed job. The equivalent model, together with the local batch size information and the data-loader is fed into the profiler which evaluates the computation time of the model for forward and backward computation. At the same time, the partitioner provides information about the collective operations required for the distributed job to the MPI estimator which, using system and strategy information, will estimate the collective completion time for each operation. The computation time information from the profiler and the collective completion time from the MPI estimators are fed into a training time estimator, together with the number of steps, to evaluate the overall training time. 

\subsection{NN Partitioner}
\label{sec:NN_partitioner}
In this paper, we evaluate the distributed training performance of two types of models: Megatron-based \cite{megatron-lm_2019} encoder-only transformer and Deep Learning Recommendation Model (DLRM) \cite{naumov2019DLRM}. For each one of them, a custom equivalent model has been developed and open-sourced. 
\subsubsection{Megatron NN partitioner}
For the Megatron model, we develop a library capable of creating the equivalent model for using the Megatron-LM partitioning strategy for transformer-based encoder-only models using data and tensor parallelism. The goal is to evaluate the difference in time to accuracy of Megatron models for different target cross-entropy (CE) loss when performing distributed training on multiple systems. The partitioner takes as input the target CE loss and the maximum number of workers to perform the computation. This information is fed into a scaling law processing block which calculates the model parameters and training requirements for reaching the target loss. The scaling law processing block uses the formulations of \cite{scaling_language_models} to determine the number of parameters, number of attention heads, hidden dimension, number of layers, global batch size and number of training steps required to reach the target accuracy. The number of parameter information, together with the number of available workers, is fed into a model parallel partitioning block which partitions the model such that the number of parameters available in a single device is as close as possible to the upper limit acceptable by an NVIDIA A100 GPU (1.6B \cite{zero-offloading_2021}). Using this approach the model parallel level for the partitioning is selected and the equivalent tensor partitioned model is generated. The equivalent model is then fed to the Data Parallel Partitioning block together with the global batch size. In this step, the data parallel level and local batch size are selected to maximise total memory utilisation, including activation memory, for a single GPU. Maximising memory utilisation for both model and data parallelism minimises the overall data transferred across the system and maximises the effective compute time. After the data parallel partition, the number of training steps to accuracy is recalculated for the effective global batch size.

The partitioned model, with either data or model parallelism, requires collective communication between workers. From the partitioned models and the batch size, the collective operations required are determined such that they can be evaluated by the MPI estimator. Following Megatron \cite{megatron-lm_2019} partitioning, the tensor partitioned model includes two main building blocks which determine different collective operations: the partitioned multi-head attention and corresponding backward path all-reduce, and partitioned feed-forward MLP with corresponding forward path all-reduce. The message size for both collective operations is determined by the local batch size and the hidden model dimension. Data parallelism requires a gradient all-reduce before the optimiser step with message size dependent on the number of parameters.

The equivalent model and batch size will be given as input to the profiler and the number of training steps will be used by the training estimator to evaluate the overall time to accuracy.

The partitioner library and profiled partitioned Megatron models for the results have been made public at \cite{mp_encoder} and \cite{mp_encoder_data} respectively.

\begin{figure}[h]
    \centering
    \includegraphics[width=\linewidth]{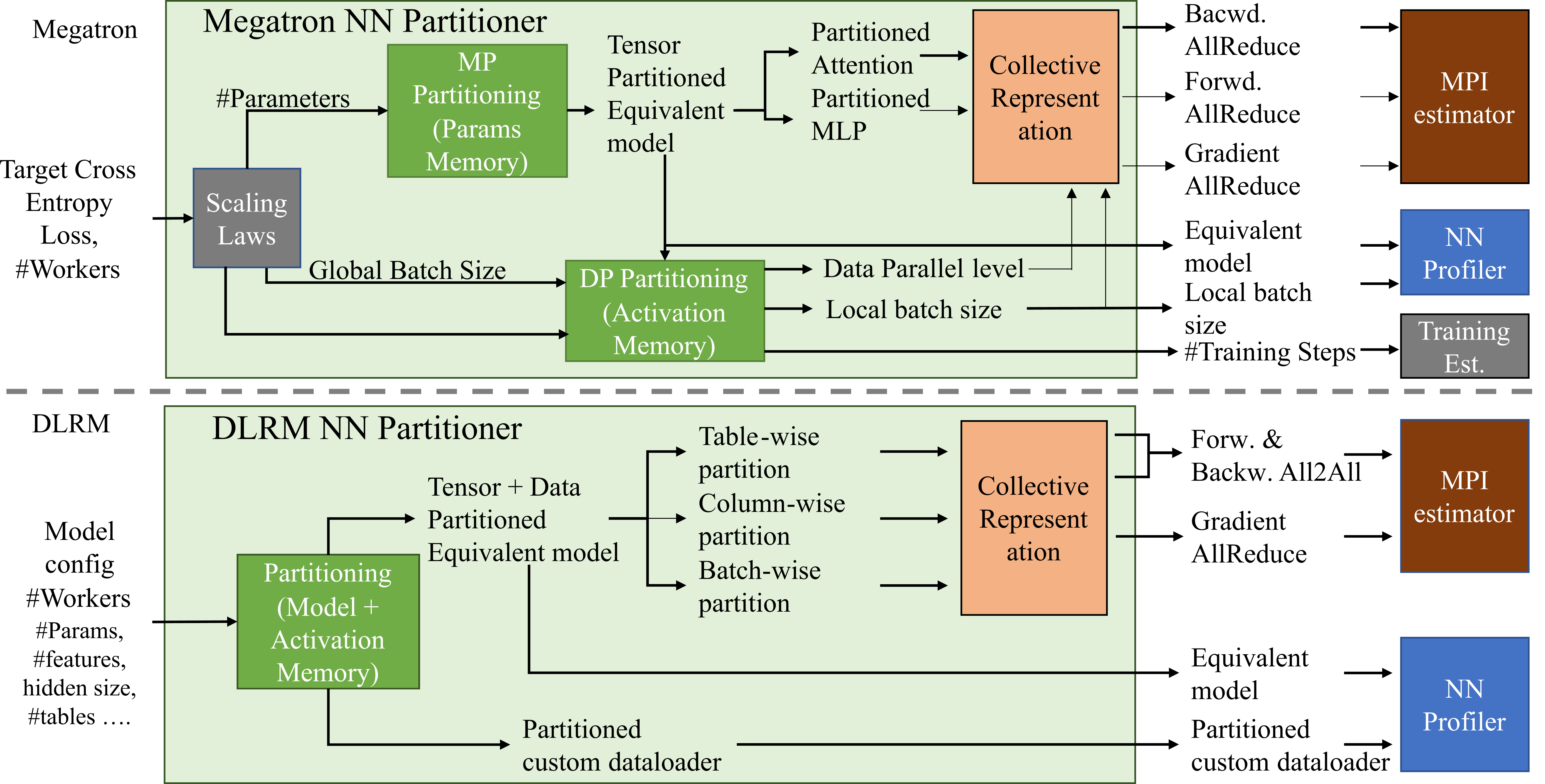}
    \caption{Diagram for the NN Partitioner block for Megatron and DLRM Models}
    \label{fig:NN_partitioner}
\end{figure}

\subsubsection{DLRM NN Partitioning}
\label{sec:DLRM_part}
Differently from Megatron, the goal of the DLRM partitioner is to identify the different individual training iteration times for models with different numbers of parameters. The partitioner takes as input the total number of parameters, the number of maximum workers and additional DLRM model architecture features such as the number of sparse and dense features, hidden size, number of tables et cetera. These parameters are fed to the partitioning blocks which will partition the model such that the overall memory utilisation of the GPU is maximised. The partitioning strategy used follows the 3D partitioning strategy for DLRM described in \cite{mudigere2022DLRM_partitioning}, in terms of table-wise, column-wise and data parallelism. For embedding tables, the partitioning method first prioritises the table-wise parallelism, as it has better activation and input data efficiency, and if greater partitioning (due to memory constraints) is required implements column-wise parallelism. For the MLP layers, data parallelism is used and depending on the available memory the local batch size is selected. 

In addition to the equivalent model, the partitioning generates a custom data-loader for the specific partitioning. If a simple distributed data-loader would be used, before the actual data feeding process, an all-to-all (with possible replication) operation would be needed for the categorical features between workers for any tensor-partitioned mode such that it could be correctly used by the equivalent model. The custom data-loader generates random batches with the correct data attributes and dimensions for a particular equivalent model such that only necessary computation and communication are included for communication and computation time analysis.

As for Megatron, the DLRM partitioned model requires multiple collective communications. Both table and column-wise parallelism require forward and backward pass all-to-all collective communication, with message size dictated by the hidden dimension, local batch size and parallelism level. The data parallelism for the dense computations requires gradient all-reduce. The collective operations are then fed as inputs to the MPI estimator.

The equivalent model and the partitioned custom data-loader are fed as input to the profiler to calculate the training step computation time. A diagram describing the partitioning procedure is shown in Fig.\ref{fig:NN_partitioner}.

The DLRM-3D partitioning library and the corresponding tested models for the results have been made public at \cite{dlrm_part} and \cite{dlrm_data} respectively.

\subsection{NN Profiler}
\label{sec:NN_Profiler}
To estimate the computation time of a training step for either Megatron or DLRM we use the built-in Pytorch Profiler. Each model is profiled for a minimum of 150 training iterations on a NVidia A100 GPU to estimate the GPU computation time correctly. 

For Megatron, a single-layer equivalent transformer block model has been profiled, and the results are generalised for the total depth of the model. The training has been performed under mixed precision \cite{mixed_precision_training_2017}, using Adam optimiser \cite{adam} and implementing activation checkpointing \cite{activation_checkpointing_2016, megatron-lm_2019} and activation offloading \cite{zero-offloading_2021} to represent large scale models. The activation checkpointing forces the model to recompute the forward pass before calculating the backward pass, incurring additional communication. A random-data data-loader has been selected which generates random embedding vectors as input to the transformer block. It needs to be noticed that the sequence length selected for all trained models is 1024 tokens.

For DLRM, the equivalent data-tensor partitioned model is profiled assuming half-precision and sparse SGD optimiser. As a data ingestion mechanism, the custom data-loader (described in sec.\ref{sec:DLRM_part}) for each model is used. The profiled models are available at \cite{mp_encoder_data, dlrm_data}.

\begin{figure}[b]
    \centering
    \includegraphics[width=\linewidth]{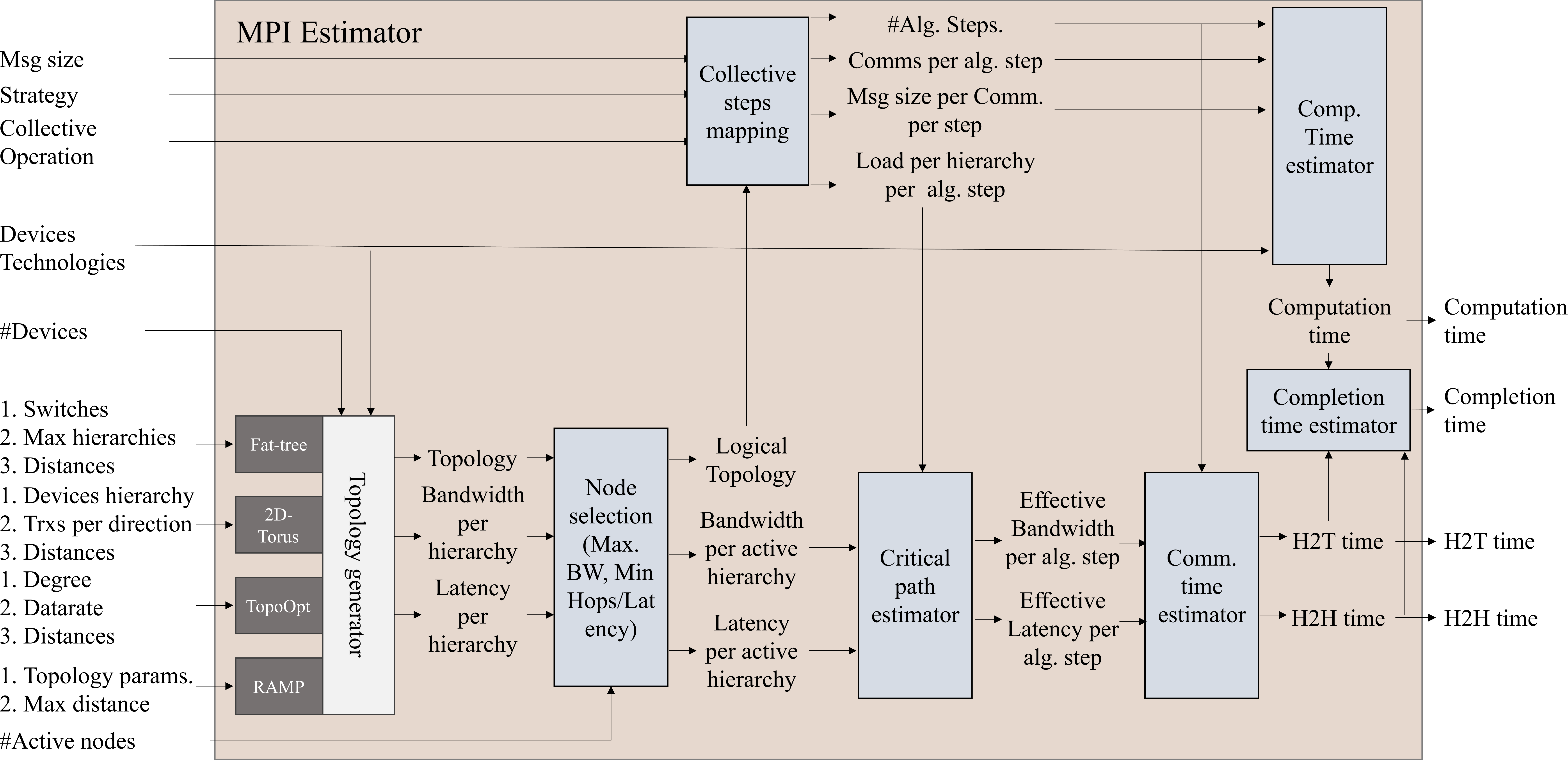}
    \caption{Block diagram of the MPI Estimator}
    \label{fig:MPI Estimator}
\end{figure}
\subsection{MPI Estimator}
\label{sec:MPI Estimator}
To evaluate the collective completion time of different MPI collective operations using different strategies on different systems a custom MPI Estimator block is developed and made accessible at \cite{ramp_library}. The block diagram of the MPI estimator can be seen in Fig.\ref{fig:MPI Estimator}. 

The first step of the MPI estimator is to generate a physical topology for the given number of workers, device technologies (\#Trxs per device, BW per Trx ...) and architecture-specific parameters. The Fat-Tree architecture requires information about the switch type per hierarchy, oversubscription per hierarchy, the maximum number of hierarchies and distances between hierarchies, in such a way topology with the corresponding characteristics and number of nodes is generated. For 2D-Toruses and Ring topologies, the information required is the maximum ring size in the first dimension, the number of transceivers and data-rate per direction and latency. For TopoOpt the communication degree latency and bandwidth per transceiver information are required. Whereas for the RAMP system, the topological parameters (\#Communition Groups, Racks, Nodes per Rack)  are required together with the maximum distance.

The topology generator creates the topology and provides information about the available capacity and effective latency at each hierarchy and path across the topology. This information together with the number of active nodes is fed to a node selection block which selects which of the workers in the physical topology will be used for the job. The nodes are selected in a greedy fashion such that high-bandwidth interconnected nodes are prioritised and at bandwidth parity, the lowest overall latency is minimised. This means that in oversubscribed Fat-Tree systems, the intra-node device utilisation will be maximised while trying to minimise the number of hops (logical diameter) between the furthest nodes. For 2D-Torus leads to choosing when possible only connectivity in the highest bandwidth direction and when needed minimising the logical diameter. For TopoOpt it consists of minimising the number of logical circuits needed such that the effective degree is one and the bandwidth for collective operation is maximised. In RAMP, as it is a single hop full-bandwidth all-to-all topology, the nodes have been selected such that the minimum number of algorithmic steps is minimised.
It needs to be noticed that the proposed approach only considers a single job and the placement approach would not be ideal for multi-job scenarios.

Once the active nodes have been selected, the logical topology is found and the maximum available bandwidth and latency for each active hierarchy and path are measured.
The logical circuit information is then used to map the devices to specific collective operation strategy steps. The mapping is made such that, for collectives and topologies which allow it, the largest message is transmitted between nodes which communicate at the highest effective bandwidth and the smallest to the ones with the lowest data rate. At parity of bandwidth, the steps are selected such that the minimum number of communication steps are used for the highest propagation latency paths. After the mapping, the information about the number of algorithmic steps, number of communication steps per algorithmic step, message size per communication step and the effective load (number of devices using same switching resources) per active hierarchy is found. 

The effective load information and the bandwidth and latency per active hierarchy are then used to estimate the critical path for each algorithmic step, which allows us to find the effective bandwidth and latency per algorithmic step. A more detailed explanation of what is considered in the critical path estimator can be found in sec.\ref{sec:critical_path}.

Using the information from the collective mapping and critical mapping estimator the propagation, idling and processing latency components, head-to-head time (H2H), and data-transfer components, head-to-tail time (H2T) are calculated, while at the same time the computation time is estimated. Using all this information the overall completion time can be found.

The MPI estimator developed has been modelled on and validated by experiments performed using the Cambridge Wilkes2 GPU-Cluster running the NVidia communication benchmark NCCL Tests \cite{NCCL, NCCL_Tests}, which performs operations using ring strategy and verified with respect to previous literature \cite{Evaluating-GPU-interconnects}. The experiments have been performed on up to 4 nodes (16 workers) using both NVidia Pascal P100 and Ampere A100 GPUs. The proposed estimator gives a lower bound for the collective communication as it assumes ideal switching, computing and load characteristics.

\begin{figure}[t]
    \centering
    \includegraphics[width=\linewidth]{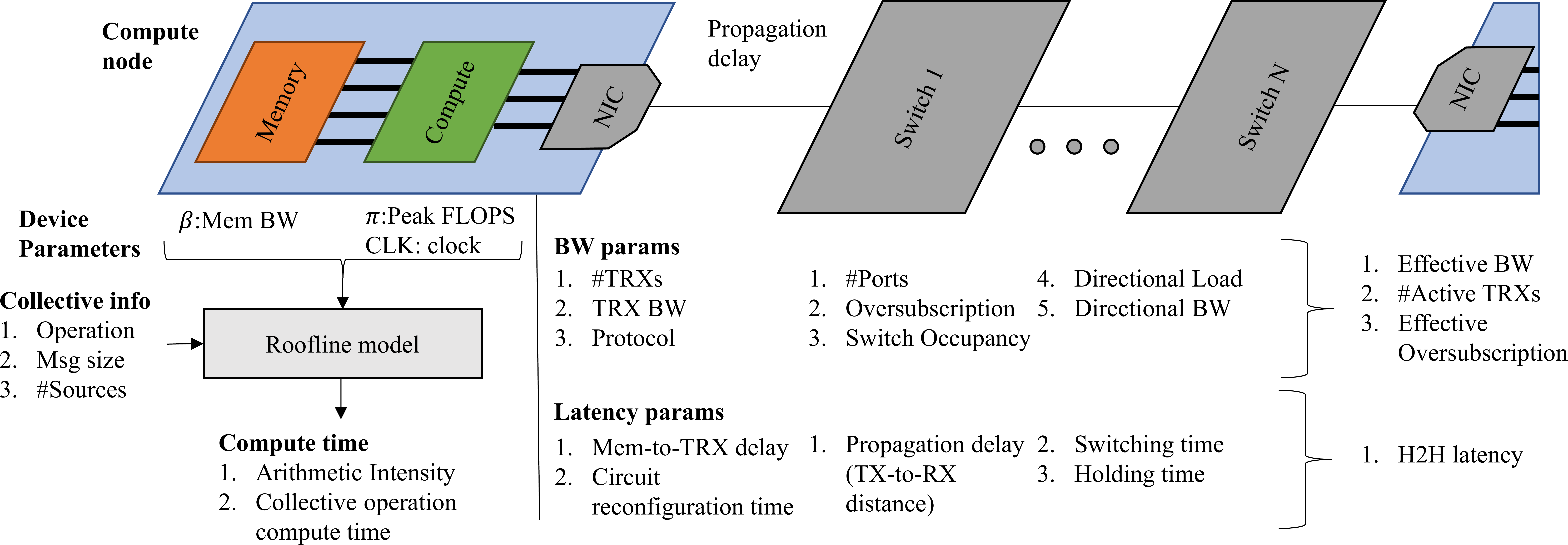}
    \caption{Diagram of critical path}
    \label{fig:critical_path}
\end{figure}
\subsubsection{Critical Path estimator}
\label{sec:critical_path}
To evaluate the effective bandwidth and latency for each algorithmic step of any collective operation strategy, the properties of the critical path, the worst performing path in either term of data-rate or propagation delay, needs to be found. In fact, due to the synchronicity and pipelining of collective strategies, the effective communication parameters will be dictated by the worst-performing link. 

A visualisation of the components taken into consideration when measuring the critical path is shown in Fig.\ref{fig:critical_path}. There are four main components considered for the calculation: the compute node, the network interface card (NIC), the interconnect and the switches.
The compute node is used to estimate the compute time of a collective operation step and the intra-node latency. For the compute time calculation the memory bandwidth ($\beta$), peak computational intensity ($\pi$) and clock period ($clk$) information from the compute node are used as parameters of the roofline model, which depending on the operation, number of sources and message size of the collective operation estimates the arithmetic intensity of the operation and the collective step computation time. Between the compute node and NIC we consider the memory-to-transceiver delay, which represents the time taken for the first bit of data to be transmitted from the memory and pushed out of the transceiver or vice-versa. 

The network components (NIC and switches) parameters can be separated into the ones related to bandwidth and latency calculations. 
To evaluate the effective bandwidth the parameters considered for the NIC are the number of available transceivers, the bandwidth per transceiver and the communication protocol used. For each switch, for EPS systems, multiple parameters are taken into account, such as the number of ports per direction, oversubscription ratio, switch occupancy, directional load (number of requests from and to each destination hierarchy) and directional bandwidth (bandwidth per transceiver per destination hierarchy). Each algorithmic step will involve multiple nodes sharing the same network resources by assigning a path for each source-destination pair. Each request that uses the same switches will determine the switch occupancy. By measuring the number of directional requests, the number of available ports, their data rate and oversubscription, the effective number of active input transceivers per switch is found. This process is repeated for all transceivers in a hierarchical fashion such that the effective network oversubscription and the number of active transceivers per NIC are found. By using the effective number of active transceivers, bandwidth per transceiver and protocol, the effective bandwidth per algorithmic step is found. For a RAMP system, the effective bandwidth and number of active transceivers is found using the formulation described in sec.\ref{sec:network_transcoder}.

For latency calculation, NIC parameters considered are the memory-to-transceiver delay and circuit reconfiguration time for OCS systems. For network components, we consider a propagation delay for the links and switching and holding time for the switches. The effective latency will be the sum of all these contributions across all network parameters.

\subsection{Networks comparison}
\label{sec:eps}
\noindent To analyse the performance of the RAMP algorithms and architecture, we compare them to electrical packet and optical circuit switching baselines. 

As a baseline for EPS, we use a Fat-Tree Topology inspired by the Nvidia DGX-A100 SuperPod~\cite{superpod} architecture which we scale to match the number of nodes of RAMP (65,536). To do so, the Fat-Tree hierarchy has been increased to a 4 tier system. The Super-Pod architecture can be viewed as a heterogeneous network,
where intra-node and inter-node communications are handled by NVLink~\cite{nvswitch} and InfinBand~\cite{mellanox_infiniband} interconnects, respectively.

This heterogeneity in links is also reflected in bandwidth, allowing intra-node GPUs unidirectional bandwidth of up to 2.4Tbps and inter-node of up to 200Gbps, leading to an effective over-subscription ratio of 1:12 between inter-node and intra-node communication. However, to fairly compare the performance of the algorithm we assume a network with 1:1 subscription ratio.

For every communication between pair of nodes, we assume the shortest path is taken and assume deterministic switching latency equal to the minimum switching time. The intra-node switching time (NVSwitch~\cite{nvswitch}) is considered to be 100$ns$ and each inter-node switch (NVIDIA Quantum QM8790~\cite{mellanox_technologies_qm8790_2021}) latency equal to 350$ns$. The intra-node propagation latency is considered to be $20ns$ and the latency between tiers $10ns, 50ns, 1.25\mu s$ respectively.

For both algorithmic and system comparison, the propagation latency between nodes in the proposed architecture is considered to be $1.3\mu s$. On the other hand, the total unidirectional capacity per node is considered 2.4Tbps for the algorithmic comparison and 12.8Tbps for the system analysis.

For completeness, even though it does not allow all-to-all communication, we also compare the system and strategies to a 2D-Torus topology, assuming a total node capacity of 2.4Tbps. In this case, we assume 128 and 512 nodes per dimension, with worst-case propagation latency of $156ns, 520ns$ respectively.

We compare RAMP with \textsc{TopoOpt} \cite{TopoOpt_2022} a recently proposed OCS network for DDL. We scale \textsc{TopoOpt} to match the number of nodes (above actual scalability of 384 direct node connectivity) and we assume the 3D-MEMs-based system to allow circuit reconfiguration (necessary for full data-rate hybrid parallelism in DDL). We assume a data-rate of 1.6Tbps per node (maximum considered in \cite{TopoOpt_2022}) and a maximum latency between nodes of 260ns (when the circuit has been established). For applications and collective completion time analysis, we assume already defined circuits and discard any additional contribution due to the circuit reconfiguration time.

Independently of the architecture, the minimum in-out latency per node (intra-GPU) is considered to be 100$ns$. To simulate the computation time, we assume for all topologies a Nvidia A100 GPU node~\cite{A100} following the roofline model~\cite{roofline}.

\subsection{MPI strategies comparison}
\label{sec:mpi_strat_comp}
\begin{figure}
    \centering
    \includegraphics[width=\linewidth]{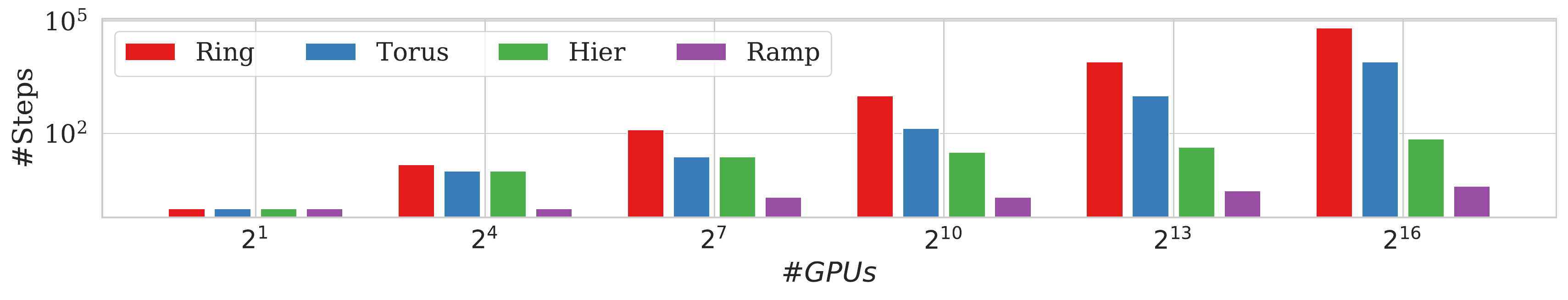}
    \caption{Number of algorithmic steps for different strategies at different scales for reduce-scatter operation.}
    \label{fig:n_steps}
\end{figure}

\noindent We compare the proposed strategy and topology with electrical counterparts running Ring based collective operation algorithms. The Ring-based strategies have been chosen because of their popularity in distributed deep learning operations as they are implemented by the Nvidia NCCL library~\cite{NCCL}. 

For each collective operation, we consider multiple strategies based on logical rings implemented in literature. Apart from the Ring all-reduce based operations~\cite{ring_reduce}, we include operations derived from hierarchical counterparts, which are 2D-Torus~\cite{torus_allreduce} and Hierarchical Ring all-reduce~\cite{hierarchical_allreduce}. For each of the strategies, the inner steps of the operations have been modified to accommodate all MPI collectives. 

For all MPI operations, apart from the reduce, the same number of data is transferred by the hierarchical counterparts as in the single ring-based collectives. However, in the hierarchical collectives, the number of algorithmic steps is reduced from being dependent on the total number of nodes to the number of active nodes in each dimension of the hierarchy. 

This reduction can be seen in Fig.\ref{fig:n_steps}, where the number of steps required by all strategies to complete a scatter-gather operation is shown as the number of active nodes in the system is increased.

For each collective, we place the devices such that the minimum number of tiers are involved for any communication and head-to-head latency (H2H: propagation, setup and  I/O latency) and head-to-tail latency (H2T: data transfer time) per communication is minimised. 

It is important to note that for \textsc{TopoOpt} only single ring-based strategies can be considered as the large circuit reconfiguration time ($>$10ms) would lead to significant overhead in communication time. In fact, for this topology, the logical connectivity is selected before application runtime and never changed till the job completion time is reached. Due to the static approach of the proposed architecture, the ring-based algorithms are the most suitable and best performing as maximum bandwidth communication can be achieved without incurring additional network overhead due to circuit and hardware reconfiguration time (for e.g. when using recursive doubling or hierarchical ring algorithms).

\section{Large-Scale Simulation Results}
\label{result}

\subsection{Distributed Deep Learning Training}

\begin{figure}[t]

    \centering
    \includegraphics[width=\linewidth]{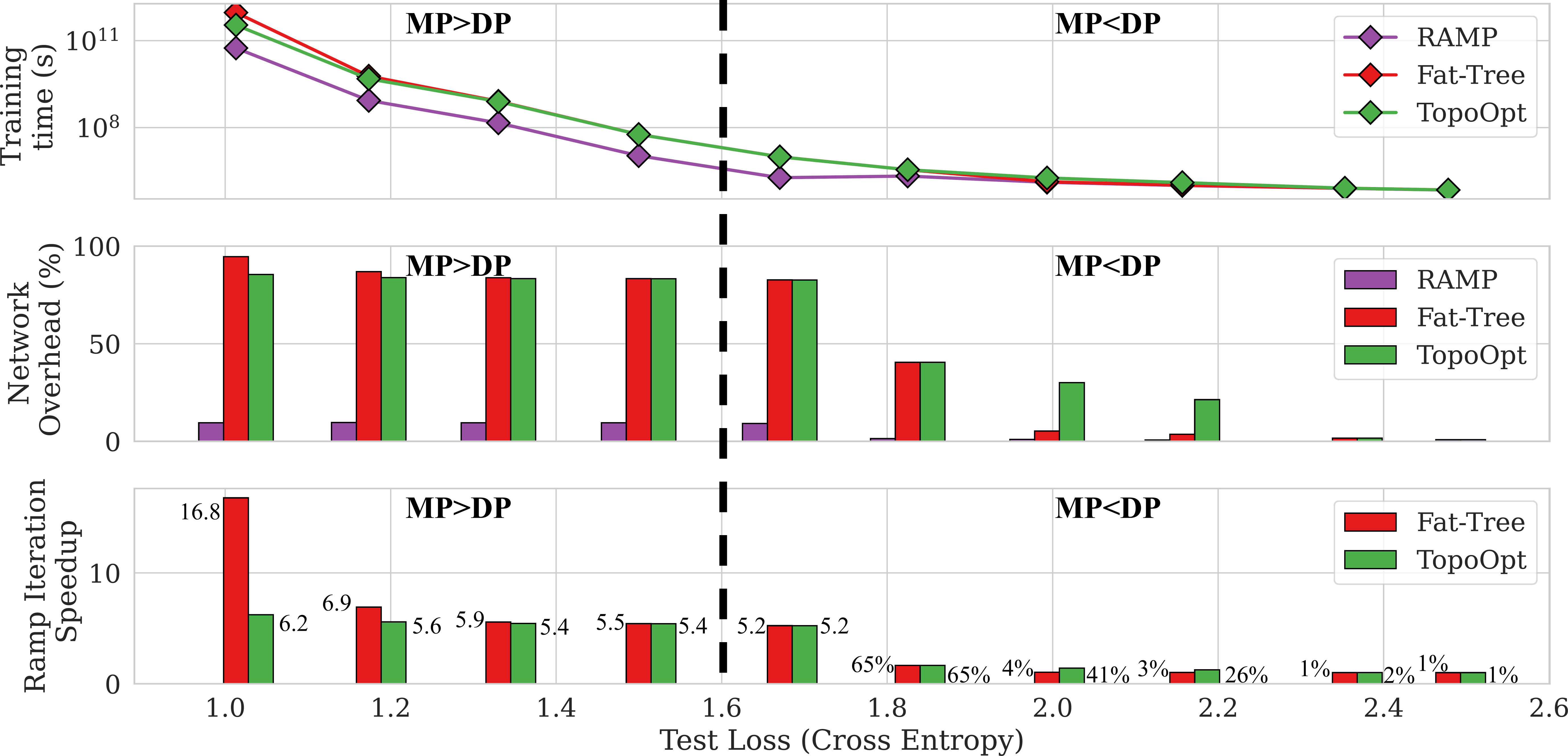}

    \caption{Training time (lines), communication contribution (bars in the second plot) and equivalent RAMP speed-up (bars third plot) for distributed Transformer encoder models with different target loss. The numbers on the left and right sides of each bar of the speed-up plot represent the speed-up values for the RAMP system in respect of Fat-Tree and TopoOpt, respectively. }
    \label{fig:training}

\end{figure}

\begin{sidewaystable}
\sidewaystablefn%
\begin{center}
\begin{minipage}{\textheight}
\begin{tabular}{|l|l|l|l|l|l|l|l|l|l|l|l|l|}
\hline
\multicolumn{1}{|c|}{\cellcolor[HTML]{EFEFEF}} &
  \multicolumn{1}{c|}{\cellcolor[HTML]{EFEFEF}} &
  \multicolumn{1}{c|}{\cellcolor[HTML]{EFEFEF}} &
  \multicolumn{1}{c|}{\cellcolor[HTML]{EFEFEF}} &
  \multicolumn{1}{c|}{\cellcolor[HTML]{EFEFEF}} &
  \multicolumn{1}{c|}{\cellcolor[HTML]{EFEFEF}} &
  \multicolumn{1}{c|}{\cellcolor[HTML]{EFEFEF}} &
  \multicolumn{1}{c|}{\cellcolor[HTML]{EFEFEF}} &
  \multicolumn{1}{c|}{\cellcolor[HTML]{EFEFEF}} &
  \multicolumn{1}{c|}{\cellcolor[HTML]{EFEFEF}} &
  \multicolumn{1}{c|}{\cellcolor[HTML]{EFEFEF}} &
  \multicolumn{1}{c|}{\cellcolor[HTML]{EFEFEF}} &
  \multicolumn{1}{c|}{\cellcolor[HTML]{EFEFEF}} \\
\multicolumn{1}{|c|}{\multirow{-2}{*}{\cellcolor[HTML]{EFEFEF}\begin{tabular}[c]{@{}c@{}}CE\\ Loss\end{tabular}}} &
  \multicolumn{1}{c|}{\multirow{-2}{*}{\cellcolor[HTML]{EFEFEF}\begin{tabular}[c]{@{}c@{}}Embed.\\ Dim.\end{tabular}}} &
  \multicolumn{1}{c|}{\multirow{-2}{*}{\cellcolor[HTML]{EFEFEF}\#Heads}} &
  \multicolumn{1}{c|}{\multirow{-2}{*}{\cellcolor[HTML]{EFEFEF}\#Lrs}} &
  \multicolumn{1}{c|}{\multirow{-2}{*}{\cellcolor[HTML]{EFEFEF}\#Steps}} &
  \multicolumn{1}{c|}{\multirow{-2}{*}{\cellcolor[HTML]{EFEFEF}\begin{tabular}[c]{@{}c@{}}Global\\ batch\end{tabular}}} &
  \multicolumn{1}{c|}{\multirow{-2}{*}{\cellcolor[HTML]{EFEFEF}\#Params}} &
  \multicolumn{1}{c|}{\multirow{-2}{*}{\cellcolor[HTML]{EFEFEF}\begin{tabular}[c]{@{}c@{}}\#Params \\ per GPU\end{tabular}}} &
  \multicolumn{1}{c|}{\multirow{-2}{*}{\cellcolor[HTML]{EFEFEF}\#GPUs}} &
  \multicolumn{1}{c|}{\multirow{-2}{*}{\cellcolor[HTML]{EFEFEF}\begin{tabular}[c]{@{}c@{}}DP \\ level\end{tabular}}} &
  \multicolumn{1}{c|}{\multirow{-2}{*}{\cellcolor[HTML]{EFEFEF}\begin{tabular}[c]{@{}c@{}}MP\\ level\end{tabular}}} &
  \multicolumn{1}{c|}{\multirow{-2}{*}{\cellcolor[HTML]{EFEFEF}\begin{tabular}[c]{@{}c@{}}DP \\ msg\end{tabular}}} &
  \multicolumn{1}{c|}{\multirow{-2}{*}{\cellcolor[HTML]{EFEFEF}\begin{tabular}[c]{@{}c@{}}MP \\ msg\end{tabular}}} \\ \hline
 &
   &
   &
   &
   &
   &
   &
   &
   &
   &
   &
   &
   \\
\multirow{-2}{*}{2.5} &
  \multirow{-2}{*}{1152} &
  \multirow{-2}{*}{12} &
  \multirow{-2}{*}{36} &
  \multirow{-2}{*}{65.6K} &
  \multirow{-2}{*}{2480} &
  \multirow{-2}{*}{574M} &
  \multirow{-2}{*}{574M} &
  \multirow{-2}{*}{16} &
  \multirow{-2}{*}{16} &
  \multirow{-2}{*}{1} &
  \multirow{-2}{*}{1.14GB} &
  \multirow{-2}{*}{0B} \\ \hline
 &
   &
   &
   &
   &
   &
   &
   &
   &
   &
   &
   &
   \\
\multirow{-2}{*}{2.4} &
  \multirow{-2}{*}{1536} &
  \multirow{-2}{*}{16} &
  \multirow{-2}{*}{40} &
  \multirow{-2}{*}{70.5K} &
  \multirow{-2}{*}{3424} &
  \multirow{-2}{*}{1.13B} &
  \multirow{-2}{*}{1.13B} &
  \multirow{-2}{*}{32} &
  \multirow{-2}{*}{32} &
  \multirow{-2}{*}{1} &
  \multirow{-2}{*}{2.27GB} &
  \multirow{-2}{*}{0B} \\ \hline
 &
   &
   &
   &
   &
   &
   &
   &
   &
   &
   &
   &
   \\
\multirow{-2}{*}{2.2} &
  \multirow{-2}{*}{2304} &
  \multirow{-2}{*}{24} &
  \multirow{-2}{*}{56} &
  \multirow{-2}{*}{78.9K} &
  \multirow{-2}{*}{4896} &
  \multirow{-2}{*}{3.57B} &
  \multirow{-2}{*}{893M} &
  \multirow{-2}{*}{128} &
  \multirow{-2}{*}{32} &
  \multirow{-2}{*}{4} &
  \multirow{-2}{*}{1.78GB} &
  \multirow{-2}{*}{150MB} \\ \hline
 &
   &
   &
   &
   &
   &
   &
   &
   &
   &
   &
   &
   \\
\multirow{-2}{*}{2} &
  \multirow{-2}{*}{4096} &
  \multirow{-2}{*}{32} &
  \multirow{-2}{*}{50} &
  \multirow{-2}{*}{87.5K} &
  \multirow{-2}{*}{7168} &
  \multirow{-2}{*}{10.1B} &
  \multirow{-2}{*}{1.2B} &
  \multirow{-2}{*}{512} &
  \multirow{-2}{*}{64} &
  \multirow{-2}{*}{8} &
  \multirow{-2}{*}{2.52GB} &
  \multirow{-2}{*}{268MB} \\ \hline
 &
   &
   &
   &
   &
   &
   &
   &
   &
   &
   &
   &
   \\
\multirow{-2}{*}{1.8} &
  \multirow{-2}{*}{6144} &
  \multirow{-2}{*}{64} &
  \multirow{-2}{*}{71} &
  \multirow{-2}{*}{98.1K} &
  \multirow{-2}{*}{10880} &
  \multirow{-2}{*}{32.2B} &
  \multirow{-2}{*}{1B} &
  \multirow{-2}{*}{2048} &
  \multirow{-2}{*}{64} &
  \multirow{-2}{*}{32} &
  \multirow{-2}{*}{2.01GB} &
  \multirow{-2}{*}{402MB} \\ \hline
 &
   &
   &
   &
   &
   &
   &
   &
   &
   &
   &
   &
   \\
\multirow{-2}{*}{1.7} &
  \multirow{-2}{*}{8192} &
  \multirow{-2}{*}{128} &
  \multirow{-2}{*}{128} &
  \multirow{-2}{*}{111k} &
  \multirow{-2}{*}{16896} &
  \multirow{-2}{*}{103.1B} &
  \multirow{-2}{*}{811M} &
  \multirow{-2}{*}{32768} &
  \multirow{-2}{*}{256} &
  \multirow{-2}{*}{128} &
  \multirow{-2}{*}{1.62GB} &
  \multirow{-2}{*}{1.11GB} \\ \hline
 &
   &
   &
   &
   &
   &
   &
   &
   &
   &
   &
   &
   \\
\multirow{-2}{*}{1.5} &
  \multirow{-2}{*}{16384} &
  \multirow{-2}{*}{512} &
  \multirow{-2}{*}{132} &
  \multirow{-2}{*}{191K} &
  \multirow{-2}{*}{14080} &
  \multirow{-2}{*}{425.2B} &
  \multirow{-2}{*}{843M} &
  \multirow{-2}{*}{65536} &
  \multirow{-2}{*}{128} &
  \multirow{-2}{*}{512} &
  \multirow{-2}{*}{1.69GB} &
  \multirow{-2}{*}{3.69GB} \\ \hline
 &
   &
   &
   &
   &
   &
   &
   &
   &
   &
   &
   &
   \\
\multirow{-2}{*}{1.3} &
  \multirow{-2}{*}{32768} &
  \multirow{-2}{*}{2048} &
  \multirow{-2}{*}{160} &
  \multirow{-2}{*}{3.7M} &
  \multirow{-2}{*}{1024} &
  \multirow{-2}{*}{2.06T} &
  \multirow{-2}{*}{1.03B} &
  \multirow{-2}{*}{65536} &
  \multirow{-2}{*}{32} &
  \multirow{-2}{*}{2048} &
  \multirow{-2}{*}{2.08GB} &
  \multirow{-2}{*}{2.15GB} \\ \hline
 &
   &
   &
   &
   &
   &
   &
   &
   &
   &
   &
   &
   \\
\multirow{-2}{*}{1.2} &
  \multirow{-2}{*}{131072} &
  \multirow{-2}{*}{8192} &
  \multirow{-2}{*}{52} &
  \multirow{-2}{*}{68M} &
  \multirow{-2}{*}{64} &
  \multirow{-2}{*}{10.7T} &
  \multirow{-2}{*}{1.35B} &
  \multirow{-2}{*}{65536} &
  \multirow{-2}{*}{8} &
  \multirow{-2}{*}{8192} &
  \multirow{-2}{*}{2.7GB} &
  \multirow{-2}{*}{2.15GB} \\ \hline
 &
   &
   &
   &
   &
   &
   &
   &
   &
   &
   &
   &
   \\
\multirow{-2}{*}{1} &
  \multirow{-2}{*}{262144} &
  \multirow{-2}{*}{65536} &
  \multirow{-2}{*}{90} &
  \multirow{-2}{*}{2.49B} &
  \multirow{-2}{*}{4} &
  \multirow{-2}{*}{74.2T} &
  \multirow{-2}{*}{1.27B} &
  \multirow{-2}{*}{65536} &
  \multirow{-2}{*}{1} &
  \multirow{-2}{*}{65536} &
  \multirow{-2}{*}{2.55GB} &
  \multirow{-2}{*}{2.15GB} \\ \hline
\end{tabular}
\caption{Parameters for the Megatron model}
\label{tab:megatorn}
\end{minipage}
\end{center}
\end{sidewaystable}

We analyse the capabilities of the proposed technologies for DDL training applications by comparing the training time of Megatron \cite{megatron-lm_2019} and DLRM models \cite{naumov2019DLRM} on different networks. We compare the proposed system and strategy with DGX-SuperPod-based Fat-Tree architecture with oversubscription (as described in sec.\ref{sec:eps}) and TopoOpt for equivalent scale and partitioning. The computation time of each model has been estimated as described in sec.\ref{sec:NN_Profiler} and partitioned as described in sec.\ref{sec:NN_partitioner}.

In Fig.\ref{fig:training}, we show the expected training time and communication time contribution for different target cross-entropy (CE) loss Megatron-based Transformer-encoder models \cite{transformer_2017}. The number of parameters, batch size and training steps for each model have been derived from the target loss following the laws described in \cite{scaling_language_models}.
The models have been trained using hybrid data parallelism (DP) and model parallelism (MP) without pipelining. The model partitioning strategy used is equivalent to the one described in \cite{megatron-lm_2019}.

The level of parallelism varies for each model (500M-566T parameters) and goes from DP:MP=16:1, through 256:128 to 1:65,536. The total number of devices used for each model is equal to $\text{MP}\times\text{DP}$.

The parameters and assumptions used for the tested models are described in Table.\ref{tab:megatorn} and the profiled models are accessible at \cite{mp_encoder_data}.





Fig.\ref{fig:training} shows a reduction in training time varying from a factor of 1.01-16.7$\times$. Two independent behaviours can be noticed, where the partitioning is DP dominant (CE$\geq$1.67) and where it is MP dominant (CE$<$1.67) and all 65K nodes are used. In the first case, the speed-up increases significantly (1.01-5.2$\times$) with a decrease in CE due to an increase in the number of devices used. This is due to the fact that by increasing the model parallel level and the number of devices used, the network overhead increases proportionally, till it becomes dominant (network overhead $>$ 50\%).
When CE$<$1.67, another exponential increase in speed-up is achieved. However, this happens at a much slower rate as the network overhead for most topologies increases only slightly (82-87\%) since the number of devices used stays constant (maximum scalability is achieved). The Fat-Tree topology is the only one that does not follow this behaviour due to the higher latency contribution (described in detail in sec.\ref{sec:architectural_comp}).

It is important to note that the improvement in overall training time is the same as the one for a single iteration. This has been calculated using real profiled computational data and the number of training steps and respective batch size were found using OpenAI's "law of scaling large language models" formulas \cite{scaling_language_models} (as described in sec.\ref{sec:NN_partitioner} and sec.\ref{sec:NN_Profiler}). 

\begin{figure}[b]

    \centering
    \includegraphics[width=\linewidth]{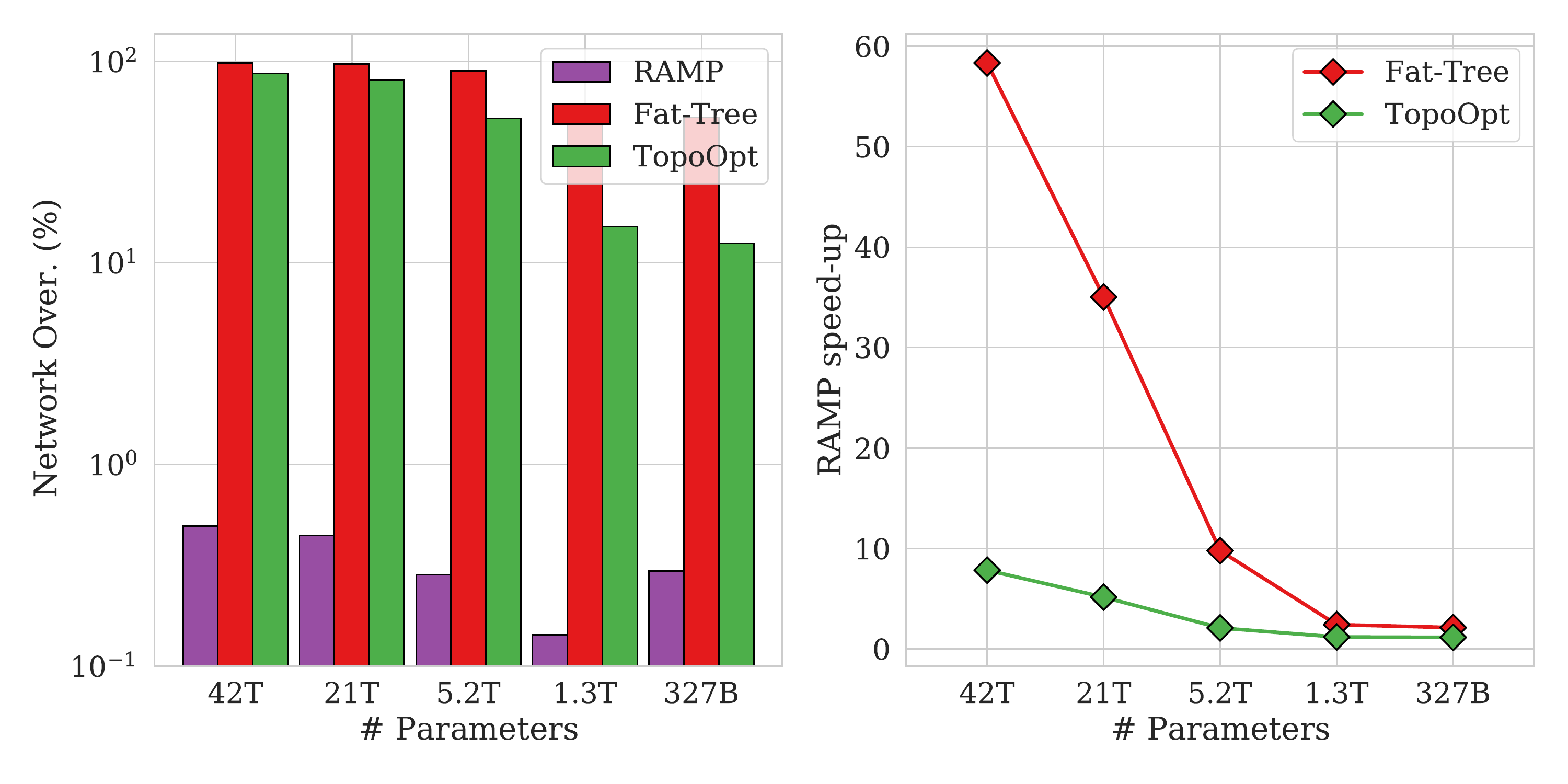}

    \caption{{Network overhead \% (bars) and RAMP offered training time reduction (lines) for different DLRM model sizes partitioned across  256 to 65k servers}}
    \label{fig:DLRM}

\end{figure}

\begin{sidewaystable}
\sidewaystablefn%
\begin{center}
\begin{minipage}{\textheight}
\begin{tabular}{|l|l|l|l|l|l|l|l|l|l|l|l|l|}
\hline
\rowcolor[HTML]{EFEFEF} 
\#GPUs &
  \#Tables &
  \#Rows &
  \begin{tabular}[c]{@{}l@{}}Sparse\\ Feature \\ size\end{tabular} &
  \begin{tabular}[c]{@{}l@{}}Part.\\ sparse\\ feature\\ size\end{tabular} &
  \begin{tabular}[c]{@{}l@{}}Batch \\ size\\ per \\ GPU\end{tabular} &
  \begin{tabular}[c]{@{}l@{}}Global \\ Batch \\ size\end{tabular} &
  \begin{tabular}[c]{@{}l@{}}Dense\\ feature\\ size\end{tabular} &
  \begin{tabular}[c]{@{}l@{}}MLPs\\ hidden \\ size\end{tabular} &
  \begin{tabular}[c]{@{}l@{}}\#Top\\ MLP\\ Layers\end{tabular} &
  \begin{tabular}[c]{@{}l@{}}\#Bottom\\ MLP\\ Layers\end{tabular} &
  \#Params &
  \begin{tabular}[c]{@{}l@{}}\#Part.\\ Params\end{tabular} \\ \hline
256   & 8   & 8E7    & 4096  & 128 & 8192 & 65536 & 16 & 1024 & 5 & 4 & 328B  & 1.3B \\ \hline
1024  & 16  & 1.6E8  & 8192  & 128 & 4096 & 65536 & 16 & 1024 & 5 & 4 & 1.3T  & 1.3B \\ \hline
4096  & 32  & 3.2E8  & 16384 & 128 & 3072 & 65536 & 16 & 1024 & 5 & 4 & 5.2T  & 1.3B \\ \hline
16384 & 128 & 1.28E9 & 16384 & 128 & 512  & 65536 & 16 & 1024 & 5 & 4 & 21T   & 1.3B \\ \hline
65536 & 256 & 2.56E9 & 16384 & 64  & 256  & 65536 & 16 & 1024  & 5 & 4 & 41.9T & 0.7B \\ \hline
\end{tabular}
\caption{Parameters DLRM model tested}
\label{tab:dlrm_tested}
\end{minipage}
\end{center}
\end{sidewaystable}

The speed-up is due to the smaller communication time contribution to the overall training time. The communication contribution of RAMP ranges between 7.7-34$\times$ smaller when compared to the other described systems and leads to an absolute change of 23.1-85 percentage points. As shown in figure \ref{fig:training}, RAMP's communication contribution is between 0.6-11\%.  For the other systems, the communication contribution ranges from 23.8-94.6\% effectively making the communication time the main bottleneck in training. If future xPUs would decrease the computation time, RAMP would also effectively decrease training time, whereas this would stay approximately constant for other systems. For example, a 2$\times$ speed-up in computation would lead to 1.76-1.94$\times$ and 1.02-1.6$\times$ reduction in training time for RAMP and the other networks respectively.

Similar behaviour can be noticed for DLRM \cite{naumov2019DLRM} models. Fig.~\ref{fig:DLRM} shows the expected training iteration completion time for DLRM models of different sizes partitioned onto 256-65K nodes using the partitioning strategies proposed in \cite{mudigere2022DLRM_partitioning}. It can be noted that RAMP achieves a speed-up in iteration time with a reduction factor of up to 7.8-58$\times$ when compared to \textsc{TopoOpt} and Fat-Tree, respectively. RAMP achieves this speedup while keeping a sub 1\% network overhead, whereas the OCS and EPS baselines suffer from overhead ranging from 12.5-87\%  and  52-98\% respectively. The low overhead for RAMP is due to the high bandwidth and fast reconfiguration of the topology which allow for efficient all-to-all collective operation (which dominates the data-transfer for this model).

The hyper-parameters and partitioning assumptions for the tested DLRM model are shown in Table.\ref{tab:dlrm_tested} and the profiling data for each of the models are accessible at \cite{dlrm_data}.



\subsection{Architectural comparison}
\label{sec:architectural_comp}

\begin{figure*}[t]

    \centering
    \includegraphics[width=\linewidth]{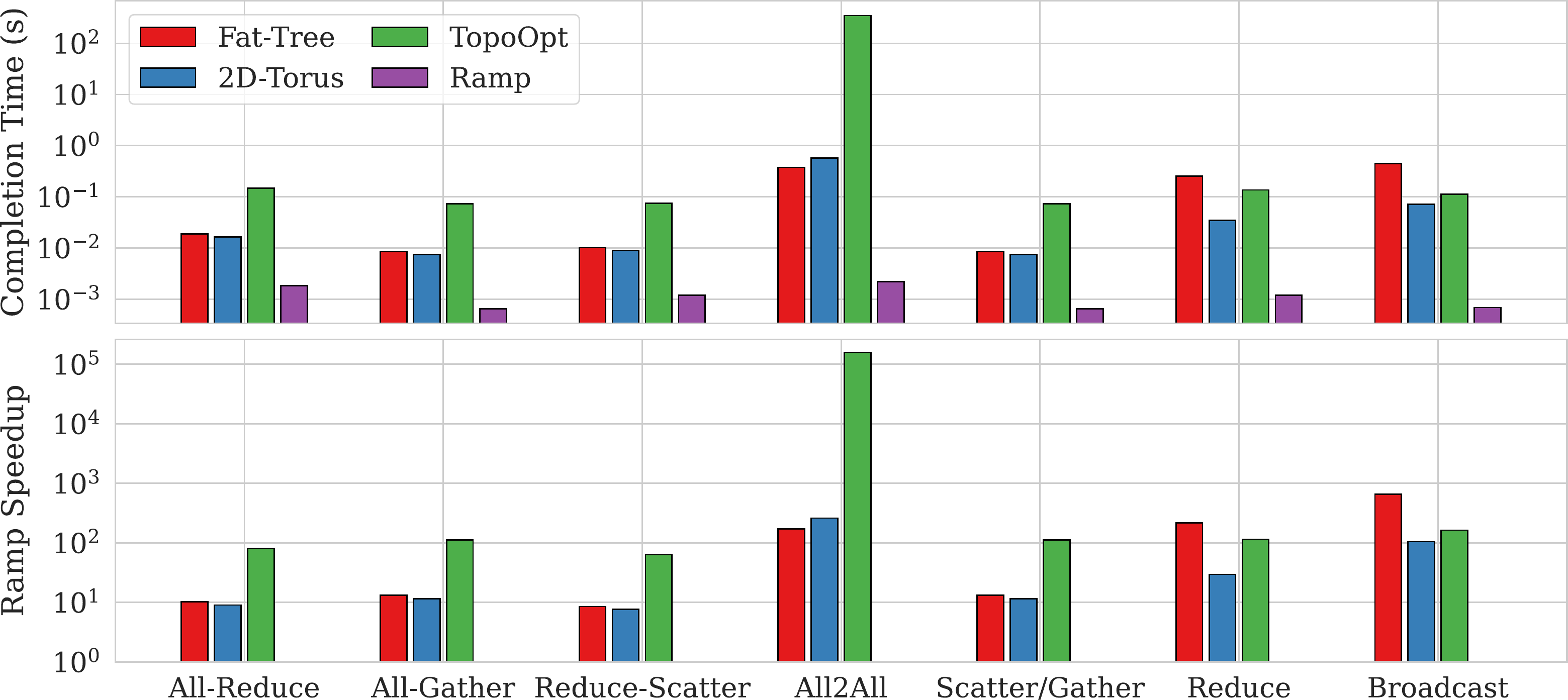}

    \caption{\centering {Time taken to perform collective operations with 1GB message using the best performing strategy on different systems and its corresponding RAMP speed-up at maximum scalability.}}
    \label{fig:collective_comp}

\end{figure*}

\noindent We compare the overall performance gain of the proposed  RAMP architecture and strategies at full potential with realistic electrical and optical systems described in sec. \ref{sec:eps}.

We perform a complete analysis of the speed-up for all the relevant MPI collective operations in Fig.\ref{fig:collective_comp}. This figure shows the total completion time for the best performing strategy in the baselines topologies and their RAMP equivalent for each collective operation and shows the minimum speed-up that RAMP achieves at maximum scale. RAMP offers a speed-up range of 7.6-171$\times$ (reduce-scatter and all-to-all) between different operations. This range in speed-up is due to the different data size transmitted in each algorithmic step between collective operations.
For the reduce-scatter operation, the data transmitted decreases with steps (or hierarchies for electrical systems), which reduces the over-subscription impact on the completion time due because the majority of data is transferred at high capacity between neighbouring devices. In contrast, for all-to-all, the data size stays constant with the steps and therefore, for electrical networks, it is limited by the lower data-rate between inter-rack/system connections. 
RAMP's reduce-scatter speed-up is smaller than other operations where data size decreases because the computation time reduction is smaller than the node capacity ratio ($2.8\times<5.3\times$), therefore, increasing the computation time contribution.

These properties show the capability of the proposed architectures and strategy to be employed for multiple HPC applications, leading to significant improvement.

\subsection{Architecture, MPI-strategy and bandwidth contributions}
\begin{figure*}[h]

    \centering
    \includegraphics[width=\textwidth]{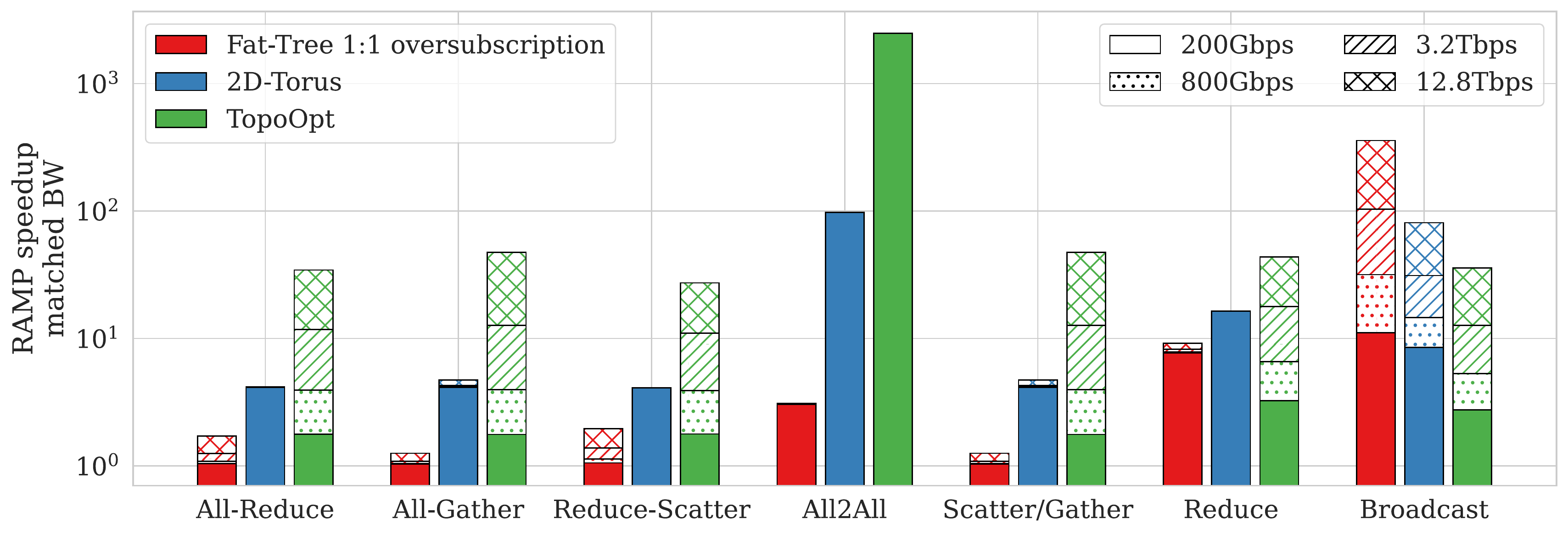}

    \caption{{Minimum RAMP speed-up for each collective operation at maximum scale in respect of the baselines when operating at matched node-to-node capacity and scale. The message size for each collective is 1GB. The best performing strategy for each system is chosen as comparison.}}
    \label{fig:matched_bw}
\end{figure*}

 We aim to disentangle the joint network architectural/MPI-strategy contributions and the effect of bandwidth on collective completion times. We compare RAMP to bandwidth matched (same node bandwidth from 0.2 to 12.8 Tbps and no over-subscription ratio) baselines described in sec.\ref{sec:eps}. Fig.\ref{fig:matched_bw} shows the speedup of the RAMP system and algorithms with respect to the best-performing strategy on all bandwidth-matched baseline systems. It can be noted that RAMP produces a speedup factor 1.04-2.24E3 $\times$ depending on collective and data-rate. Apart from All2All and Broadcast, a similar pattern can be seen on the RAMP speed-up against Fat-Tree and TopoOpt at low data-rate (200 Gbps) and the speed-up further increases when the data-rate increases. In fact, the speed-up increases up to 32$\times$ for Fat-Tree (Broadcast) and 27$\times$ for TopoOpt (All-Gather and Scatter) when the data-rate is increased from 200Gbps to 12.8Tbps. This is because the head-to-head (H2H) latency contribution becomes more significant at higher rates and the significantly reduced communication steps of RAMP lead to negligible overheads. It can be noticed that for the All-to-All collective, the speed-up (3-2.2E4$\times$) is purely due to algorithmic  efficiencies as it is bandwidth dominated. The best-performing baseline (disregarding Broadcast) is always Fat-Tree due its fast reconfiguration, which allows the use of more efficient strategies (hierarchical). However, as discussed in sec.\ref{EPS_limitation}sec.\ref{power_sec}, scaling Fat-Tree systems to such performances is unfeasible in terms of power and cost.

 \begin{figure*}[h]
    \centering
    \includegraphics[width=\linewidth]{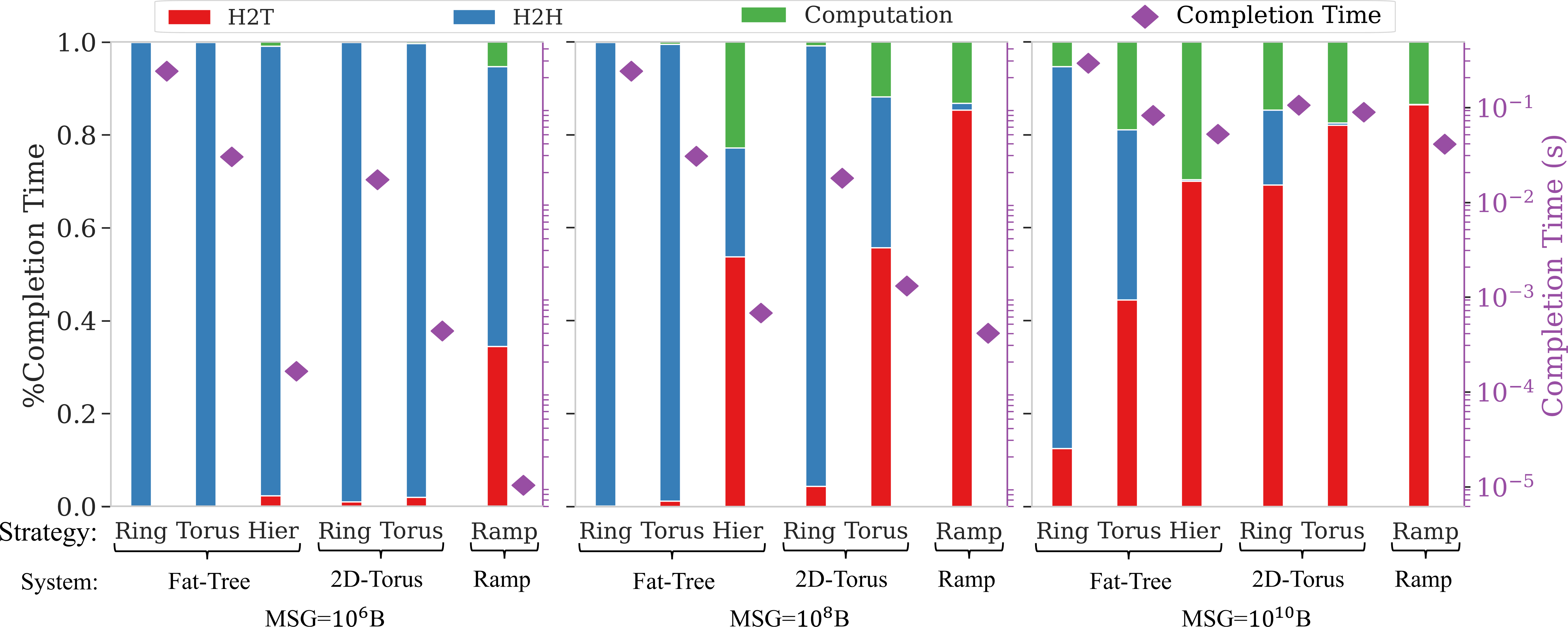}
    \caption{Analysis of how each individual component affects the total collective completion time for the all-reduce operation for different strategies, systems and message sizes.}
    \label{fig:individual_components}
\end{figure*}
Fig.\ref{fig:individual_components} shows all-reduce completion time as the breakdown percentage of individual contributors (H2T, H2H and computation time) and magnitude for the electrical architectures and RAMP with their corresponding collective strategies at maximum scale. It can be seen that the RAMP collective strategy on the proposed architectures outperforms the collectives on electrically switched networks with a speed-up ranging from 9-3.26$\times 10^4$, depending on the data-size and strategy.
At small message sizes ($<$10GB), the RAMP speed-up is larger due to its relatively smaller H2H latency contribution, which becomes insignificant with larger messages for all systems.
Even when the completion time is dominated by computation and H2T time for electrical counterparts (torus and hierarchical strategy on a 2D-Torus and Fat-Tree system respectively for 10GB message), the proposed architecture leads to a speed-up (7.3$\times$) above the total node-capacity ratio ($12.8\textrm{Tbps}/2.4\textrm{Tbps}=5.3\times$). This is due to the elimination of over-subscription in RAMP, which allows high-capacity communication between any set of devices. The reduction in computation time leads to significant improvement in completion time as the high data-rate leads to similar portions of H2T and compute time (44\% and 56\% respectively).

\subsection{Algorithmic comparison}
\begin{figure*}[h]
    \centering
    \includegraphics[width=\linewidth]{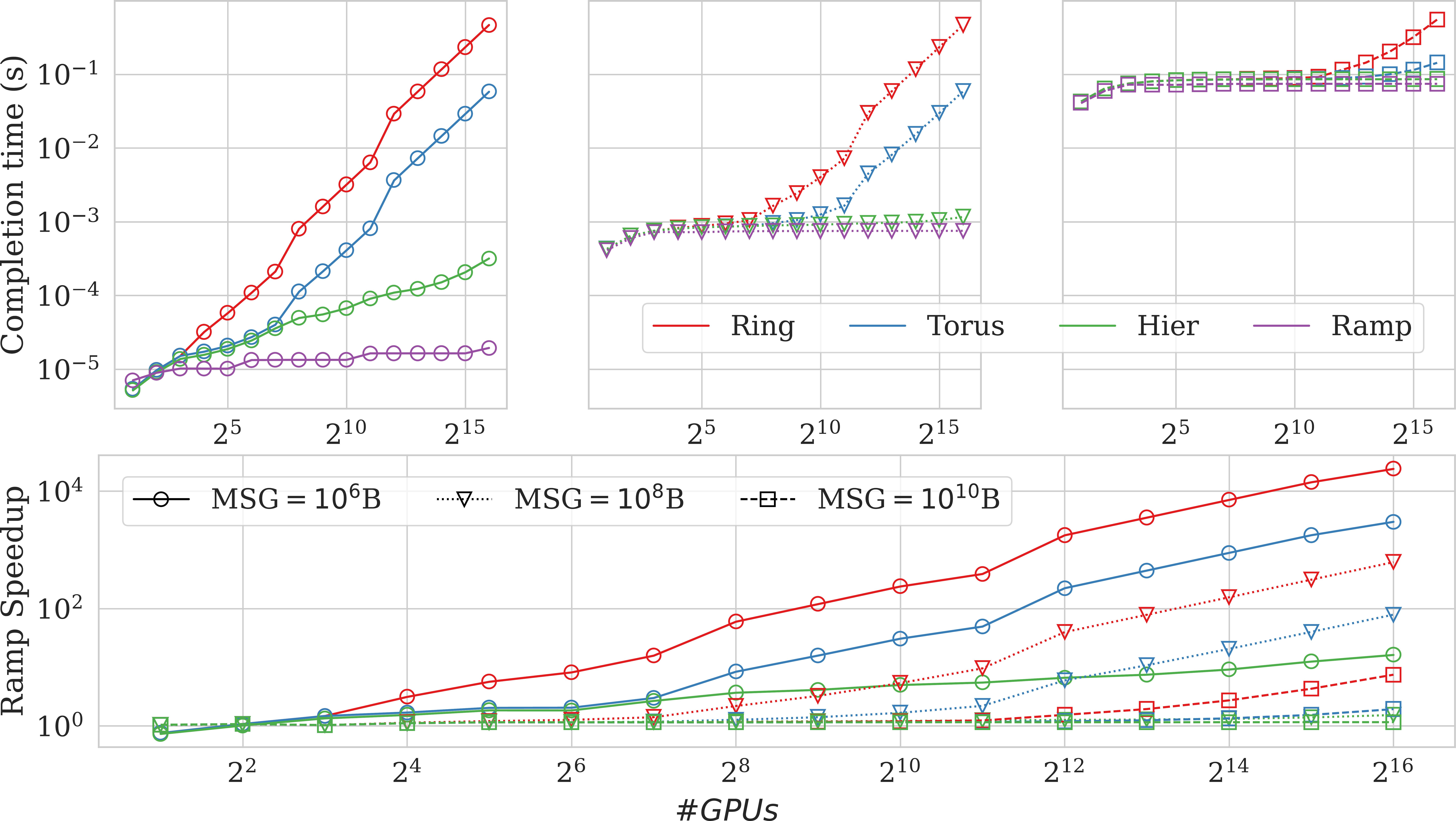}

    \caption{All-Reduce completion time with different number of GPUs and different message sizes, for Ring, 2D-Torus and Hierarchical strategies on Fat-Tree topology vs proposed strategy and architecture. The marker represents the message size and the colour the strategy. Dashed lines are only used to better differentiate curves. The bottom plot represents the completion time of each strategy and message size divided by the equivalent RAMP ones.}
    \label{fig:all-red-alg}
\end{figure*}

To showcase the algorithmic performance of the proposed strategies, we compare RAMP architectures with reduced node capacity (from the theoretical 12.8~Tbps down to 2.4~Tbps) to a Fat-Tree SuperPod-based architecture in which no oversubscription occurs between intra-DGX nodes and inter-DGX nodes (from 1:12 to 1:1). 

Fig.~\ref{fig:all-red-alg} shows that the proposed algorithm performs better than the ring-based counterparts. It can be noted that the speed-up increases with the number of devices involved, especially for small message sizes. 

There are two main factors which influence the speed-up of RAMP with respect to the ring-based counterparts: algorithmic steps required and computational speed-up.

\subsubsection{Algorithmic steps analysis}

\begin{figure}[h]
    \centering
    \includegraphics[width=\linewidth]{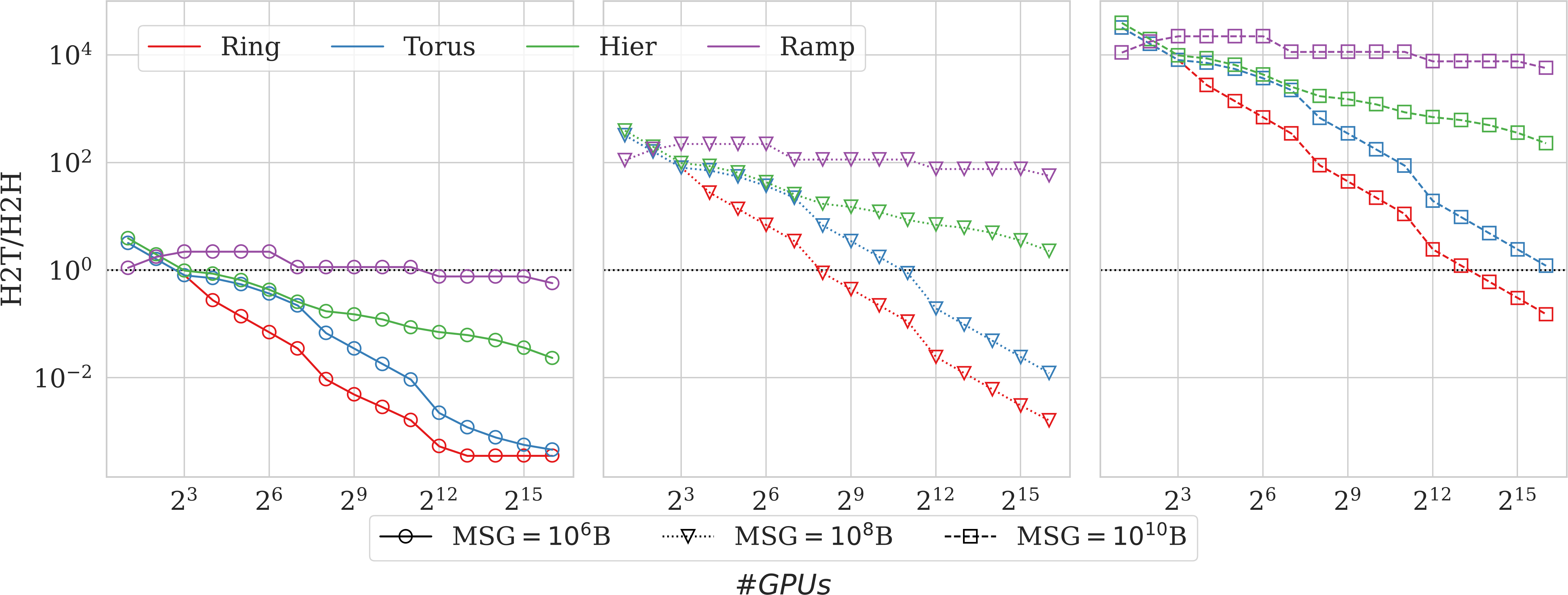}
    \caption{Head-to-Tail / Head-to-Head (H2T/H2H) ratio for an all-reduce operation for different scale and message size on Fat-Tree SuperPod using ring-based strategy and the RAMP system and strategy.}
    \label{fig:h2h_h2t}
\end{figure}


As shown in Fig. \ref{fig:all-red-alg}, 
RAMP's largest speed-up, with a factor ranging from 9-10,000$\times$, is with respect to the ring all-reduce at maximum scale.
For the hierarchical counterparts, the speed-up is reduced to a range varying from 1.16 to 10. This is due to the fact that the ring-based reduction time increases with the number of devices (at a different rate for each strategy), whereas the RAMP collective stays almost flat with the increase in size. This is a result of the different number of algorithmic steps required by each strategy.

From Fig. \ref{fig:n_steps}, it can be noted that the number of algorithmic steps for the ring collectives increases at a faster rate than the one for the RAMP collective. Each algorithmic step will incur additional H2H, which will affect the total completion time.  

\noindent The analysis of how the steps and H2H latency influence the collective completion time is shown in Fig \ref{fig:h2h_h2t} where the H2T/H2H latency ratio for an all-reduce operation using different message sizes varying the number of nodes is displayed. 

\noindent In Fig. \ref{fig:all-red-alg}, every point below $10^0$ corresponds to a collective operation which is H2H latency limited whereas everything above is data transfer limited. To minimise the collective operation time, the strategy should be data transfer limited.
The H2H latency becomes negligible when the H2T/H2H ratio is $>10$.
Fig. \ref{fig:all-red-alg} shows a similar behaviour between all curves for different messages. This trend is due to the fact that the H2H latency is independent of the message size but dependent on the number of algorithmic steps. In contrast, the opposite is valid for data transfer time, where increasing the data size, the starting point of the curves shifts upwards (data transfer limited region). Due to the larger number of algorithmic steps, the ring-based strategies (apart from the hierarchical with a message of 10GB), even though starting in a data transfer limited region, as the number of devices increases, the strategies find themselves in portions of the space where the H2H latency is not negligible. Differently, the small number of algorithmic steps in RAMP, keeps the ratio approximately constant with the number of nodes. 

For the largest message size, both the hierarchical and RAMP strategy is in the H2H latency negligible regime, minimising the gain of the RAMP collective. 

\begin{figure}[h]
    \centering
    \includegraphics[width=\linewidth]{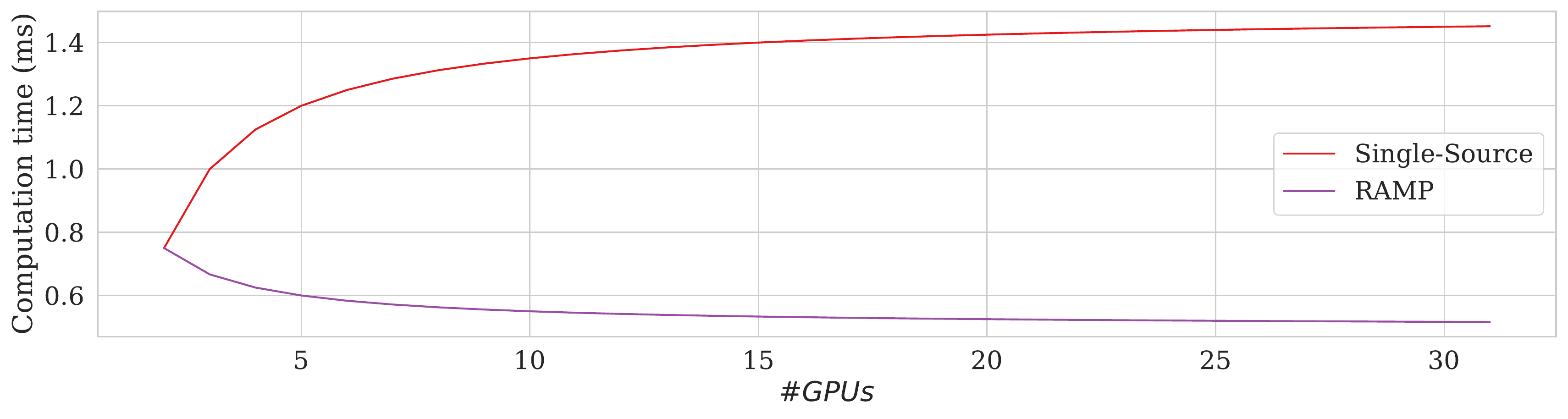}
    \caption{Computational time taken to sum 1GB of information scattered onto $\#GPUs$ workers, using algorithm having a single source and RAMP (multiple sources).}
    \label{fig:Compute-reduction}
\end{figure}

\subsubsection{Computational analysis}
It can be noted that even when the ring-based collectives are in data-transfer limited region the RAMP architecture still produces a speed-up of a factor above 16\% in respect of its ring-based counterpart as shown in Fig \ref{fig:all-red-alg}. This advantage is given not only by the smaller H2H contribution but also by a difference in the computation time of the reduce operation. 

Differently from algorithms that use a single source per step, the RAMP strategies receive information from up to $x-1$ devices, transforming the local operation from a 2-to-1 reduction to an x-to-1, leading to higher arithmetic intensity. This leads to a speedup factor of up to 2.8$\times$ considering the $x$ for maximum scale. The improvement is shown in Fig. \ref{fig:Compute-reduction} in which the computation time required to perform a reduce scatter operation on a message of 1~GB using $n$ workers is plotted for sequential algorithms and parallel RAMP network. Note that for all our calculations, we assume half-precision operations (16~bit) as it is widely used in DDL and the roofline model.

\section{Conclusion}

\noindent Current EPS networks for HPC and distributed deep learning systems are affected by over-subscription ratios between intra- and inter-rack/system nodes, effectively reducing the node-to-node capacity, and increasing MPI operations completion and job completion times. Scaling EPS systems to a large number of nodes and high node-to-node bandwidth leads to unsustainable energy consumption (306-400 MW for a 65k node system with an upper power budget of {\raise.17ex\hbox{$\scriptstyle\sim$}}30~MW). 

We proposed near-Exascale (Ebps) optical network architectures, RAMP, that provide high scalability (65,536 nodes), all-to-all connections with a single-hop, ultra-low and synchronous latency, while being highly energy efficient (8-9.5~pJ/bit/path or 6.4-8 MW total network power). It supports nanosecond-speed re-configuration ($<$20~ns) and high-node bandwidth (12.8~Tbps) as well as total network capacity (0.84 Ebps).
Moreover, the proposed networks significantly reduce the network cost to \$1.62-6.12/Gbit, offering a 6.4-26.5$\times$ improvement over matched-bandwidth and scale state-of-the-art EPS fat-tree networks. RAMP is capable of achieving 64 times higher capacity for 33-66\% increase in power consumption compared to EPS systems. Compared to EPS networks with similar cost, the proposed architecture achieves 10$\times$ higher capacity with a $\geq3.6\times$ lower power consumption.

We co-designed the collective operation strategies called RAMP-x (x=MPI operation type) and system-level reconfiguration algorithms to enable all MPI operations. These collective operations allow collective completion using up to 4 (8 for reduce and all-reduce) algorithmic steps even on large-scale systems, making the completion time almost scale-independent. We develop the RAMP Engine, which includes the MPI Engine and Network Transcoder, capable of handling the overall MPI process. The MPI Engine uses topological, application  MPI operation information to handle the data processing in a distributed fashion and generates instructions that will be used by the Network Transcoder to transmit data in the physical layer in a schedule-less and contention-less manner.

By evaluating the performance of our network and strategies, we show a speed-up of 1.16-10,000$\times$ compared to equivalent fully connected (no over-subscription) Fat-tree and Nvidia DGX-A100 SuperPod-based networks at matched node capacity. The combination of the proposed architecture and algorithms achieves 7.6-171$\times$ speed-up in completion times across all MPI operations with respect to realistic electronic and OCS network counterparts at maximum scale. 
Moreover, we compare the DDL training application performance using the proposed system against EPS and OCS baselines. We show that RAMP achieves 1.01-16.7$\times$ {and 7.8-58$\times$ speed-up in Transformer and DLRM model} training time and a reduction of 23.8-84 percentage points in communication time contribution, making the proposed system a suitable candidate to meet future DDL jobs requirements.

All opto-electronic devices and communication methods have been demonstrated in the past in lab experiments. Future work will focus on building a complete system.

\section{Declarations}
\textbf{Competing interests} The authors have no competing interests as defined by Springer, or other interests that might be perceived to influence the results and/or discussion reported in this paper.
\\
\textbf{Ethical Approval} Not applicable.
\\
\textbf{Funding} All authors are funded through the EPSRC OptoCloud grant with ID EP/T026081/1.
\\
\textbf{Authors Contribution} Alessandro Ottino wrote the main manuscript and Joshua Benjamin contributed to the writing of the optical components section. All figures were created by Alessandro Ottino, with the exception of figures 1 and 7, which have been adapted from the corresponding sources (refer to relevant figures) by Joshua Benjamin. All authors reviewed the paper.
\\
\textbf{Availability of data and materials} Code to reproduce the results for the MPI collectives are available at \cite{ramp_library} and additional data is available at \cite{Ottino2022_additional_data}. Code for the DDL simulations is available at \cite{mp_encoder,dlrm_part} and the profiled data at \cite{mp_encoder_data, dlrm_data}.

\section{Disclaimer}
Paper currently under consideration of SpringerNature "The Journal of Supercomputing".

\bibliography{reference} 



\end{document}